\documentclass[11pt,a4paper,twoside]{report}
\usepackage{graphicx}
\usepackage{amssymb}
\usepackage{amsmath}
\usepackage{fancyhdr}
\usepackage{psfrag}
\usepackage[section]{placeins}
\usepackage[hang]{caption2}

\begin{document}

\clubpenalty=10000
\widowpenalty=10000

\fancyhead{} \fancyfoot{} \fancyhead[LO]{\rightmark} \fancyhead[RE]{\leftmark} \fancyhead[RO,LE]{\thepage} \pagestyle{fancy}
\renewcommand{\sectionmark}[1]{\markright{\thesection.\ #1}{}}
\renewcommand{\chaptermark}[1]{\markboth{\chaptername\ \thechapter .\ #1}{}}

\newcommand{\bra}[1]{\langle #1|}
\newcommand{\ket}[1]{|#1\rangle}
\newcommand{\braket}[2]{\langle #1|#2\rangle}

\newcommand{\ons}{on-shell}
\newcommand{\ofs}{off-shell}
\newcommand{\trl}{tree-level}

\newcommand{\eins}{1\!\! 1}

\newcommand{\frontmatter}{}
\newcommand{\mainmatter}{}
\newcommand{\backmatter}{}

\newcommand{\orders}[1]{\ensuremath{\mathcal{O}\left(s^{#1}\right)}}
\newcommand{\ordershat}[1]{\mathcal{O}\left(\hat{s}^{#1}\right)}
\newcommand{\orderslambda}[1]{\mathcal{O}\left(s_\Lambda^{#1}\right)}
\newcommand{\ordereps}[1]{\ensuremath{\mathcal{O}\left(\epsilon^{#1}\right)}}
\newcommand{\order}[1]{\mathcal{O}\left(#1\right)}

\newcommand{\dk}{d^{D-2}{\bf k}}
\newcommand{\dka}{d^{D-2}{{\bf k}_a}}\newcommand{\dkb}{d^{D-2}{{\bf k}_b}}
\newcommand{\dkl}{d^{D-2}{{\bf k}_l}}\newcommand{\dki}{d^{D-2}{{\bf k}_i}}
\newcommand{\dkone}{d^{D-2}{{\bf k}_1}}\newcommand{\dktwo}{d^{D-2}{{\bf k}_2}}
\newcommand{\dkjet}{d^{D-2}{{\bf k}_J}}
\newcommand{\dkjetone}{d^{D-2}{{\bf k}_{J_1}}}\newcommand{\dkjettwo}{d^{D-2}{{\bf k}_{J_2}}}
\newcommand{\dkpure}{d^{D-2}{{\bf k}}}\newcommand{\dkprime}{d^{D-2}{{\bf k}'}}
\newcommand{\ki}{{\bf k}_i}\newcommand{\kim}{{\bf k}_{i-1}}\newcommand{\kip}{{\bf k}_{i+1}}
\newcommand{\kj}{{\bf k}_j}\newcommand{\kjm}{{\bf k}_{j-1}}\newcommand{\kjp}{{\bf k}_{j+1}}
\newcommand{\km}{{\bf k}_m}\newcommand{\kmm}{{\bf k}_{m-1}}\newcommand{\kmp}{{\bf k}_{m+1}}
\newcommand{\kl}{{\bf k}_l}\newcommand{\klm}{{\bf k}_{l-1}}\newcommand{\klp}{{\bf k}_{l+1}}
\newcommand{\kn}{{\bf k}_n}\newcommand{\knm}{{\bf k}_{n-1}}\newcommand{\knp}{{\bf k}_{n+1}}
\newcommand{\ka}{{\bf k}_a}\newcommand{\kb}{{\bf k}_b}
\newcommand{\kzero}{{\bf k}_0}\newcommand{\kone}{{\bf k}_1}\newcommand{\ktwo}{{\bf k}_2}
\newcommand{\kjet}{{\bf k}_J}
\newcommand{\kjetone}{{\bf k}_{J_1}}\newcommand{\kjettwo}{{\bf k}_{J_2}}
\newcommand{\kpure}{{\bf k}}\newcommand{\kprime}{{\bf k}'}
\newcommand{\ddkone}{d^{2}{{\bf k}_1}}\newcommand{\ddktwo}{d^{2}{{\bf k}_2}}
\newcommand{\ddka}{d^{2}{{\bf k}_a}}\newcommand{\ddkb}{d^{2}{{\bf k}_b}}
\newcommand{\ddkjet}{d^{2}{{\bf k}_J}}

\newcommand{\dqone}{d^{D-2}{{\bf q}_1}}\newcommand{\dqtwo}{d^{D-2}{{\bf q}_2}}
\newcommand{\dqa}{d^{D-2}{{\bf q}_a}\;}\newcommand{\dqb}{d^{D-2}{{\bf q}_b}\;}
\newcommand{\dqi}{d^{D-2}{{\bf q}_i}\;}
\newcommand{\dqt}{d^{D-2}{\tilde{\bf q}}\;}
\newcommand{\qi}{{\bf q}_i}\newcommand{\qim}{{\bf q}_{i-1}}\newcommand{\qip}{{\bf q}_{i+1}}
\newcommand{\qj}{{\bf q}_j}\newcommand{\qjm}{{\bf q}_{j-1}}\newcommand{\qjp}{{\bf q}_{j+1}}
\newcommand{\ql}{{\bf q}_l}\newcommand{\qlm}{{\bf q}_{l-1}}\newcommand{\qlp}{{\bf q}_{l+1}}
\newcommand{\qn}{{\bf q}_n}\newcommand{\qnm}{{\bf q}_{n-1}}\newcommand{\qnp}{{\bf q}_{n+1}}\newcommand{\qnpp}{{\bf q}_{n+2}}
\newcommand{\qpure}{{\bf q}}\newcommand{\qprime}{{\bf q}'}
\newcommand{\qone}{{\bf q}_1}\newcommand{\qtwo}{{\bf q}_2}
\newcommand{\qa}{{\bf q}_a}\newcommand{\qb}{{\bf q}_b}
\newcommand{\qt}{\tilde{\bf q}}
\newcommand{\ddqone}{d^{2}{{\bf q}_1}}\newcommand{\ddqtwo}{d^{2}{{\bf q}_2}}
\newcommand{\ddqprime}{d^{2}{{\bf q}'}}\newcommand{\ddqpure}{d^{2}{{\bf q}}}
\newcommand{\ddqa}{d^{2}{{\bf q}_a}}\newcommand{\ddqb}{d^{2}{{\bf q}_b}}
\newcommand{\ddqi}{d^{2}{{\bf q}_i}}

\newcommand{\dktwoeps}{\frac{d^{D-2}{{\bf k}_2}}{\mu^{2\epsilon}(2\pi)^{D-4}}}
\newcommand{\dlambdeps}{\frac{d^{D-2}{{\bf \Lambda}}}{\mu^{2\epsilon}(2\pi)^{D-4}}}

\newcommand{\oma}{\omega_0({\bf k}_a)}
\newcommand{\om}[1]{\omega_0(#1)}
\newcommand{\omall}{\omega_0^{LL}({\bf k}_a)}
\newcommand{\omll}[1]{\omega_0^{LL}(#1)}
\newcommand{\omhat}{\hat{\omega}_0}\newcommand{\omhati}{\hat{\omega}_i}\newcommand{\omhatim}{\hat{\omega}_{i-1}}\newcommand{\omhatl}{\hat{\omega}_l}\newcommand{\omhatn}{\hat{\omega}_n}\newcommand{\omhatlm}{\hat{\omega}_{l-1}}

\newcommand{\del}[1]{\delta^{(2)}\left(#1\right)}
\newcommand{\non}{\nonumber\\}
\newcommand{\asquare}[4]{\left|\mathcal{A}(#1,#2,#3,#4)\right|^2}
\newcommand{\bssquare}[4]{\left|\mathcal{B}_s(#1,#2,#3,#4)\right|^2}
\newcommand{\btssquare}[4]{\left|{\widetilde{\mathcal{B}}}_s(#1,#2,#3,#4)\right|^2}
\newcommand{\bsquare}[4]{\left|\mathcal{B}(#1,#2,#3,#4)\right|^2}
\newcommand{\agsquare}[4]{\left|\mathcal{A}_{2g}(#1,#2,#3,#4)\right|^2}
\newcommand{\aqsquare}[4]{\left|\mathcal{A}_{2q}(#1,#2,#3,#4)\right|^2}

\newcommand{\asbar}{\bar\alpha_s}
\newcommand{\seplog}[2]{\ln\frac{s_\Lambda}{\sqrt{#1^2 #2^2}}}

\newcommand{\shat}{\hat{s}}\newcommand{\that}{\hat{t}}\newcommand{\uhat}{\hat{u}}
\newcommand{\lambd}{{\bf{\Lambda}}}\newcommand{\delt}{{\bf{\Delta}}}

\newcommand{\opqone}{\hat{\bf q}_1}\newcommand{\optwo}{\hat{\bf q}_2}
\newcommand{\opk}{\hat{\mathcal{K}}}
\newcommand{\opf}{\hat{f}}
\newcommand{\opjetone}{\hat{\Phi}_{\rm jet,1}}\newcommand{\opjettwo}{\hat{\Phi}_{\rm jet,2}}
\newcommand{\oplept}{\hat{\Phi}_{\rm leptonic}}

\newcommand{\plusinu}[1]{\left(#1\right)^{i\nu-\frac{1}{2}}}
\newcommand{\minusinu}[1]{\left(#1\right)^{-i\nu-\frac{1}{2}}}
\newcommand{\plusinup}[1]{\left(#1\right)^{i\nu'-\frac{1}{2}}}
\newcommand{\minusinup}[1]{\left(#1\right)^{-i\nu'-\frac{1}{2}}}
\newcommand{\infint}{\int_{-\infty}^{\infty}}

\setlength{\oddsidemargin}{46pt} \setlength{\headheight}{14pt}
\frontmatter

\title{Phenomenology of jet physics\\in the BFKL formalism at NLO}

\author{Dissertation\\zur Erlangung des Doktorgrades\\
des Departments Physik\\
der Universit\"at Hamburg}
\date{vorgelegt von\\Florian Schwennsen\\aus Schleswig\vspace{2cm}\\Hamburg 2007}
\maketitle

\pagestyle{empty}

\begin{tabular}[bl!]{ll}
\vspace{16cm}\\
Gutachter der Dissertation:             & Prof.~Dr. Jochen Bartels\\
                                        & Prof.~Dr. Bernd A. Kniehl\\
Gutachter der Disputation:              & Prof.~Dr. Jochen Bartels\\
                                        & Dr. habil. Markus Diehl\\
Datum der Disputation:                  & 30. Januar 2007\\
Vorsitzender des Pr\"ufungsausschusses:  & Dr. Hans Dierk R\"uter\\
Vorsitzender des Promotionsausschusses: & Prof.~Dr. G\"unter Huber\\
Departmentleiter:                       & Prof.~Dr. Robert Klanner\\
Dekan der Fakult\"at f\"ur Mathematik,  &  \\
 Informatik und Naturwissenschaften:    &  Prof.~Dr. Arno Fr\"uhwald
\end{tabular}

\clearpage{\pagestyle{empty}\cleardoublepage}

~
\vspace{2cm}

\begin{center}\bf Abstract \end{center}

We study jet physics in the high energy regime of QCD. Based on the NLO BFKL equation, we construct a vertex for the production of a jet at central rapidity in $k_T$-factorization. A jet algorithm is introduced, and we take special care of the separation of multi-Regge and quasi-multi-Regge kinematics. The connection with the energy scale of the evolution is investigated in detail. The result is discussed for two situations: scattering of highly virtual photons, which requires a symmetric energy scale to separate the impact factors from the gluon Green's function, and hadron-hadron collisions, where a non-symmetric scale choice is needed. For the second case we are able to define a NLO unintegrated gluon density, valid in the small-$x$ regime, and give the evolution equation for this density as well.

In the second part, we examine the angular decorrelation of Mueller-Navelet jets.  Using an operator formalism in the space of anomalous dimension and conformal spin, we implement the NLO BFKL Green's function to study the rapidity dependence of angular decorrelations. We incorporate the necessary summation of collinearly enhanced corrections beyond NLO accuracy. We compare our results with data from the Tevatron $p\bar p$-collider and provide predictions for the LHC as well. We also  extend our study to the angular decorrelation between a forward jets and the electron in deep inelastic $ep$ scattering. The angular decorrelation has not been measured in DIS so far, but we give theoretical results for this observable which already implement the experimental cuts.

\clearpage
\pagestyle{empty}

~
\vspace{2cm}

\begin{center}\bf Zusammenfassung\end{center}

Wir untersuchen Jet-Physik im Hochenergie-Regime der QCD. Basierend auf der NLO BFKL Gleichung konstruieren wie einen Vertex f\"ur die Produktion eines Jets in zentraler Rapidit\"at in $k_T$-Faktorisierung. Ein Jetalgorithmus wird eingef\"uhrt, und wir verwenden besondere Sorgfalt auf die Trennung von Multi-Regge und Quasi-Multi-Regge Kinematik. Die Verbindung zur Energieskala der Evolution wird detailiert untersucht. Das Ergebnis wird f\"ur zwei Situationen diskutiert: der Streuung von hoch virtuellen Photonen, welche eine symmetrische Energieskala erfordert, um die Impaktfaktoren von der Gluon-Greenfunktion zu separieren, sowie Hadron-Hadron Kollisionen, bei denen eine asymmetrische Wahl der Skala erforderlich ist. F\"ur den zweiten Fall sind wir in der Lage, eine NLO unintegrierte Gluonendichte zu definieren, die im Bereich kleiner $x$ g\"ultig ist, und geben die Evo\-lu\-tions\-glei\-chung f\"ur diese Dichte an.

Im zweiten Teil betrachten wir die Winkeldekorrelation von Mueller-Navelet Jets. Unter Verwendung eines Operatorformalismus im Raum von anomaler Dimension und  konformem Spin implementieren wir die NLO BFKL Greenfunktion, um die Rapidit\"atsabh\"angigkeit der Winkelkorrelation zu studieren. Wir ber\"ucksichtigen die notwendige Summation von kollinear verst\"arkten Korrekturen jenseits der NLO Genauigkeit. Wir vergleichen unsere Ergebnisse mit Daten vom $p\bar{p}$-Beschleuniger TEVATRON und stellen Vorhersagen f\"ur den LHC zur Verf\"ugung. Zudem dehnen wir unsere Untersuchung auf die Winkelkorrelation zwischen einem Vorw\"artsjet und dem Elektron in tiefinelastischer $ep$-Streuung aus. Die Winkelabh\"angigkeit in diesem Kontext wurde noch nicht gemessen wurde, aber wir  geben theoretische Resultate f\"ur diese Observable an, die bereits die experimentellen Schnitte ber\"ucksichtigen.

\clearpage{\pagestyle{empty}\cleardoublepage}

\setlength{\evensidemargin}{22pt}

\setlength{\oddsidemargin}{71pt}

\tableofcontents

\fancyhead{} \fancyfoot{} \fancyhead[LO]{Content} \fancyhead[RE]{Content} \fancyhead[RO,LE]{\thepage} \pagestyle{fancy}

\clearpage{\pagestyle{empty}\cleardoublepage}

\mainmatter

\fancyhead{} \fancyfoot{} \fancyhead[LO]{\rightmark} \fancyhead[RE]{\leftmark} \fancyhead[RO,LE]{\thepage} \pagestyle{fancy}
\renewcommand{\sectionmark}[1]{\markright{\thesection.\ #1}{}}
\renewcommand{\chaptermark}[1]{\markboth{\chaptername\ \thechapter .\ #1}{}}

\chapter{Introduction}

Today,  Quantum Chromodynamics (QCD) has established itself as the adequate  theory to describe the strong force between fermionic quarks and bosonic gluons. One of its striking features that led directly to its discovery as a candidate theory of the strong interactions is asymptotic freedom. Asymptotic freedom implies that at short distances, i.e. whenever a hard momentum scale is present, quarks behave as almost free particles and, hence, we can apply perturbation theory. The other side of the coin is confinement, i.e. the coupling increases with the distance, and at large distances  only bound states of quarks and gluons exist which cannot be described by means of a perturbative expansion. This interplay of perturbative versus nonperturbative physics, arising from a quite frugal appearing Lagrangian,  makes QCD a challenging theory with a very rich phenomenology.

For the soft scale sector of QCD we have to rely on purely phenomenological models, QCD sum rules or on lattice calculations which have made impressing progress over the last decade. However, 
lattice calculations cannot answer all our question on QCD, and due to its dependence on vast computer resources lattice calculations are still restricted in its possibilities. 
Even though every hard scale process is accompanied by soft scale physics in a real experiment, it is possible to disentangle these parts using factorization theorems which allow to describe the hard part within perturbative QCD by a fixed order expansion in the coupling $\alpha_s$. These theorems are strictly proven for a limited number of processes, but have shown its usability in a large number of processes.

A manifestation of these theorems is the well known {\it collinear factorization} in the context of a hadron involved in a collision process. Due to this factorization, cross sections of hadronic interactions can be written in terms of a process-dependent hard matrix element convoluted with universal parton density functions which are described  by the Dokshitzer-Gribov-Lipatov-Altarelli-Parisi (DGLAP) equation \cite{Gribov:1972ri,Lipatov:1974qm,Altarelli:1977zs,Dokshitzer:1977sg}. Because of the strong ordering of virtuality in the DGLAP evolution, the virtualities of the partons entering the hard interaction could be neglected compared with the large scale of the probe, i.e. they are considered as being collinear to the proton.

Whenever one considers kinematic regions characterized by two different large scales, this fixed order calculation is not sufficient since logarithms of these two scales appear at each order of the perturbation series. If the two scales are ordered, the logarithm of the ratio of the two scales becomes large and compensates the smallness of the coupling. Therefore, these logarithms have to be resummed to all orders to justify a perturbative treatment. One famous example is the case of high energy scattering  with fixed momentum transfer. If the center of mass energy $s$ is much larger than the momentum transfer $|t|$ -- the so-called Regge asymptotics of the process -- the gluon exchange in the crossed channel dominates and logarithms of the type $[\alpha_s\ln(s/|t|)]^n$ have to be resummed. This is realized by the leading logarithmic (LL) Balitsky-Fadin-Kuraev-Lipatov (BFKL) \cite{Fadin:1975cb,Kuraev:1976ge,Kuraev:1977fs,Balitsky:1978ic} equation for the gluon Green's function describing the momentum exchange in the $t$-channel.

In the context of hadron collisions such a two-scale situation is easily realized if the scale of the hard process is large compared to $\Lambda_{\rm QCD}$, but nevertheless small compared to the center of mass energy $s$. We are then in the region of small longitudinal momentum fractions $x$ of the partons entering the hard matrix element. Therefor, there are no longer grounds for neglecting their transverse momenta $k_T$. It is believed that in this case a better descriptions is given by the BFKL evolution equation. Here large logarithms of the form $[\alpha_s\ln(1/x)]^n$  are taken into account. Another evolution equation which resums these type of logarithms is the Ciafaloni-Catani-Fiorani-Marchesini (CCFM) evolution equation \cite{Ciafaloni:1987ur,Catani:1989yc,Catani:1989sg,Marchesini:1994wr}. Just as for DGLAP, it is possible to factorize the cross section into a a convolution of process-dependent hard matrix elements with universal parton distributions. But as the virtualities and transverse momenta are no longer ordered (as it is the case in DGLAP evolution), the matrix elements have to be taken \ofs\  and the convolution  made also over transverse momenta with the so-called {\it unintegrated parton densities}. This factorization scheme is called {\it $k_T$-factorization} \cite{Catani:1990eg,Collins:1991ty} or {\it semihard approach} \cite{Gribov:1984tu,Levin:1991ry}.

The BFKL equation as it stands is aimed to describe inclusive quantities. 
If we consider for instance deep inelastic scattering of a virtual photon on a proton the BFKL equation predicts the observed steep rise of the proton structure function $F_2(x)$ at small Bjorken-$x$. 
For this observable -- measured at the electron proton collider HERA -- logarithms of the type $[\alpha_s\ln(1/x)]^n$ have to be resummed. However, for $F_2$ the hard scale of the photon has to be connected with the soft scale of the proton such that it is not clear, whether $F_2$ can be completely described by pure BFKL dynamics. Moreover, the existing HERA data can be described by conventional DGLAP evolution as well. But also the region of large $x$ provides the opportunity to observe BFKL dynamics. 
If one considers the production of forward  jets instead of the proton structure function $F_2(x)$, large logarithms of the form  $[\alpha_s\ln(x_{\rm Bj}/x_{\rm FJ})]^n$ appear, with $x_{\rm Bj}$ being the Bjorken-$x$ of the photon and $x_{\rm FJ}$ the longitudinal momentum fraction of the forward jet.
In the case of two jets produced in the very forward direction at a proton proton collision, logarithms of the form $[\alpha_s\ln(s/Q^2)]^n$ appear, where $Q$ is the scale of the forward jets. Mueller and Navelet \cite{Mueller:1986ey} proposed this dijet cross section as an ideal observable to be described by the BFKL equation.

One also has to deal with BFKL dynamics if one 
aims to understand actually
the physics behind jet production. This knowledge is an essential ingredient in phenomenological studies at present and future colliders. At high center of mass energies the theoretical study of multijet events becomes an increasingly important task. In the context of collinear factorization the calculation of multijet production is complicated because of the large number of contributing diagrams. There is, however, a region of phase space where it is indeed possible to describe the production of a large number of jets: the Regge asymptotics (small--$x$ region) of scattering amplitudes. If the jets are well separated in rapidity, the according matrix element factorizes with effective vertices for the jet production connected by a chain of $t$-channel Reggeons.

It turned out to be very fruitful to investigate more exclusive observables then the total cross section. Considering again the case of two forward jets produced in a hadron hadron collision, not only the cross section grows with increasing rapidity distance between both jets, but the additional emission of gluons between the two jets tagged in the Mueller-Navelet scenario lead to an angular decorrelation between those two jets. 
Since the angular decorrelation stems from the additional emissions between the colliding particles, this decorrelation should be also observable in other reactions which can be described by the BFKL equation.

One of the corner stones of the BFKL approach is the \emph{Reggeization} of the gluon, i.e. a process involving  color octet exchange in the $t$-channel is described at lowest order  by the exchange of one gluon, but asymptotically behaves as $\sim (s/t)^{j(t)}$. Furthermore, Reggeization implies that the \emph{trajectory} $j(t)=1+\omega(t)$ passes through 1 (the spin of the gluon) at $t=0$ (the mass of the gluon). Although the trajectory itself is not infrared finite, the divergences are canceled if real emissions, using gauge invariant Reggeon-Reggeon-gluon vertices, are taken into account. It is then possible to describe scattering amplitudes with any number of well separated particles (jets) in the final state. The $(\alpha_s \ln{s})^n$ resummation is known as leading-order (LO) approximation and provides a simple picture of the underlying physics. Nevertheless, it suffers from some drawbacks. One of them is the complete indetermination of the energy scale $s_0$ scaling the energy $s$ in the resummed logarithms. Another handicap concerns the coupling $\alpha_s$ which, at LO, is just a global parameter with its scale being  not restricted as well. These limitations can be removed if the accuracy in the calculation is increased, and next-to-leading (NLO) terms of the form $\alpha_s \left(\alpha_s \ln{s}\right)^n$ are taken into account \cite{Fadin:1998py,Ciafaloni:1998gs}. Diagrams of higher order then include those which contribute to the running of the coupling, and the correct matching of the different higher order contributions fixes the energy scale $s_0$.

While at LO the only emission vertex -- the Reggeon-Reggeon-gluon vertex -- can be identified with the production of one jet, at NLO also Reggeon-Reggeon-gluon-gluon and Reggeon-Reggeon-quark-antiquark vertices enter the game. The first part of this thesis is dedicated to the construction of an inclusive one-jet-production vertex from these different contributions. Due to these new emission vertices at NLO we have to introduce a jet definition discriminating between the production of one or two jets by two particles. It is not sufficient to simply start from the fully integrated emission vertex available in the literature \cite{Fadin:1998py,Ciafaloni:1998gs}. Rather, we have to carefully separate all the different contributions in its unintegrated form first before we can combine them. By this procedure we will also be able to determine the right choice of energy scales relevant for the process. Particular attention is given to the separation of multi-Regge and quasi-multi-Regge kinematics. 

As it will turn out, the scale of the two projectiles in the scattering process has a large impact on the structure of the result. The jet vertex can not be constructed without properly defining the interface to the scattering objects. To show this, we will perform this study for two different cases: the jet production in the scattering of two photons with large and similar virtualities, and in hadron-hadron collisions. In the former case the cross section has a factorized form in terms of the photon impact factors and of the  gluon Green's function which is valid in the Regge limit. In the latter case, since the momentum scale of the hadron is substantially lower than the typical $k_T$ entering the production vertex, the gluon Green's function for hadron-hadron collisions has a slightly different BFKL kernel which, in particular, also incorporates some $k_T$-evolution from the nonperturbative, and model dependent, proton impact factor to the perturbative jet production  vertex. 

Our final expression for the cross section of the jet production in hadron-hadron scattering contains an \emph{unintegrated gluon density}. This density depends on the longitudinal momentum fraction -- as typical to the  conventional collinear factorization -- and on the transverse momentum $k_T$. This scheme is known as $k_T$-factorization and has been considered up to now only at LO. In fact our result, valid in the small-$x$ limit, shows that it is possible to extend the $k_T$-factorization to NLO. Nevertheless, we have to state that it is not obvious how to give a more general formulation of unintegrated parton densities at NLO for general $x$.

One of the most famous testing ground for BFKL physics are the already mentioned Mueller-Navelet jets \cite{Mueller:1986ey}. The predicted powerlike rise of the cross section with increasing energy has been observed at the Tevatron $p\bar{p}$-collider \cite{Abbott:1999ai}, but the measurements revealed an even stronger rise then predicted by BFKL calculations.
As another option, we direct our attention to a more exclusive observable within this process. In the second half of this thesis we study the azimuthal correlation between these jets. 
When we consider hadron-hadron scattering in the common parton model to describe two jet production, we have the following picture in mind:  a gluon or quark is struck out of each of the initial hadrons which then scatters off the other one and -- after the process of hadronization -- is observed as a jet of particles. 
In this back-to-back reaction we expect the azimuthal angles of the two jets always to be $\pi$ and hence completely correlated. But when we rise the rapidity difference between these jets, the phase space allows for more and more emissions leading to an angular decorrelation between the jets.
In the academical limit of infinite rapidity, the angles should be completely uncorrelated. In the regime of large, but realizable rapidity differences the resummation of large logarithms calls for a description within the BFKL theory. Unfortunately, the leading logarithmic approximation  \cite{DelDuca:1993mn,Stirling:1994zs} overestimates this decorrelation by far. Improvements have been obtained by taking into account some corrections of higher order like the running of the coupling \cite{Orr:1997im,Kwiecinski:2001nh}. A full NLO description would incorporate the NLO Green's function -- in fact not in its angular averaged form -- and the NLO jet vertices for Mueller-Navelet jets \cite{Bartels:2001ge,Bartels:2002yj}. Due to the complexity of both pieces a fully analytical treatment seems cumbersome. Even on the numerical level investigations so far have been performed only for studies of the NLO kernel itself \cite{Andersen:2003an,Andersen:2003wy,Andersen:2006sp,Vera:2006rp}. 

We make an approximation to a full NLO calculation and study the angular correlation on the basis of the NLO kernel. To consider the jet vertices just at LO accuracy is justified by the fact that the angular correlation is mainly driven by the evolution kernel, especially if one considers the rapidity dependence of the correlation. Nevertheless, we will discuss the correct implementation of NLO jet vertices and its consequences. Using the framework of an operator formalism, developed in Refs.~\cite{Ivanov:2005gn,Vera:2006un,Vera:2006xa}, we study the important NLO feature, namely,  that it determines the energy scale $s_0$ and the running of the coupling. Furthermore, we carefully study the influence of different renormalization schemes and in this context also the need of a  partial resummation of collinearly enhanced higher order terms. 

Since the angular correlation of Mueller-Navelet jets was measured at the Tevatron collider by the {D$\emptyset$} collaboration \cite{Abachi:1996et}, we can compare our results with these data. We find that our NLO calculation including the resummation of collinearly enhanced terms improves the LO description significantly, but still misses the data.
Furthermore, we formulate our results as a prediction for the forthcoming Large Hadron Collider.

The studies of Mueller-Navelet jets always have been related to forward jet experiments in deep inelastic scattering (DIS). The analogies can be drawn for the angular decorrelation as well. Therefore, we transfer the machinery developed in the framework of Mueller-Navelet jets to DIS. Some details change due to the different projectile, namely the electron, but the main features -- as expected from theoretical side -- appear to be the same as in the Mueller-Navelet case. Unfortunately, there are no experimental studies dealing with angular correlation in DIS, but we hope that our work draws the attention to this problem also from the experimental side.

This thesis is organized as follows: In chapter \ref{sec:bfkl} we introduce and explain the BFKL equation at LO and NLO accuracy. We especially work out the different contributions to the NLO kernel and discuss its properties. Furthermore we elaborate on the resummation of terms beyond NLO accuracy for general conformal spin which are necessary to stabilize the NLO kernel. Chapter \ref{sec:jetproduction} shows how to construct the inclusive jet production cross section from the NLO BFKL equation. We give an explicit formula for the jet vertex. Furthermore, we show how, in the context of hadron scattering, we are led to an NLO unintegrated gluon density, and give an evolution equation for this quantity. The discussion on angular decorrelation is contained in chapter \ref{sec:decorrelation}. There, we derive analytic expressions for the angular correlation in proton-proton and electron-proton scattering. We perform a resummation of collinearly enhanced terms of higher order and compare our calculations with existing data. In chapter \ref{sec:summary} we draw our conclusions. In appendix \ref{sec:alternative} we present an alternative subtraction term for the jet vertex, as derived in chapter \ref{sec:jetproduction}. The appendix \ref{sec:resummation} comprises the technical details on the resummed kernel.

\clearpage{\pagestyle{empty}\cleardoublepage}
\chapter{BFKL equation}
\label{sec:bfkl}

This first chapter is dedicated to the derivation of the BFKL equation in leading and next-to-leading logarithmic approximation. Starting from its original derivation \cite{Fadin:1975cb,Kuraev:1976ge,Kuraev:1977fs,Balitsky:1978ic} a number of different approaches to the BFKL equation has been presented. Our presentation is by no means extensive, but instead focuses on those points which will be most relevant for the rest of this study. In the first section we present the derivation of the leading order BFKL equation and in the second chapter we show its generalization to next-to-leading order. We take special care of the features, which are new compared to LO and, hence,  specific for NLO. In the third section we discuss the properties of the NLO kernel of the BFKL equation. The fourth section contains an overview over the phenomenology of BFKL dynamics.

\section{BFKL equation at LO}

Let us consider the case of the total cross section $\sigma_{\rm AB}$ in the 
scattering of two particles A and B. It is convenient to work 
with the Mellin transform
\begin{equation}
  \label{eq:news84}
  {\cal F}(\omega,s_0)=\int_{s_0}^\infty \frac{ds}{s}\left(\frac{s}{s_0}\right)^{-\omega}\sigma_{\rm AB}(s),
\end{equation}
acting on the center--of--mass energy $s$. The dependence on the scaling 
factor $s_0$ belongs to the NLO approximation since the LO calculation is 
formally independent of $s_0$.

If we denote the matrix element for 
the transition ${\rm A} + {\rm B} \rightarrow {\tilde {\rm A}} + {\tilde{\rm  B}} + {\rm n}$ produced particles with momenta $k_i$ ($i=1,\ldots,n$) as 
$\mathcal{A}_{\tilde{\rm A}\tilde{\rm B}+{\rm n}}$, and the corresponding 
element of phase space as $d\Phi_{\tilde{\rm A}\tilde{\rm B}+{\rm n}}$, 
 we can write the total cross section as
\begin{equation}
  \label{eq:news9}
  \sigma_{\rm AB}=\frac{1}{2s}\sum_{n=0}^\infty \int d\Phi_{\tilde{\rm A}\tilde{\rm B}+{\rm n}} |\mathcal{A}_{\tilde{\rm A}\tilde{\rm B}+ {\rm n}}|^2.
\end{equation}
As we mentioned in the introduction we are interested in the Regge limit 
where $s$ is asymptotically larger than any other scale in the scattering 
process. In this region the 
scattering amplitudes are 
dominated by the production of partons widely separated in rapidity from 
each other. This particular configuration of phase space is known as 
multi--Regge kinematics (MRK). Particles produced in MRK are strongly 
ordered in rapidity, but there is no ordering of the transverse momenta which 
are only assumed not to be growing with energy.

\begin{figure}[htbp]
  \centering
  \includegraphics[width=8cm]{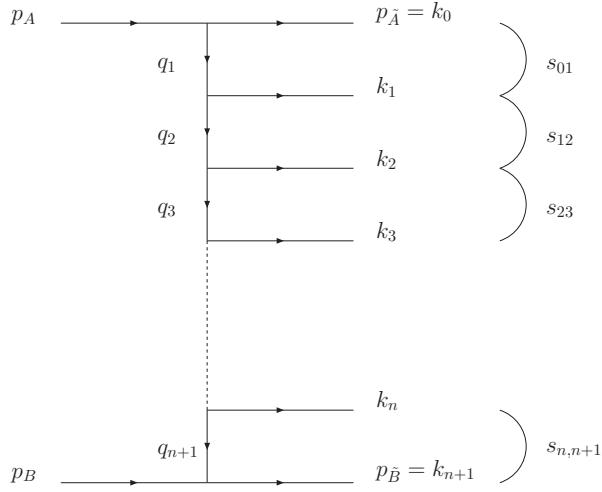}
  \caption{Notation for particle production in MRK.}
  \label{fig:kinematics}
\end{figure}

We fix our notation in Fig.~\ref{fig:kinematics}: $q_i$ correspond to the 
momenta of those particles exchanged in the $t$--channel while the subenergies 
$s_{i-1,i}=(k_{i-1}+k_i)^2$ are related to the rapidity difference between  
consecutive $s$--channel partons. Euclidean two--dimensional transverse 
momenta are denoted in bold. For future discussion we use the Sudakov 
decomposition $k_i = \alpha_i \,p_A + \beta_i \, p_B + {k_i}_\perp$ for the 
momenta of emitted particles.

In MRK the center--of--mass energy for the incoming external particles can be 
expressed in terms of the internal subenergies as
\begin{gather}
  s \simeq\left[\prod_{i=1}^{n+1} s_{i-1,i}\right]\left[\prod_{i=1}^{n} 
\ki^2\right]^{-1}\simeq\sqrt{\qone^2\qnp^2}\,\prod_{i=1}^{n+1} \frac{s_{i-1,i}}{\sqrt{\kim^2\ki^2}},
\label{eq:ssi}
\end{gather}
where we have used the fact that in Regge kinematics $s$ is much larger than 
$-t$ and, therefore, $\alpha_0 \simeq \beta_{n+1} \simeq 1$, 
$\kzero^2 \simeq \qone^2 $ and $\knp^2 \simeq \qnp^2 $. 
To write down the measure of phase space we use dimensional 
regularization with $D=4 + 2 \, \epsilon$, i.e.
\begin{gather}
ds\; d\Phi_{\tilde{\rm A}\tilde{\rm B}+{\rm n}} = 2\pi\prod_{i=1}^{n+1}\frac{ds_{i-1,i}}{2 \, s_{i-1,i}}\frac{\dqi}{(2\pi)^{D-1}}.
\label{eq:phasespace}
\end{gather}
At tree level, the matrix element $\mathcal{A}_{\tilde{\rm A}\tilde{\rm B}+{\rm n}}$ of 
Eq.~(\ref{eq:news9}) can be written in MRK in the factorized form
\begin{equation}
  \label{eq:news17tree}
  \frac{\mathcal{A}_{\tilde{\rm A}\tilde{\rm B}+{\rm n}}}{2\,s}= \Gamma_A\left[\prod_{i=1}^n \frac{1}{q_i^2}\gamma(q_i,q_{i+1})\right]\frac{1}{q_{n+1}^2}\Gamma_B,
\end{equation}
with  $\Gamma_{P}$ being the couplings of the gluon to the external particles, and $\gamma(q_i,q_{i+1})$ the gauge invariant effective (non-local) vertices of two $t$-channel gluons to a produced $s$-channel gluon. 
These vertices are called non-local since they encode the propagators of the connected $t$-channel gluons.

A calculation of the loop corrections in leading logarithmic approximation does not break the factorized form of the amplitude. Instead it leads to the Reggeization of the $t$-channel gluons, i.e. the gluon propagator of $q_i$ gets a multiplicative factor $\left(\frac{s_{i-1,i}}{s_{R}}\right)^{\omega_i}$ with $\omega_i=\omega(q_i^2)$ the gluon Regge trajectory depending on the momentum carried by the Reggeon. At LO the Reggeon scale $s_R$ is a free parameter. 

The statement that the gluon reggeizes means that the Reggeon carries the same quantum numbers as the gluon, that the so-called \emph{trajectory} $j(t)=1+\omega(t)$ passes through 1 at $t=0$, and that this Reggeon gives the leading contribution in each order of perturbation theory. 
The consistency of this picture demands that if we start from the assumption of Reggeization the calculation of  all real and virtual corrections to the elastic scattering of two particles with color octet exchange in the $t$-channel has to result in the exchange of simply one Reggeon. This consistency condition and other related ones are called \emph{bootstrap conditions} and play an important role. They have been proven at LO \cite{Balitsky:1979ap} and NLO \cite{Fadin:2006bj} as well.

Hence we can write the matrix element $\mathcal{A}_{\tilde{\rm A}\tilde{\rm B}+{\rm n}}$ in its resummed form at leading logarithmic accuracy as
\begin{equation}
  \label{eq:news17}
  \frac{\mathcal{A}_{\tilde{\rm A}\tilde{\rm B}+{\rm n}}}{2\,s}= \Gamma_A\left[\prod_{i=1}^n \frac{1}{q_i^2}\left(\frac{s_{i-1,i}}{s_{R}}\right)^{\omega_i}\gamma(q_i,q_{i+1})\right]\frac{1}{q_{n+1}^2}\left(\frac{s_{n,n+1}}{s_{R}}\right)^{\omega_{n+1}}\Gamma_B,
\end{equation}

Gathering all these elements together it is possible to write the Mellin 
transform of Eq.~(\ref{eq:news84}) as the sum
\begin{equation}
 \label{eq:fns}
 {\cal F}(\omega,s_0) = \sum_{n=0}^\infty {\cal F}^{(n)}(\omega,s_0),
\end{equation}
with the contributions from the emission of n $s$--channel gluons being
\begin{multline}
\frac{{\cal F}^{(n)}(\omega,s_0)}{(2\pi)^{2-D}} = 
\int \left[\prod_{i=1}^{n+1}\dqi\frac{ds_{i-1,i}}{s_{i-1,i}}
\left( \frac{s_{i-1,i}}{s_R}\right) ^{2\omega_i}\left(\frac{s_{i-1,i}}{\sqrt{\kim^2\ki^2}}\right)^{-\omega}\right]\\
\times \left(\frac{s_0}{\sqrt{\qone^2{\qnp^2}}}\right)^{\omega}\frac{\Phi_A(\qone)}{\qone^2}\left[\prod_{i=1}^n{\cal K}_r(\qi,\qip)\right]\frac{\Phi_B(\qnp)}{\qnp^2} .
\label{eq:lla}
\end{multline}
The impact factors $\Phi_P$ and the real emission kernel for Reggeon--Reggeon into a $s$--channel gluon ${\cal K}_r$ can be written in terms of the square of the vertices $\Gamma_{P}$ and $\gamma$, respectively. The  kernel
${\cal K}_r\left(\qi,\qip\right)$ 
 is defined such that it includes one gluon propagator on each side:
$\left(\qi^2 \qip^2\right)^{-1}$. The integration over 
$s_{i-1,i}$ in Eq.~\eqref{eq:lla} takes place from a 
finite $s_0$ to infinity. At LO terms of the form $\omega\ln\ki^2$ or 
$\omega_i \ln s_R$ can be neglected when the integrand is expanded in 
$\alpha_s$. Therefore, at this accuracy, Eq.~\eqref{eq:lla} gives
\begin{equation}
 \frac{{\cal F}^{(n)}(\omega,s_0)}{(2\pi)^{2-D}} =
\int\left[\prod_{i=1}^{n+1}\frac{\dqi}{\omega-2\,\omega_i}\right]\frac{\Phi_A(\qone)}{\qone^{~2}}\left[\prod_{i=1}^n{\cal K}_r(\qi,\qip)\right]\frac{\Phi_B(\qnp)}{\qnp^2},
\label{eq:lla2}
\end{equation}
where the poles in the complex $\omega$--plane correspond to Reggeon 
propagators. This simple structure is a consequence of the linearity of the 
integral equation for the gluon Green's function. 
To see this connection explicitly we can introduce the gluon Green's function
\begin{multline}
  \label{eq:ggf}
  f_\omega(\qa,\qb) = \sum_{n=0}^\infty\int\left[\prod_{i=1}^{n+1}\frac{\dqi}{\omega-2\,\omega_i}\right]\\
\times\left[\prod_{i=1}^n{\cal K}_r(\qi,\qip)\right]\del{\qa-\qone}\del{\qb-\qnp}
\end{multline}
being a solution of the BFKL equation
\begin{equation}
  \label{eq:bfklequation}
  \omega\, f_\omega(\qa,\qb) = \del{\qa-\qb}+\int\dqt \mathcal{K}(\qa,\qt)\, f_\omega(\qt,\qb).
\end{equation}
The integration kernel $\mathcal{K}$ contains the real emission kernel $\mathcal{K}_r$ and the trajectory $\omega(\qa)^2$ which in this context often is called the virtual part of the kernel.
\begin{equation}
  \mathcal{K}(\qa,\qt) = 2\omega(\qa^2)\,\del{\qa-\qt} + \mathcal{K}_r(\qa,\qt) 
\end{equation}

For the sake of clarity we have to say that Eq.~\eqref{eq:bfklequation} as derived here is only the \emph{forward} BFKL equation. The name stems from the fact that the total cross section is related to the imaginary part of the elastic forward scattering amplitude by the optical theorem. The basis of the optical theorem is unitarity of the $S$-matrix which can be  applied to the non forward scattering as well. The non-forward version of Eq.~\eqref{eq:news9} connects the imaginary part of the elastic scattering ${\rm A}+{\rm B}\to {\rm A'}+{\rm B'}$ with the sum of products of two different production amplitudes:
\begin{equation}
  \label{eq:news9nonforward}
  \Im {\rm m} \,\mathcal{A}_{{\rm AB}\to{\rm A'B'}}
=\frac{1}{2}\sum_{n=0}^\infty \int d\Phi_{\tilde{\rm A}\tilde{\rm B}+{\rm n}} 
\mathcal{A}_{{\rm AB}\to\tilde{\rm A}\tilde{\rm B}+ {\rm n}}\left(\mathcal{A}_{{\rm A'B'}\to\tilde{\rm A}\tilde{\rm B}+ {\rm n}}\right)^*.
\end{equation}
Following the same logic and steps leads to the BFKL equation with momentum transfer.

Coming back to the case of the total cross section and summing the contributions of Eq.~\eqref{eq:lla2} we can write Eq.~\eqref{eq:fns} as a convolution of this Green's function with the impact factors
\begin{equation}
  \mathcal{F}(\omega,s_0) = \frac{1}{(2\pi)^{D-2}}\iint\frac{\dqa}{\qa^2}\frac{\dqb}{\qb^2}\Phi_A(\qa)f_\omega(\qa,\qb)\Phi_B(\qb) .
\end{equation}
We will see below that Eq.~\eqref{eq:lla2} holds very similarly at NLO. 

After this brief introduction to the structure of BFKL cross 
sections and its iterative expression we now turn to the NLO case. The 
factorization into impact 
factors and Green's function will remain, while the kernel and trajectory 
will be more complex than at LO. We discuss these points 
in the next section.


\section{Different contributions at NLO}

To discuss the various contributions to NLO BFKL cross sections we follow 
Ref.~\cite{Fadin:1998sh}. We comment in more detail those points which 
will turn out to be more relevant for our later discussion of inclusive jet 
production. 
Our starting point are Eqs.~\eqref{eq:news84} to \eqref{eq:phasespace}, which 
remain unchanged. Since at NLO the $s_R$ scale is no longer a free parameter, 
we should modify Eq.~\eqref{eq:news17} to read
\begin{align}
  \label{eq:news17nlo}
  \frac{\mathcal{A}_{\tilde{\rm A}\tilde{\rm B}+{\rm n}}}{2 \,s} =& 
\Gamma_A^{(s_{R;0,1})}\left[\prod_{i=1}^n \frac{1}{q_i^2}\left(\frac{s_{i-1,i}}{s_{R;i-1,i}}\right)^{\omega_i}\gamma^{(s_{R;i-1,i},s_{R;i,i+1})}(q_i,q_{i+1})\right]\non
&\times \frac{1}{q_{n+1}^2}\left(\frac{s_{n,n+1}}{s_{R;n,n+1}}\right)^{\omega_{n+1}}\Gamma_B^{(s_{R;n,n+1})}.
\end{align}
The propagation of a Reggeized gluon with momentum $q_i$ in MRK takes place 
between two emissions with momenta $k_{i-1}$ and $k_i$ 
(see Fig.~\ref{fig:kinematics}). Therefore, at NLO, the term $s_R$, which 
scales the invariant energy $s_{i-1,i}$, does depend on these two consecutive emissions 
and, in general, will be written as $s_{R;i-1,i}$. It is important to note 
that the production amplitudes should be independent of the energy scale 
chosen and, therefore,
\begin{equation}
\Gamma_{A}^{(s_{R;0,1})} = \Gamma_A^{(s_{R;0,1}')} \left(\frac{s_{R;0,1}}{s_{R;0,1}'}\right)^{\frac{\omega_1}{2}},\,\,\,\,\Gamma_{B}^{(s_{R;n,n+1})} = \Gamma_B^{(s_{R;n,n+1}')} \left(\frac{s_{R;n,n+1}}{s_{R;n,n+1}'}\right)^{\frac{\omega_{n+1}}{2}}
\end{equation}
for the particle--particle--Reggeon vertices and
\begin{multline}
\gamma^{(s_{R;i-1,i},s_{R;i,i+1})} \left(q_i,q_{i+1}\right)\\
= \gamma^{(s_{R;i-1,i}',s_{R;i,i+1}'')} \left(q_i,q_{i+1} \right)
\left(\frac{s_{R;i-1,i}}{s_{R;i-1,i}'}\right)^{\frac{\omega_{i}}{2}}
\left(\frac{s_{R;i,i+1}}{s_{R;i,i+1}''}\right)^{\frac{\omega_{i+1}}{2}}
\end{multline}
for the Reggeon--Reggeon--gluon production vertices.

At NLO, besides the two--loop corrections to the gluon Regge trajectory \cite{Fadin:1996tb}, 
there are four other contributions which affect the real emission vertex. 
The first one consists of 
virtual corrections to the one gluon production vertex \cite{Fadin:1993wh,Fadin:1994fj,Fadin:1996yv}. 
The second stems from the fact that in a chain of emissions widely separated in rapidity 
two of them are allowed to be nearby in this variable which  is known as 
{\it quasi--multi--Regge} kinematics (QMRK) \cite{Fadin:1996nw,Fadin:1997zv}. A third source is 
obtained by perturbatively expanding the Reggeon propagators in 
Eq.~\eqref{eq:news17nlo} while keeping 
MRK and every vertex at LO. 
A final fourth contribution is that of the production of 
quark--antiquark pairs \cite{Catani:1990xk,Catani:1990eg,Camici:1996st,Camici:1997ta,Fadin:1997hr}. 
The common feature of all of these new NLO elements is that they 
generate an extra power in the coupling constant without building up a corresponding 
logarithm of energy so that $\alpha_s \left(\alpha_s \ln{s}\right)^n$ terms 
are taken into account.

With the idea of introducing a jet definition later on, it is important to 
understand the properties of the production vertex which we will describe now
in some detail. 

Let us u start with the virtual corrections to the single--gluon 
emission vertex. These are rather simple and correspond to Eq.~\eqref{eq:lla2} 
with the insertion of a single kernel or impact factor 
with NLO virtual contributions (noted as ($v$)) 
while leaving the rest of the expression 
at Born level (written as ($B$)). More explicitly we have
\begin{multline}
\frac{{\cal F}^{(n)}_{\rm virtual}(\omega,s_0)}{(2\pi )^{2-D}}=\int\left[\prod_{i=1}^{n+1}\frac{\dqi}{(\omega -2\omega_i)}\right]\\
\Bigg\{\frac{\Phi^{(B)}_A(\qone)}{\qone^2}\left[\prod_{i=1}^n{\cal K}_r^{(B)}(\qi,\qip)\right]\frac{\Phi^{(v)}_B(\qnp)}{\qnp^2} \\
+\frac{\Phi^{(v)}_A(\qone)}{\qone^2}\left[\prod_{i=1}^n{\cal K}_r^{(B)}(\qi,\qip)\right]\frac{\Phi^{(B)}_B(\qnp)}{\qnp^2}\\
+\frac{\Phi^{(B)}_A(\qone)}{\qone^2}\sum_{j=1}^n 
\left[\prod_{i=1}^{j-1}{\cal K}_r^{(B)}(\qi,\qip)\right]
{\cal K}_r^{(v)}(\qj,\qjp)\\
\times\left[\prod_{i=j+1}^{n}{\cal K}_r^{(B)}(\qi,\qip)\right]
\frac{\Phi^{(B)}_B(\qnp)}{\qnp^2}\Bigg\}\label{eq:nllavirtual}.
\end{multline}

Now we turn to the discussion of how to define QMRK. For this purpose the 
introduction of an extra scale is mandatory 
in order to define a separation in rapidity space between different emissions. 
Following Ref.~\cite{Fadin:1998sh}, we call this new scale 
$s_\Lambda$. At LO MRK implies that all $s_{ij}=(k_i+k_j)^2$ are larger than $s_\Lambda$. 
In rapidity space this means that their rapidity difference $|y_i-y_j|$ is larger than 
$\ln(s_\Lambda/\sqrt{\ki^2\kj^2})$. As stated earlier, in QMRK one single pair of  emissions is allowed to 
be close in rapidity. When any of these two 
emissions is one of the external particles 
$\tilde{\rm A}$ or $\tilde{\rm B}$, it contributes as a real correction to the 
corresponding impact factor. If this is 
not the case it qualifies as a real correction to the kernel. 
This is summarized in the following 
expression where we denote real corrections  to the impact factors by ($r$):
\begin{multline}
\frac{{\cal F}^{(n+1)}_{\rm QMRK}(\omega,s_0)}{(2\pi )^{2-D}} =
\int\left[\prod_{i=1}^{n+1}\frac{\dqi}{(\omega -2\omega_i)}\right]\\
\times\Bigg\{\frac{\Phi^{(B)}_A(\qone)}{\qone^2}\left[\prod_{i=1}^n{\cal K}_r^{(B)}(\qi,\qip)\right]\frac{\Phi^{(r)}_B(\qnp)}{\qnp^2} \\
+\frac{\Phi^{(r)}_A(\qone)}{\qone^2}\left[\prod_{i=1}^n{\cal K}_r^{(B)}(\qi,\qip)\right]\frac{\Phi^{(B)}_B(\qnp)}{\qnp^2}\\
+\frac{\Phi^{(B)}_A(\qone)}{\qone^2}\sum_{j=1}^n 
\left[\prod_{i=1}^{j-1}{\cal K}_r^{(B)}(\qi,\qip)\right]{\cal K}_{\rm QMRK}(\qj,\qjp)\\
\times\left[\prod_{i=j+1}^{n}{\cal K}_r^{(B)}(\qi,\qip)\right]\frac{\Phi^{(B)}_B(\qnp)}{\qnp^2}\Bigg\}.
\label{eq:nllaqmrk}
\end{multline}
The modifications due to QMRK belonging to the kernel or to the impact 
factors are, respectively, ${\cal K}_{\rm QMRK}$ and $\Phi^{(r)}_P$, 
i.e.
\begin{align}
{\cal K}_{\rm QMRK}(\qi,\qip) =& (N_c^2-1)\int d \shat  \, \frac {I_{RR} \, \sigma_{RR\rightarrow GG}(\shat ) \, \theta(s_{\Lambda}-\shat )}{(2\pi)^D \,\qi^2 \, \qip^2}\label{eq:kqmrk},\\
\Phi^{(r)}_P(\kpure) =& \sqrt{N_c^2-1}\int d \shat  \, \frac {I_{PR} \, \sigma_{PR\rightarrow PG}(\shat )\, \theta(s_{\Lambda}-\shat )}{(2\pi) \, s}
\label{eq:phiqmrk}.
\end{align}
In both cases $\shat $ denotes the invariant mass of the two emissions in QMRK. The 
Heaviside functions are used to separate the regions of phase space where the emissions 
are at a relative rapidity separation smaller than $s_\Lambda$. It is within 
this region where the LO emission kernel is modified. $\sigma_{RR\rightarrow GG}$ and 
$\sigma_{PR\rightarrow PG}$ are the total cross sections for scattering of two Reggeons into two 
gluons, and an external particle and a Reggeon into an external particle and a gluon, 
respectively. $I_{RR}$ and $I_{PR}$ are  the corresponding invariant fluxes, and $N_c$ is the number of colors.

For those sectors remaining in the MRK we use a Heaviside function to ensure that
$s_{i-1,i} > s_\Lambda$. In this way, MRK is clearly separated from QMRK. We 
then follow the same steps as at LO and use 
Eq.~\eqref{eq:lla} with the modifications already introduced in 
Eq.~\eqref{eq:news17nlo}, i.e.
\begin{multline}
 \frac{{\cal F}_{\rm MRK}^{(n+1)}(\omega,s_0)}{(2\pi)^{2-D}}\\
= \int\left[\prod_{i=1}^{n+2}\dqi\frac{ds_{i-1,i}}{s_{i-1,i}}\left( \frac{s_{i-1,i}}{s_{R;i-1,i}}\right) ^{2\omega_i}\left(\frac{s_{i-1,i}}{\sqrt{\kim^2\ki^2}}\right)^{-\omega}\theta(s_{i-1,i}-s_\Lambda)\right]\\
\times \left(\frac{s_0}{\sqrt{\qone^2\qnpp^2}}\right)^{\omega}\frac{\Phi_A^{(B)}(\qone)}{\qone^2}\left[\prod_{i=1}^{n+1}{\cal K}_r^{(B)}(\qi,\qip)\right]\frac{\Phi_B^{(B)}(\qnpp)}{\qnpp^2} .
\label{eq:nlla}
\end{multline}
After performing the integration over the $s_{i-1,i}$ variables the following 
interesting dependence on $s_\Lambda$ arises:
\begin{multline}
 \frac{{\cal F}_{\rm MRK}^{(n+1)}(\omega,s_0)}{(2\pi )^{2-D}} = \int\left[\prod_{i=1}^{n+2}\frac{\dqi}{(\omega-2\omega_i)}\left( \frac{s_\Lambda}{s_{R;i-1,i}}\right)^{2\omega_i}\left(\frac{s_\Lambda}{\sqrt{\kim^2 \ki^2}}\right)^{-\omega}\right]\\
\times \left(\frac{s_0}{\sqrt{\qone^2\qnpp^2}}\right)^{\omega}\frac{\Phi_A^{(B)}(\qone)}{\qone^2}\left[\prod_{i=1}^{n+1}{\cal K}_r^{(B)}(\qi,\qip)\right]\frac{\Phi_B^{(B)}(\qnpp)}{\qnpp^2}.
\end{multline}

It is now convenient to go back to Eq.~(\ref{eq:news84}) and write the lower 
limit $s_0$ of the Mellin transform as a 
generic product of two 
scales related to the external impact factors, i.e. 
$s_0=\sqrt{s_{0;A} \, s_{0;B}}$. By expanding in $\alpha_s$ the factors 
with powers in $\omega$ and $\omega_i$ it is then possible to identify the 
NLO terms:
\begin{multline}
 \frac{{\cal F}_{\rm MRK}^{(n+1)}(\omega,s_0)}{(2\pi )^{2-D}}\\
 =  \int\left[\prod_{i=1}^{n+2}\frac{\dqi}{(\omega-2\omega_i)}\right]\frac{\Phi_A^{(B)}(\qone)}{\qone^2}\left[\prod_{i=1}^{n+1}{\cal K}_r^{(B)}(\qi,\qip)\right]\frac{\Phi_B^{(B)}(\qnpp)}{\qnpp^2}\\
\times\Bigg\{ 1 -\frac{\omega}{2}\ln\frac{s_\Lambda^2}{\kone^2 s_{0;A}}+\omega_1\ln\frac{s_\Lambda^2}{s_{R;0,1}^2}-\sum_{i=2}^{n+1}\left[\frac{\omega}{2}\ln\frac{s_\Lambda^2}{\kim^2 \ki^2}-\omega_i\ln\frac{s_\Lambda^2}{s_{R;i-1,i}^2}\right]\\
\phantom{\times\Bigg[ 1}-\frac{\omega}{2}\ln\frac{s_\Lambda^2}{\knp^2 s_{0;B}}+\omega_{n+2}\ln\frac{s_\Lambda^2}{s_{R;n+1,n+2}^2}\Bigg\}.
\end{multline}
To combine this expression with that of the QMRK contribution we should make a choice 
for $s_R$. The most convenient one is $s_{R;i,j}=\sqrt{s_{R;i} \, s_{R;j}}$, where  
for intermediate Reggeon propagation we use $s_{R;i}=\ki^2$, and for the connection with the 
external particles $s_{R;0}= s_{0;A}$ and $s_{R;n+2}=s_{0;B}$. We can then 
write
\begin{multline}
 \frac{{\cal F}_{\rm MRK}^{(n+1)}(\omega,s_0)}{(2\pi )^{2-D}}\\
 =  \int\left[\prod_{i=1}^{n+2}\frac{\dqi}{(\omega-2\omega_i)}\right]\frac{\Phi_A^{(B)}(\qone)}{\qone^2}\left[\prod_{i=1}^{n+1}{\cal K}_r^{(B)}(\qi,\qip)\right]\frac{\Phi_B^{(B)}(\qnpp)}{\qnpp^2}\\
\times\Bigg\{ 1 -\frac{(\omega-2\omega_1)}{2}\ln\frac{s_\Lambda^2}{\kone^2 s_{0;A}}-\sum_{i=2}^{n+1}\left[\frac{(\omega-2\omega_i)}{2}\ln\frac{s_\Lambda^2}{\kim^2 \ki^2}\right]\\
-\frac{(\omega-2\omega_{n+2})}{2}\ln\frac{s_\Lambda^2}{\knp^2 s_{0;B}}\Bigg\}.
\label{MRKfinal}
\end{multline}
This corresponds to the LO result for ${\cal F}^{(n+1)}$ plus additional 
terms where the $\omega-2\omega_i$ factor cancels in such a way that they 
can be combined with the LO result of ${\cal F}^{(n)}$. 

The quark contribution can be included in a straightforward manner since 
between the quark--antiquark emissions there is no propagation of a Reggeized 
gluon. In this way one can simply write
\begin{multline}
\frac{{\cal F}^{(n+1)}_{Q\bar{Q}}(\omega,s_0)}{(2\pi )^{2-D}}=\int\left[\prod_{i=1}^{n+1}\frac{\dqi}{(\omega -2\omega_i)}\right]\frac{\Phi^{(B)}_A(\qone)}{\qone^2}\frac{\Phi^{(B)}_B(\qnp)}{\qnp^2}\\
\times\sum_{j=1}^n \left[\prod_{i=1}^{j-1}{\cal K}_r^{(B)}(\qi,\qip)\right]{\cal K}_{Q\bar{Q}}(\qj,\qjp)\left[\prod_{i=j+1}^{n}{\cal K}_r^{(B)}(\qi,\qip)\right].
\label{eq:nllaqqbar}
\end{multline}
The production kernel can be written as
\begin{align}
{\cal K}_{Q\bar{Q}}(\qi,\qip) ~=&~ (N_c^2-1)\int d \shat  \, \frac {I_{RR} \, 
\sigma_{RR\rightarrow Q\bar{Q}}(\shat )}{(2\pi)^D \, \qi^2 \, \qip^2},
\label{eq:krrqq}
\end{align}
with $\sigma_{RR\rightarrow Q\bar{Q}}$ being the total cross section for two Reggeons 
producing the quark--antiquark pair with an invariant mass ${\hat s}$.

The combination of all the NLO contributions together generates the 
following expression for the NLO cross section: 
\begin{multline}
{\cal F}(\omega ,s_0)_{AB}=\sum_{n=0}^{\infty}\frac{1}{(2\pi )^{D-2}}\int\left[\prod_{i=1}^{n+1}\frac{\dqi}{(\omega -2\omega_i)}\right]\\
\times\frac{\Phi_A(\qone;s_{0;A})}{\qone^2}\left[\prod_{i=1}^n{\cal K}_{r}(\qi,\qip)\right]\frac{\Phi_B(\qnp;s_{0;B})}{\qnp^2},
\end{multline}
where the NLO real emission kernel contains several terms:
\begin{align}
{\cal K}_{r}(\qi,\qip) =& 
\left( {\cal K}_r^{(B)} + {\cal K}_r^{(NLO)} \right) (\qi,\qip) \non
=& \left( {\cal K}_r^{(B)} + {\cal K}_r^{(v)} + 
{\cal K}_{GG} + {\cal K}_{Q\bar Q} \right) (\qi,\qip)
\label{eq:kernelnlo},
\end{align}
with ${\cal K}_{Q\bar Q}$ given by Eq.~\eqref{eq:krrqq}. The two gluon 
production kernel ${\cal K}_{GG}$ is the combination of ${\cal K}_{\rm QMRK}$ of Eq.~\eqref{eq:kqmrk} 
and the MRK contribution in Eq.~(\ref{MRKfinal}). It explicitly reads
\begin{multline}
{\cal K}_{GG} (\qi,\qip) = (N_c^2-1)\int d \shat  \frac {I_{RR} \sigma_{RR\rightarrow GG}(\shat ) \, \theta(s_{\Lambda}-\shat )}{(2\pi)^D \, \qi^2 \, \qip^2}\\
- \int\dqt \, {\cal K}_r^{(B)}(\qi,\qt) \, {\cal K}_r^{(B)}(\qt,\qip)\frac{1}{2}\ln \left(\frac{s_{\Lambda}^2}{(\qi-\qt)^2(\qip-\qt)^2}\right). 
\label{eq:krrgg}
\end{multline}
Below we will show that in the limit $s_\Lambda\to\infty$  the second term 
of this expression subtracts the logarithmic divergence of the first one. 
When computing the total cross section, it is natural to remove the dependence 
on the parameter $s_\Lambda$ in this way. For our jet production cross 
section, however, we prefer to retain the dependence upon $s_\Lambda$.

For the impact factors a similar expression including virtual and MRK corrections as in 
Eq.~\eqref{eq:phiqmrk} arises:
\begin{multline}
\label{eq:impactfactornlo}
\Phi_P(\qone;s_{0;P}) = \Phi_P^{(B)}+\Phi_P^{(v)}+\sqrt{N_c^2-1}\int d \shat  \frac{I_{PR} \, \sigma_{PR}(\shat ) \, \theta(s_{\Lambda}-\shat)}{(2\pi) \, s}\\
-\int \dqt \, \Phi^{(B)}_P(\qt) \, {\cal K}_r^{(B)}(\qt,\qone)\frac{1}{2}\ln \left(\frac{s_{\Lambda}^2}{(\qone-\qt)^2s_{0;P}}\right).
\end{multline}
From this expression it is now clear why to choose the factorized form 
$s_0=\sqrt{s_{0;A} \, s_{0;B}}$: in this way each of the impact factors $\Phi_{A,B}$ 
carry its own $s_{0;A,B}$ term at NLO independently of the choice of scale for the other.

To conclude this section, for the sake of clarity, the different contributions to 
the NLO BFKL kernel

\begin{center}
\begin{figure}[htbp]
  \centering
  \includegraphics[width=11cm]{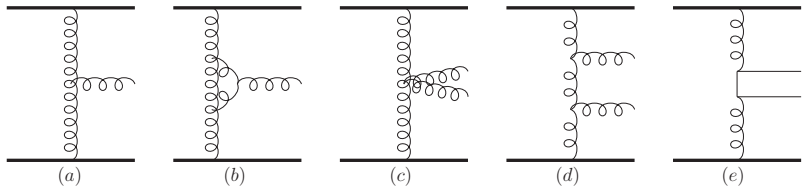}
  \caption{Contributions to real emission kernel at LO (a) and NLO (b-e).}
  \label{fig:grouping}
\end{figure}
\end{center}

\begin{center}
\begin{tabular}[h]{lcc}
CONTRIBUTION          & NUMBER OF EMISSIONS  & Fig.\ref{fig:grouping}\\
\hline
MRK @ LO              & $n$                  & (a)\\
\hline
Virtual               & $n$                  & (b)\\
QMRK                  & $n+1$                & (c)\\
MRK @ NLO             & $n+1$                & (d)\\
Quark--antiquark pair & $n+1$                & (e)\\
\hline
\end{tabular}
\end{center}
are pictorially represented in Fig.~\ref{fig:grouping}.
As a final remark we would like to indicate that the divergences present in 
the gluon trajectories $\omega_i$ (see Ref.~\cite{Fadin:1998py,Ciafaloni:1998gs}) are all canceled 
inside the inclusive terms. We will see how the soft and collinear 
divergences of the production vertex are either canceled amongst its 
different components or are 
regularized by the jet definition.

After having introduced the notation and highlighted the different 
constituents of a BFKL production kernel at NLO, we will
discuss, in the coming section, the properties of this kernel in detail.


\section[Properties of the NLO BFKL kernel, Resummation]{Properties of the NLO BFKL kernel and resummation of terms beyond NLO}
\label{sec:bfklmellin}

Shortly after the derivation of the NLO corrections to the BFKL kernel \cite{Fadin:1998py,Ciafaloni:1998gs,Ciafaloni:1998kx,Camici:1997ij} it was pointed out \cite{Ross:1998xw,Levin:1998pk} that these corrections are not only large and negative, but that they also lead to possible instabilities. After the suggestion that NNLO or even NNNLO terms might cure the problem \cite{Blumlein:1997bs,Blumlein:1997em}, various attempts have been made to guess the  higher order terms and to resum them \cite{Salam:1998tj,Ciafaloni:1998iv,Ciafaloni:1999yw,Brodsky:1998kn,Schmidt:1999mz,Forshaw:1999xm,Vera:2005jt}. Actually, these procedures have proven to be very successful. 

Although such a resummation does not influence the construction of an explicit jet vertex in the following chapter, it plays a vital role when we discuss the impact of the gluon Green's function in chapter \ref{sec:decorrelation}. Therefore, we like to motivate the resummation by a study of the kernel itself. Furthermore, we will discuss its implementation and consequences. 

\subsection{The LO kernel and its eigenfunctions}

The LO BFKL equation \eqref{eq:bfklequation} is solved if one knows a complete set of eigenfunctions $\phi_{n,\nu}(\qpure)$. At LO these eigenfunctions are well known to be
\begin{equation}
  \phi_{n,\nu}(\qpure) = \frac{1}{\pi\sqrt{2}}\plusinu{\qpure^2}e^{in\theta},
  \label{eq:eigenfunctions}
\end{equation}
where the transverse momentum is expressed in terms of its modulus $\sqrt{\qpure^2}$ and its azimuthal angle $\theta$. The eigenvalues associated with the eigenfunctions $\phi_{n,\nu}(\qpure)$ are $\omega(n,\nu)=\asbar \chi_0(|n|,1/2+i \nu)$, where $\asbar=\alpha_s N_c/\pi$ and
\begin{equation}
\chi_0(n,\gamma) =2\psi(1)-\psi\left(\gamma+\frac{n}{2}\right)-\psi\left(1-\gamma+\frac{n}{2}\right) .
\label{eq:lokernel}
\end{equation}
The function $\psi$ is given as $\psi(x)=\Gamma'(x)/\Gamma(x)$, where $\Gamma$ is the Euler gamma function.

It is natural to change from the presentation in terms of transverse momenta to those of conformal spin $n$ and anomalous dimension $\gamma=1/2+i \nu$. We use this change to introduce an operator representation used in Ref.~\cite{Ivanov:2005gn} and extended in Ref.~\cite{Vera:2006un,Vera:2006xa}. 
The transverse momentum representation is defined by
\begin{align}
  \opqone\ket{\qone} =& \qone\ket{\qone} & \braket{\qone}{\qtwo} =& \del{\qone-\qtwo},
\end{align}
the kernel of the operator $\opk$ is
\begin{equation}
  \mathcal{K}(\qone,\qtwo) = \bra{\qone}\opk\ket{\qtwo},
\end{equation}
and the BFKL equation together with its formal solution simply reads  
\begin{align}
  \label{eq:opbfkl}
  \omega\,\opf_\omega =& \opf_\omega+\opk\opf_\omega &&\Rightarrow& \opf_\omega=\frac{1}{\omega-\opk} .
\end{align}

\begin{figure}[htbp]
  \centering
  \includegraphics[height=4.75cm]{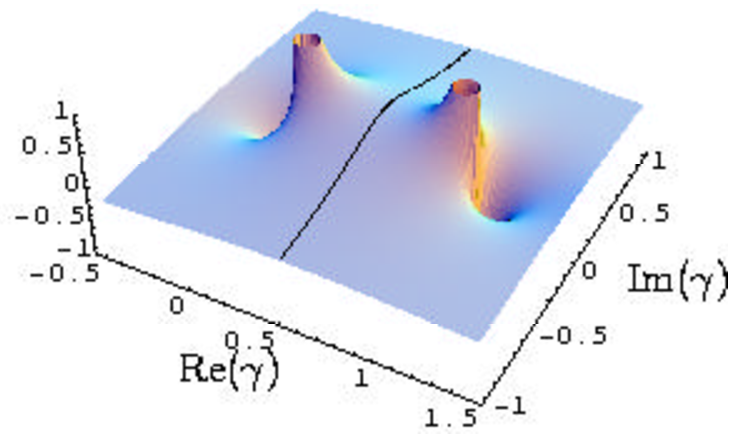}\\
  \vspace{.5cm}
  \includegraphics[height=4.75cm]{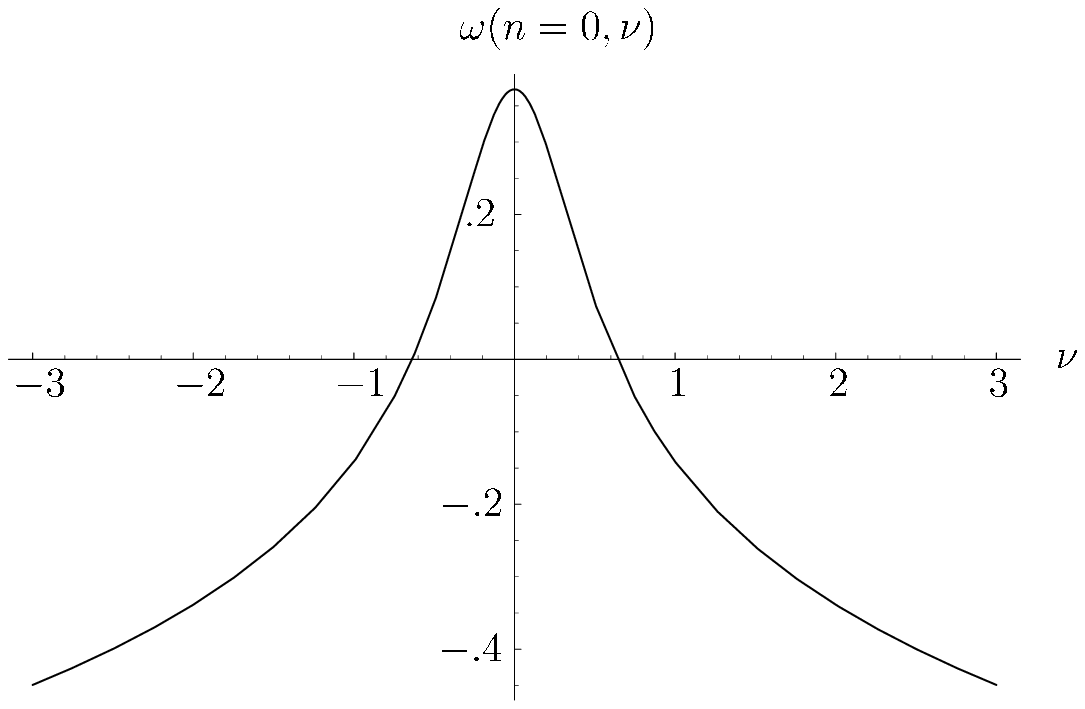}
  \caption{LO kernel for conformal spin $n=0$ at a scale $\mu=30$GeV.}
\label{fig:lokernel}
\end{figure}

As a second basis we will use $n$ and $\nu$. The eigenfunctions given in Eq.~\eqref{eq:eigenfunctions} relate both bases:
\begin{equation}
  \braket{\qone}{\nu,n}=\frac{1}{\pi\sqrt{2}}\plusinu{\qone^2}e^{in\theta_1},
\end{equation}
where $\theta_1$ is the azimuthal angle of $\qone$. The normalization of this new basis follows directly as
\begin{equation}
  \braket{n',\nu'}{\nu,n} = \int\ddqone \frac{1}{2\pi^2}\left(\qone^2\right)^{i(\nu-\nu')-1}e^{i(n-n')\theta}=\delta(\nu-\nu')\delta_{nn'} ,
\end{equation}
and the LO eigenvalue equation simply reads
\begin{equation}
  \label{eq:actionoflok}
  \opk\ket{\nu,n} = \omega\left(n,\nu\right)\ket{\nu,n} = \asbar\chi_0\left(|n|,\frac{1}{2}+i\nu\right)\ket{\nu,n}.
\end{equation}
Often the functions  $\omega(n,\nu)$ and $\chi(n,1/2+i\nu)$ are called kernel as well. 
Knowing the action of the kernel (see Eq.~\eqref{eq:actionoflok}), we can use the formal solution of the BFKL equation \eqref{eq:opbfkl} to write
\begin{equation}
  \bra{n,\nu}\opf_\omega\ket{\nu',n'}  = \frac{1}{\omega-\asbar\chi_0\left(|n|,\frac{1}{2}+i\nu\right)}\delta(\nu-\nu')\delta_{nn'}  .
\label{eq:fnnulo}
\end{equation}
Using our operator formalism, it is straight forward to express the Green's function in physical terms of energy and transverse momenta:
\begin{align}
&\hspace{-1cm}f\left(\qone,\qtwo,\frac{s}{s_0}\right)\non
=& \int\frac{d\omega}{2\pi i}\,
\sum_{n,n'}\iint d\nu \,d\nu'\,
\braket{\qone}{\nu,n}
\bra{n,\nu}\opf_\omega\ket{\nu',n'}
\braket{n',\nu'}{\qtwo} \left(\frac{s}{s_0}\right)^{\omega}\non
 =& \sum_n \int\frac{d\gamma}{2\pi i}\,\int\frac{d\omega}{2\pi i}\,\frac{1}{\omega-\asbar\chi_0\left(|n|,\gamma\right)}\left(\frac{s}{s_0}\right)^{\omega}\frac{1}{\qone^2}\left(\frac{\qone^2}{\qtwo^2}\right)^\gamma e^{in(\theta_1-\theta_2)}\non
 =& \sum_n \int\frac{d\gamma}{2\pi i}\,\left(\frac{s}{s_0}\right)^{\asbar\chi_0\left(|n|,\gamma\right)}\frac{1}{\qone^2}\left(\frac{\qone^2}{\qtwo^2}\right)^\gamma e^{in(\theta_1-\theta_2)}.
\label{eq:fnnulo2}
\end{align}
For the last step we used Cauchy's theorem to perform the $\omega$ integration. Let us, now, focus on the conformal spin $n=0$ which is the dominant contribution. We will come back to higher conformal spins later. The $\gamma$ dependence of the LO kernel is shown in Fig.~\ref{fig:lokernel}. It has a saddle point at $\gamma=1/2$, and along the contour $\gamma=1/2+i\nu$ this saddle point appears as a maximum at $\nu=0$. Since the kernel is exponentiated this pronounced maximum allows for a solution by a saddle point approximation, where one expands the kernel in $\nu$ and keeps only the first two terms.

\subsection{The NLO kernel}
\label{sec:bfklresummation}

The action of the NLO kernel on the eigenfunctions of the LO one has been calculated in Ref.~\cite{Kotikov:2000pm} and reads
\begin{multline}
  \int\ddqtwo \; \mathcal{K}(\qone,\qtwo)\plusinup{\frac{\qtwo^2}{\qone^2}}e^{in'(\theta_2-\theta_1)}\\
=\asbar(\qone^2)\left[\chi_0\left(|n'|,\frac{1}{2}+i\nu'\right)+\frac{\asbar(\qone^2)}{4}\delta\left(|n'|,\frac{1}{2}+i\nu'\right)\right],
\label{eq:KactingKotikov}
\end{multline}
where
\begin{align}
   \delta(n,\gamma) =& 4\chi_1(n,\gamma)+\frac{\beta_0}{2N_c} \left[\Psi'\left(\gamma+\frac{n}{2}\right)-\Psi'\left(1-\gamma+\frac{n}{2}\right)\right],\label{eq:definitiondelta}\\
  \beta_0=&\frac{11N_c-2n_f}{3} .
\end{align}
This means that the NLO kernel is diagonal with respect to the LO eigenfunctions up to terms associated with the running of the coupling. While $\chi_1(n,\gamma)$ is symmetric under the exchange $\gamma\leftrightarrow 1-\gamma$, the remainder of $\delta(n,\gamma)$ is antisymmetric. One could remove this part by modifying the LO eigenfunctions in the following way
\begin{equation}
  \plusinup{\frac{\qtwo^2}{\qone^2}}\quad\longrightarrow\quad\sqrt{\frac{\asbar(\qone^2)}{\asbar(\qtwo^2)}}\plusinup{\frac{\qtwo^2}{\qone^2}}.
\end{equation}
Unfortunately, this set of functions no longer forms an orthonormal basis. Therefore, we stay with the LO eigenfunctions as they are. The antisymmetric term will vanish for some other reason in our study of angular decorrelation, as we will show in Eq.~\eqref{eq:fnloacting2}. 
The function $\chi_1$ for general conformal spin $n$ is given by
\begin{align}
  \chi_1(n,\gamma) =& \phantom{+}\mathcal{S}\chi_0(n,\gamma) + \frac{3}{2}\zeta(3)-\frac{\beta_0}{8N_c}\chi_0^2(n,\gamma)\non
& +\frac{1}{4}\left[\psi''\left(\gamma+\frac{n}{2}\right)+\psi''\left(1-\gamma+\frac{n}{2}\right)-2\phi(n,\gamma)-2\phi(n,1-\gamma)\right]\non
&- \frac{\pi^2\cos(\pi\gamma)}{4\sin^2(\pi\gamma)(1-2\gamma)}\Bigg\{\left[3+\left(1+\frac{n_f}{N_c^3}\right)\frac{2+3\gamma(1-\gamma)}{(3-2\gamma)(1+2\gamma)}\right]\delta_{n,0}\non
&\hspace{2cm}-\left(1+\frac{n_f}{N_c^3}\right)\frac{\gamma(1-\gamma)}{2(3-2\gamma)(1+2\gamma)}\delta_{n,2}\Bigg\},
\label{eq:nlokernel}
\end{align}
with the constant  ${\mathcal S} = (4 - \pi^2 + 5 {\beta_0}/{N_c})/12$.  $\zeta(n)=\sum_{k=1}^\infty k^{-n}$ is the Riemann zeta function  while the function $\phi$ reads
\begin{align}
  \phi(n,\gamma) =& \sum_{k=0}^\infty \frac{(-1)^{k+1}}{k+\gamma+\frac{n}{2}}\Bigg(
\phantom{+}\psi'(k+n+1)-\psi'(k+1)\non
&\hphantom{\sum_{k=0}^\infty \frac{(-1)^{k+1}}{k+\gamma+\frac{n}{2}}\Bigg(} 
+ (-1)^{k+1}\left[\beta'(k+n+1)+\beta'(k+1)\right]\non
&\hphantom{\sum_{k=0}^\infty \frac{(-1)^{k+1}}{k+\gamma+\frac{n}{2}}\Bigg(}
 +\frac{\psi(k+1)-\psi(k+n+1)}{k+\gamma+\frac{n}{2}}\Bigg),
\end{align}
with
\begin{equation}
  \beta'(\gamma) = \frac{1}{4}\left[\psi'\left(\frac{1+\gamma}{2}\right)-\psi'\left(\frac{\gamma}{2}\right)\right].
\end{equation}

\begin{figure}[htbp]
  \centering
  \includegraphics[height=4.5cm]{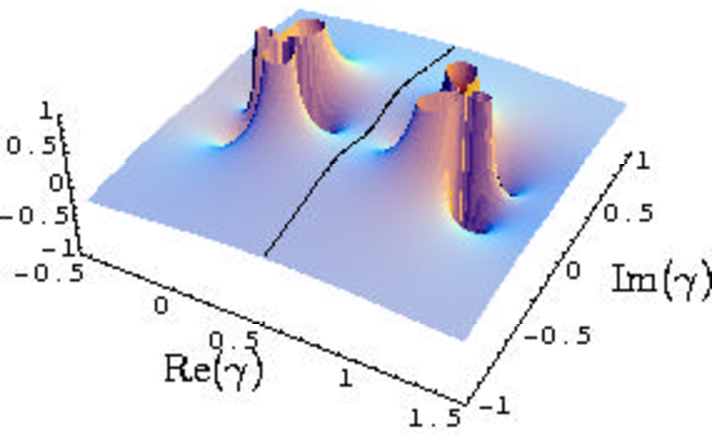}\\
  \vspace{.5cm}
  \includegraphics[height=4.5cm]{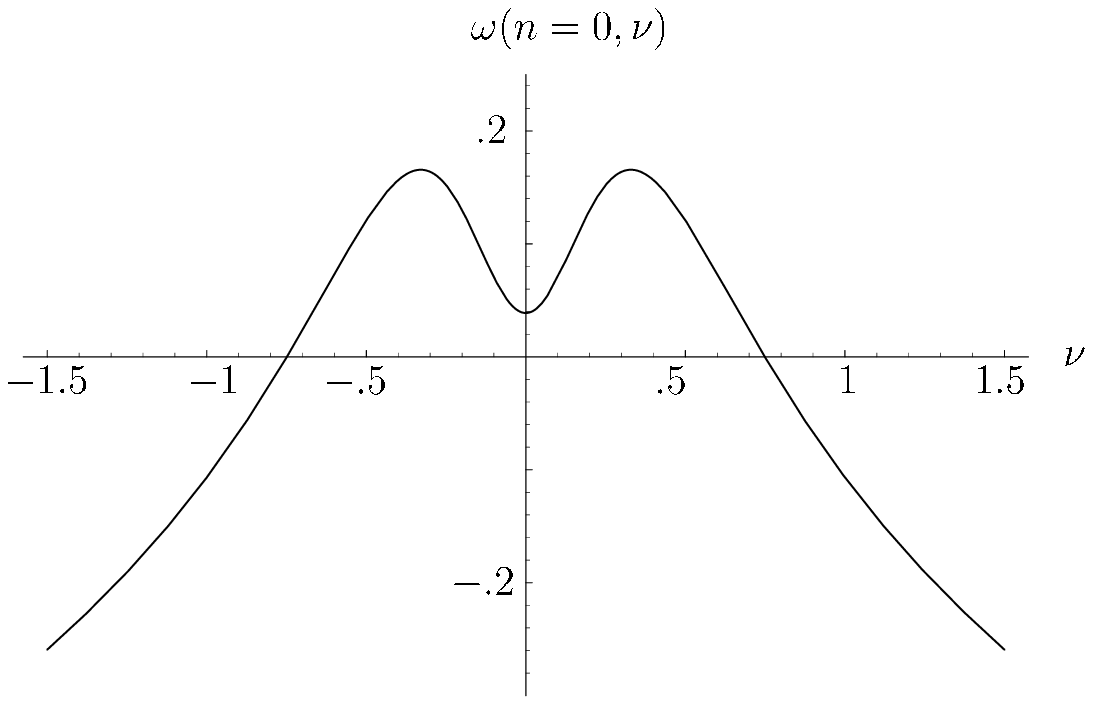}
  \caption{NLO kernel for conformal spin $n=0$ at a scale $\mu=30$GeV.}
\label{fig:nlokernel}
\end{figure}

Let us have a closer look at the conformal spin $n=0$ component of the NLO kernel, depicted in Fig.~\ref{fig:nlokernel}, and compare it with the LO one, depicted in Fig.~\ref{fig:lokernel}.

The LO kernel has a single saddle point at $\gamma=1/2$. This property is changed in the NLO case if $\asbar \gtrsim .05$. If we examine the contour $\gamma=1/2+i\nu$ for real $\nu$, we find, instead of a maximum at $\nu=0$, two maxima at $\nu\ne 0$ accompanied by a local minimum at $\nu=0$. An unwelcome consequence of the saddle points having complex values is an oscillatory behavior in the ratio of the transverse momenta, which in the end can lead to negative cross sections. To understand the origin of this strange behavior, it is fruitful to investigate the pole structure of the kernel. The LO kernel  (given in Eq.~\eqref{eq:lokernel}) has a simple pole at $\gamma=0$:
\begin{equation}
  \chi_0(0,\gamma) = \frac{1}{\gamma}+\order{\gamma^0} .
\end{equation}
In contrast, the NLO kernel (given in Eq.~\eqref{eq:nlokernel}) has a more complicated pole structure around $\gamma=0$:
\begin{equation}
   \chi_1(0,\gamma) = -\frac{1}{2\gamma^3}+\frac{a}{\gamma^2}+\frac{b}{\gamma}+\order{\gamma^0} .
\end{equation}
The cubic pole does compensate for the equivalent terms appearing when the symmetric energy scale is changed to an asymmetric one. The quadratic pole can be understood as a sequence of a small-$x$ branching and a not-small-$x$ branching, since the explicit form of $a$ turns  out to equal
\begin{equation}
  a = A_{gg}(0) + \frac{C_F}{C_A}A_{qg}(0) - \frac{\beta_0}{8N_c} ,
\end{equation}
where $A_{gg}(\omega)$ ($A_{qg}(\omega)$) is the Mellin transform of the splitting function $P_{gg}(z)$ ($P_{qg}(z)$). The $\beta_0$ term reveals that also the running of the coupling generates a quadratic pole at $\gamma=0$.

These higher-poles-terms are responsible for the problematic behavior of the kernel.   The asymmetric scale change and the not-small-$x$ branching  are related to the collinear limit. 
Imposing that the collinear limit of the kernel does not conflict with the DGLAP equations, the problem can be cured \cite{Salam:1998tj,Ciafaloni:1998iv,Ciafaloni:1999yw,Ciafaloni:2003rd} by a kernel which resums sub-leading collinear corrections. 
The focus, in these works, has been on conformal spin $m=0$. Since in our study of decorrelation in chapter~\ref{sec:decorrelation} we are interested in angular dependences, we extend the solution proposed in Ref.~\cite{Salam:1998tj} to general conformal spin. 

It can be shown \cite{Salam:1998tj,Ciafaloni:1998iv,Ciafaloni:1999yw,Ciafaloni:2003rd} that this behavior is an artifact of the truncation of the perturbative expansion. If one starts from a proper form of the kernel in the DGLAP limit with an asymmetric energy scale, then the change to a symmetric energy scale for the BFKL regime can be obtained by a $\omega$-shift in $\gamma$. An expansion in $\omega$ then gives the well known form of the NLO kernel. The resummation is based on the strategy to keep this $\omega$ dependence in the kernel leading to a transcendental equation for the kernel.

Technically the resummation is realized -- for a given accuracy $\order{\asbar^l}$ -- by replacing divergences $\asbar^n / (\gamma+\frac{m}{2})^k$ ($n\le l$) with terms $\asbar^n / (\gamma+\frac{m}{2}+\frac{\omega}{2})^k$ without changing the correct expansion to $\order{\asbar^l}$.
For a specific pole this change induces modifications of the order $\order{\asbar^{n+1}}$ since $\omega\sim\asbar\chi_0+\order{\asbar^2}$. Therefore, for all poles $n<l$ we have to compensate this modification to keep the correct expansion to $\order{\asbar^l}$.
Let us denote by $\chi(m,\gamma)$ the BFKL kernel to all orders. Its expansion reads 
\begin{equation}
  \label{eq:mcompleteexpanded}
  \chi(m,\gamma) = \sum_{n=0}^\infty \asbar^n \chi_n(m,\gamma) .
\end{equation}
So far the LO kernel $\chi_0$ and the NLO kernel $\chi_1$ are known. The fixed order contribution as an expansion in $\gamma+\frac{m}{2}$ and $1-\gamma+\frac{m}{2}$, respectively, reads
\begin{subequations}
\begin{align}
  \chi_n(m,\gamma) =& \sum_{k=0}^\infty  \frac{d_{n,k}(m)}{\left(\gamma+\frac{m}{2}\right)^k}+\order{\gamma+\frac{m}{2}} \\
  \chi_n(m,\gamma) =& \sum_{k=0}^\infty  \frac{\bar{d}_{n,k}(m)}{(1-\gamma+\frac{m}{2})^k} +\order{1-\gamma+\frac{m}{2}}
\end{align}
\end{subequations}
For the LO and NLO BFKL kernel we have $d_{n,k}(m)=\bar{d}_{n,k}(m)$ and 
\begin{subequations}
\begin{align}
  d_{0,1}(m) =& 1, \\
  d_{1,1}(m) =& \;\mathcal{S}-\frac{\pi^2}{24}
+\frac{1}{8}\left[\Psi'\left(\frac{m+1}{2}\right)-\Psi'\left(\frac{m+2}{2}\right)\right] +\frac{1}{2}\Psi'(m+1)\non
&+\frac{\beta_0}{4N_c}\big(\Psi(m+1)-\Psi(m)\big)-\frac{\delta_{m,0}}{36}\left(67+13\frac{n_f}{N_c^3}\right)\non
&-\frac{47\delta_{m,2}}{1800}\left(1+\frac{n_f}{N_c^3}\right),\\
 d_{1,2}(m) =& -\frac{\beta_0}{8N_c}-\frac{1}{2}\left(\Psi(m+1)-\Psi(1)\right)-\frac{\delta_{m,0}}{12}\left(11+\frac{2n_f}{N_c^3}\right)\non
&-\frac{\delta_{m,2}}{60}\left(1+\frac{n_f}{N_c^3}\right), \\
 d_{1,3}(m) =& -\frac{1}{2} .
\end{align}
\end{subequations}

We now introduce a class of resummed kernels $\chi^{(N)}(\gamma)$. Its perturbative expansion reproduces for the first $N$ terms the expansion coefficients of the exact BFKL kernel. The differences to the exact kernel are of higher order.
\begin{equation}
  \label{eq:mresummedexpanded}
  \chi^{(N)}(m,\gamma) = \sum_{n=0}^N \asbar^n \chi_n(m,\gamma) + \sum_{n=N+1}^\infty \asbar^n \chi_n^{(N)}(m,\gamma) 
\end{equation}
Although these resummed kernels will become $\omega$ dependent, we omit to write this dependence explicitly.
Its expansion in $\gamma$ defines the coefficients $d^{(N)}_{n,k}(m)$ in the following way
\begin{align}
  \label{eq:mchiNexpanded}
  \chi^{(N)}_n(m,\gamma) =&  \sum_{k=0}^\infty \frac{d^{(N)}_{n,k}(m)}{\left(\gamma+\frac{m}{2}\right)^k}+\order{\gamma+\frac{m}{2}} \\
  \chi^{(N)}_n(m,\gamma) =&  \sum_{k=0}^\infty \frac{\bar{d}^{(N)}_{n,k}(m)}{(1-\gamma+\frac{m}{2})^k}+\order{1-\gamma+\frac{m}{2}} .
\end{align}

It is now necessary to define a set of functions $D_k(\gamma)$ which are regular for $\gamma>1/2$ and behave at $\gamma = 0$ like $1/\gamma^k$. The concrete form of $D_k(\gamma)$ is of higher order accuracy.
Developing the ideas presented in Ref.~\cite{Salam:1998tj}, we construct from these functions a kernel $\chi^{(0)}(\gamma)$ which has its pole shifted by $\omega/2$:
\begin{multline}
\label{eq:mchinode}
  \chi^{(0)}(m,\gamma) = \chi_0(m,\gamma)+d_{0,1}(m)\left[D_1\left(\gamma+\frac{m}{2}+\frac{\omega}{2}\right)-D_1\left(\gamma+\frac{m}{2}\right)\right]\\
+\bar{d}_{0,1}(m)\left[D_1\left(1-\gamma+\frac{m}{2}+\frac{\omega}{2}\right)-D_1\left(1-\gamma+\frac{m}{2}\right)\right].
\end{multline}
The master formula for $N>0$ then is a recursion formula and  a generalization of Eq.~(3.2) in Ref.~\cite{Salam:1998tj}:
\begin{multline}
 \label{eq:mchiN}
  \chi^{(N)}(m,\gamma) = \chi^{(N-1)}(m,\gamma)+\asbar^N\left(\chi_N(m,\gamma)-\chi^{(N-1)}_N(m,\gamma)\right)\\
+\asbar^N\sum_{k=1}^{N+1}\Bigg[\left(d_{N,k}(m)-d^{(N-1)}_{N,k}(m)\right)\left[D_k\left(\gamma+\frac{m}{2}+\frac{\omega}{2}\right)-D_k\left(\gamma+\frac{m}{2}\right)\right]\\
+\left(\bar{d}_{N,k}(m)-\bar{d}^{(N-1)}_{N,k}(m)\right)\left[D_k\left(1-\gamma+\frac{m}{2}+\frac{\omega}{2}\right)-D_k\left(1-\gamma+\frac{m}{2}\right)\right]\Bigg] .
\end{multline}
The terms $-\asbar^N\chi^{(N-1)}_N$, $-\asbar^N d^{(N-1)}_{N,k}$, and $-\asbar^N {\bar{d}}^{(N-1)}_{N,k}$  are included to avoid double counting. The second and third line of Eq.~\eqref{eq:mchiN} shifts the poles of $\chi_N$ by $\omega/2$.
The solution of the transcendental equation $\asbar \chi^{(N)}(m,\gamma) = \omega$ for $\omega$ is the resummed kernel $\omega(m,\gamma)$, which replaces $\asbar\chi_0(m,\gamma)+\asbar^2\chi_1(m,\gamma)$.

\begin{figure}[htbp]
  \centering
  \includegraphics[height=4.5cm]{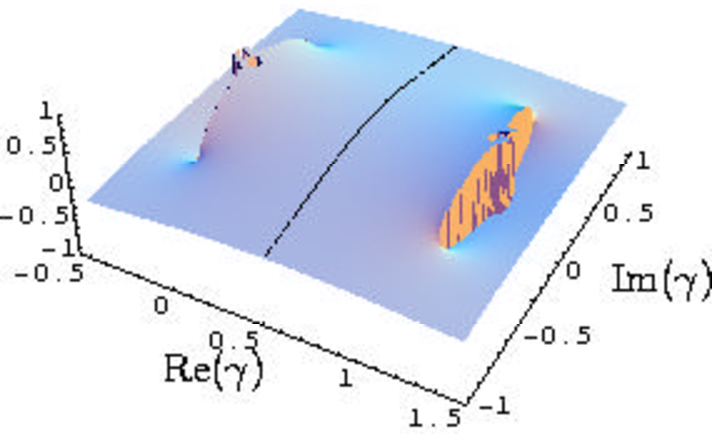}\\
  \vspace{.5cm}
  \includegraphics[height=4.5cm]{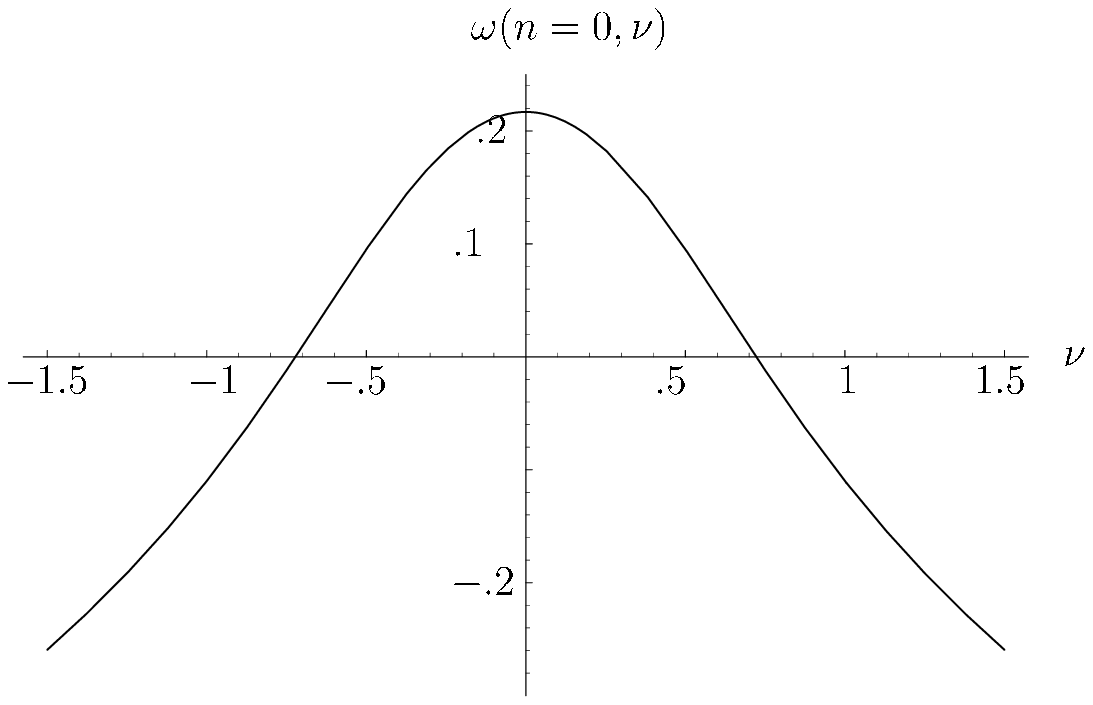}
  \caption{Resummed kernel (scheme 3) for conformal spin $n=0$ at a scale $\mu=30$GeV.}
\label{fig:resummedkernel}
\end{figure}

Even though this procedure to modify the kernel is already quite concrete, we want to keep the possibility to vary this procedure within its limitations. Consequently, we study different choices for $D_k(\gamma)$ and do this in the same line as Ref.~\cite{Salam:1998tj}. The explicit prescription of four different schemes can be found in appendix~\ref{sec:resummation}. There we will discuss the different schemes which overall are very alike. Some minor differences let us favor scheme 3 and we show the effect of the resummation, with scheme 3 as being representative,  in Fig.~\ref{fig:resummedkernel}. The change to the pure NLO kernel is obvious: the poles at $\gamma=0$ and  $\gamma=1$ disappear and the shape along the contour $\gamma=1/2+i\nu$ has one single maximum at $\nu=0$.

\section{Phenomenology}
\label{sec:phenomenology}

So far, we have introduced the BFKL equation at leading and next-to-leading order. Furthermore, we have discussed properties of the BFKL kernel and the need for a resummation of terms beyond NLO accuracy. Before we turn to consider specific observables in the following chapters, we would like to give some kind of overview over the phenomenology of BFKL dynamics.

If one is willing to accept non perturbative input, a broad field of processes may be describable by the BFKL equation. We already mentioned the proton structure function $F_2$, and diffractive scattering processes fall in this category as well. 
To study pure BFKL dynamics, it is necessary that both scattering partners provide a similar and hard scale. In the following we describe the phenomenology of processes that fulfill this     necessity.

One of the most famous predictions from the BFKL equation is the power-like rise $s^\omega$ of the total cross section with increasing energy. Due to $\omega$ being larger than 0, this behavior at some point violates unitarity, e.g. for hadronic processes the Froissart-Martin bound \cite{Froissart:1961ux,Martin:1962rt} restricts the growth of the total cross section like $\sigma_{\rm tot}<A \ln^2 s$, where $A$ is some constant. Therefore, for extremely high energies the descriptiveness by BFKL dynamics ends and different or modified models enter the game.

But up to this regime, the power like rise should be observable, and the scattering of highly virtual photons has been proposed in Refs.~\cite{Bartels:1996ke,Barger:1996kp,Brodsky:1997sd} as an appropriate environment to study this phenomenon. 
The virtual photons interact strongly when fluctuating into a quark-antiquark pair. The virtualities of the photons can be tuned to be of equal, or at least similar, scale such that DGLAP like evolution in the transverse momenta is suppressed offering a clean environment described by the BFKL equation.
Therefore, it is often called the ``gold-plated'' process to test BFKL dynamics and is an object of intense studies from theoretical \cite{Bartels:1997er,Bartels:2000sk,Kwiecinski:2000zs,Brodsky:2001ye} and experimental side \cite{Acciarri:1998ix,Abbiendi:2001tv}. The large logarithm in this context is $Y\sim\ln s_{\gamma^*\gamma^*}/\sqrt{Q_1^2 Q_2^2}$, where $Q_1^2$ and $Q_2^2$ are the photon virtualities. The data clearly show a rise with $Y$ as expected by BFKL. Nevertheless, this rise is less steep than predicted by a LO BFKL calculation. 

Another candidate, summoned as a witness, are the already mentioned Mueller-Navelet jets. To be more explicit, we think of hadron-hadron collisions and of two jets emitted in the very forward/ backward region with a similar semihard scale $p^2$: $s\gg p^2\gg\Lambda_{\rm QCD}^2$. For large rapidity differences logarithms of the form $[\alpha_s\ln( s/p^2)]^n$ have to be resummed, which can be done in the framework of the BFKL equation. Mueller and Navelet proposed this process in Ref.~\cite{Mueller:1986ey} as ideal to apply the BFKL equation and predicted a power like rise of the cross section. 
However, to see this growth of the cross section directly as a manifestation of Reggeization is hardly possible because it is drastically damped by the behavior of the parton distribution functions (PDFs) for $x\to 1$. One way out is to fix the PDFs and to vary the center of mass energy of the hadron collider itself, and thereby vary the rapidity difference $\Delta\eta$ between the both jets. BFKL theory predicts a rapidity dependence of the cross section like $\sigma\sim\exp((\alpha-1)\Delta\eta)/\sqrt{\Delta\eta}$, with $\alpha$ the so-called {\it intercept} .
Usually, the collider energy is fixed and not tunable, but the {D$\emptyset$} collaboration analyzed data taken at the Tevatron $p\bar p$-collider from two periods of measurement. From these two points at $\sqrt{s}=$ 1800GeV and 630GeV, they extracted an intercept of $1.65\pm.07$ \cite{Abbott:1999ai}. This is an even stronger growth of the cross section than predicted by a LO BFKL calculation which for the kinematics of the {D$\emptyset$} experiment yields a value of $1.45$. It has been argued \cite{Andersen:2001kt} that the exact experimental and theoretical definitions of the cross sections disagreed making an interpretation of the results difficult, and the fact that the experimental determination of the intercept is based on just two data points leaves room for more possible explanations.

The measurement of forward jets at an electron-proton collider like HERA acts a combination of both processes and has been an object of intense studies as well.
Compared to the proton-proton collider the dependence on the parton distribution functions can be singled out more easily. The large logarithm here is $Y\sim\ln x_{\rm FJ}/x_{\rm Bj}$, where $x_{\rm FJ}$ is the longitudinal momentum fraction of the forward jet, and the variable $x_{\rm Bj}=Q^2/s_{ep}$ is given by the photon virtuality. By varying the virtuality of the photon one can tune $x_{\rm Bj}$ and thus tune $Y$ without touching $x_{\rm FJ}$. 
Measurements taken by the H1 and the ZEUS collaboration \cite{Adloff:1998fa,Breitweg:1998ed} have successfully been compared to BFKL calculations \cite{Bartels:1996gr,Kwiecinski:1999hv}, while fixed order calculations underestimate the cross sections. Also the intercept has been extracted from these experiments \cite{Contreras:1998pc} and is in better agreement with the theoretical predictions then in the case of the Mueller-Navelet jets. 
Recently, new data on forward jet production at HERA have been published \cite{Chekanov:2005yb,Aktas:2005up}. With the higher integrated luminosity, it was possible to present the triple differential cross section $d\sigma/(dx_{\rm Bj}\,dQ^2\,dp_{T,\rm jet}^2)$, and first promising comparisons to NLO BFKL calculations have been presented \cite{Kepka:2006xe} even though they rely on the saddle point approximation which for the NLO BFKL case is questionable as we will discuss in section \ref{sec:mnatnlo} in detail.

A very interesting experiment at an electron proton collider, as well,  is the production of a vector meson --  like $\rho$, $\phi$ or $J/\psi$ -- 
with a large momentum transfer $-t\gg \Lambda_{\rm QCD}$. Considering the case of a large rapidity gap between the proton remanent and the vector meson, the assumption of just a two gluon exchange fails. This process was proposed in Ref.~\cite{Forshaw:1995ax} as an observable to be described by the BFKL equation. In deed, further theoretical studies \cite{Bartels:1996fs,Enberg:2003jw,Poludniowski:2003yk} are in very good agreement with the data from experiment \cite{Aktas:2006qs}. 

For all these processes a complete NLO calculation is still outstanding. For the processes involving the virtual photon the reason is that the calculation of the NLO impact factor for the virtual photon is still work in progress \cite{Bartels:2000gt,Bartels:2001mv,Bartels:2002uz,Bartels:2004bi,Fadin:2001ap,Fadin:2002tu}. The NLO Mueller-Navelet  jet vertices are in principle available \cite{Bartels:2001ge,Bartels:2002yj}, but are not yet cast in a form ready to be combined with the NLO Green's function. 

Instead, the impact factor for a virtual photon going into a light vector meson was calculated at NLO \cite{Ivanov:2004pp} up to terms suppressed by the virtuality of the photon. Quite recently, it has been used to calculate the electroproduction of two light vector mesons in NLO accuracy within the framework of BFKL theory \cite{Ivanov:2005gn}.

We will comment on the status for the angular observables in chapter~\ref{sec:decorrelation}, where we present new calculations for these observables. 
Meanwhile we describe -- in the following chapter -- how to calculate the inclusive production of jets in two different 
environments. The first one is the case of the interaction between two 
small and perturbative objects, namely highly virtual photons, and 
the second will be the collision of two large and non-perturbative 
external particles such as the ones taking place at hadron-hadron colliders.

\clearpage{\pagestyle{empty}\cleardoublepage}
\chapter{Jet production}
\label{sec:jetproduction}

In this chapter we study the inclusive production of a jet at central rapidity. We start with a LO description. With that first section we settle the notations and the framework, which we then -- in the second section -- extend to NLO. In doing so we show the necessary modifications if the scattering objects do not provide a hard scale. In the third section we discuss the separation between multi-Regge kinematics (MRK) and quasi-multi-Regge kinematics (QMRK) as well as the cancellation of divergences.

\section{Inclusive jet production at LO}
\label{sec:jetproductionlo}

As MRK relies on the transverse scales of the emissions and internal lines 
being of the same order, it is natural to think that processes characterized 
by two large and similar transverse momenta are the ideal environment for 
BFKL dynamics to  show up. Moreover, as the resummation is based on 
perturbative degrees of freedom, these large scales associated to the external 
particles should favor the accuracy of the predictions. 
An ideal scenario is the interaction between two photons with large 
virtualities $Q_{1,2}^2$ in the Regge limit $s \gg |t|\sim Q_1^2\sim Q_2^2$. 
The total cross section for this process has been investigated in a large 
number of publications in recent years. Here we are interested in the inclusive 
production of a single jet in the central region of rapidity in this process. 
We will consider the case where the transverse momentum of the jet is of the 
same order as the virtualities of the photons.

As a starting point we review single jet production at LO accuracy. As usual 
the total cross section can be written as a convolution of the photon impact 
factors with the gluon Green's function, i.e.
\begin{equation}
  \label{eq:total}
\sigma(s) = \int\frac{\ddka}{2\pi\ka^2}\int\frac{\ddkb}{2\pi\kb^2} \, 
\Phi_A(\ka) \, \Phi_B(\kb) \,\int_{\delta-i\infty}^{\delta+i\infty}\frac{d\omega}{2\pi i} \left(\frac{s}{s_0}\right)^\omega f_\omega(\ka,\kb).
\end{equation}
A  common choice for the energy scale is $s_0=|\ka|\,|\kb|$ which naturally 
introduces the rapidities $y_{\tilde A}$ and $y_{\tilde B}$ of the emitted particles with momenta $p_{\tilde A}$ and $p_{\tilde B}$ since
\begin{equation}
  \left(\frac{s}{s_0}\right)^\omega = e^{\omega(y_{\tilde A}-y_{\tilde B})}.
\end{equation}
Let us remark that a change in this scale can be treated as a redefinition
of the impact factors and, if $s_0$ is chosen to depend only on $\ka$ or
only on $\kb$, the kernel as well. This treatment lies beyond LO and will be discussed in the next section. 
The gluon Green's function $f_\omega$ corresponds to the solution of the BFKL 
equation \eqref{eq:bfklequation}.

\begin{center}
\begin{figure}[htbp]
  \centering
  \includegraphics[height=8cm]{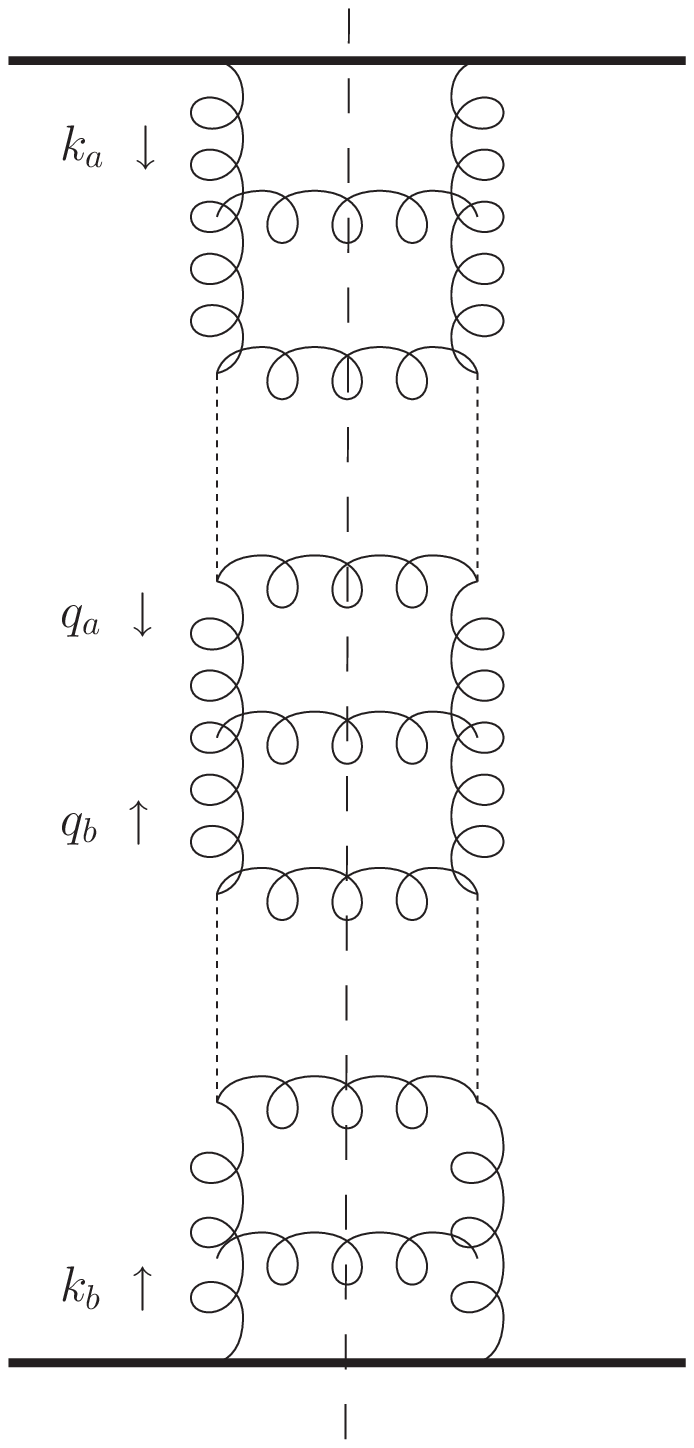}
  \hspace{1cm}
  \includegraphics[height=8cm]{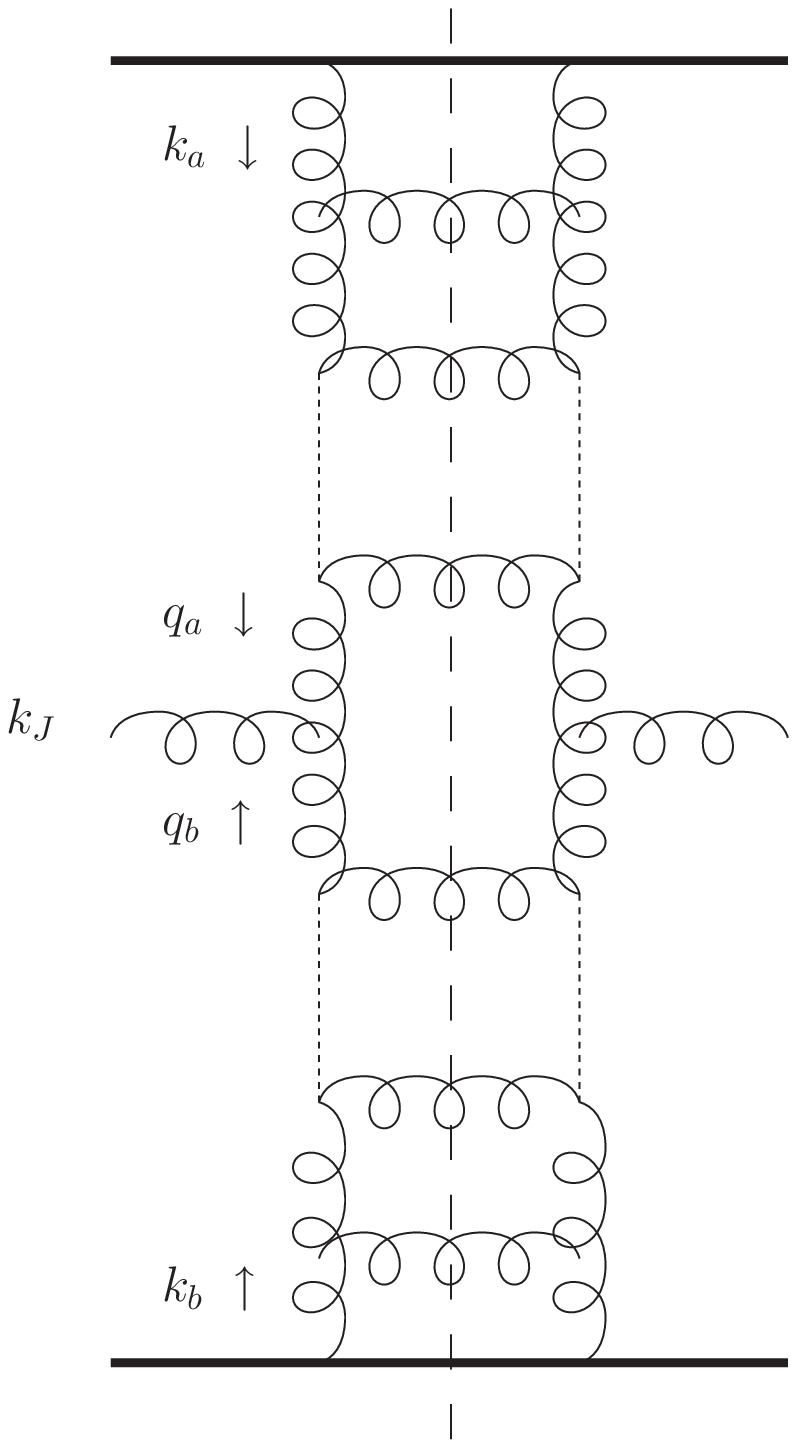}
  \caption{Total cross section and inclusive one jet production in the BFKL 
approach.}
  \label{fig:crosslo}
\end{figure}
\end{center}

For the inclusive production of a single jet we assign to it a 
rapidity $y_J$ and a transverse momentum $\kjet$, as shown in 
Fig.~\ref{fig:crosslo}. In this way, if 
$k_J=\alpha_J p_A+\beta_J p_B+k_{J \perp}$ the corresponding rapidity is 
$y_J=\frac{1}{2}\ln\frac{\alpha_J}{\beta_J}$. Using its {\ons}   condition 
we can write 
\begin{equation}
   k_J=\sqrt{\frac{\kjet^2}{s}}e^{y_J}p_A+\sqrt{\frac{\kjet^2}{s}}
e^{-y_J}p_B+k_{J \perp}.
\end{equation}

It is possible to single out one gluon emission by extracting its  
emission probability from the BFKL kernel. 
The differential cross section in terms of the jet variables can then be 
constructed in the following way:
\begin{multline}
\frac{d\sigma}{\ddkjet dy_J} =
\int\frac{\ddka}{2\pi\ka^2}\int\frac{\ddkb}{2\pi\kb^2} \, \Phi_A(\ka) 
\, \Phi_B(\kb)\\
\times\int \ddqa \int \ddqb \int_{\delta-i\infty}^{\delta+i\infty}
\frac{d\omega}{2\pi i} \left(\frac{s_{AJ}}{s_0}\right)^\omega 
f_\omega(\ka,\qa)\\
\times  
{\cal V}(\qa,\qb;\kjet,y_J)\int_{\delta-i\infty}^{\delta+i\infty}\frac{d\omega'}{2\pi i} \left(\frac{s_{BJ}}{s_0'}\right)^{\omega'}f_{\omega'}(-\qb,-\kb)
\label{eq:masterformula0}
\end{multline}
with the LO emission vertex being
\begin{equation}
{\cal V}(\qa,\qb;\kjet,y_J) = \mathcal{K}_r^{(B)}\left(\qa,-\qb\right) \, \del{\qa+\qb-\kjet}. 
\label{eq:jetvertexloprelim}
\end{equation}
By selecting one emission to be exclusive we have factorized the gluon Green's 
function into two components. Each of them connects one of the external 
particles to the jet vertex. In the notation of Eq.~\eqref{eq:masterformula0} 
the energies of these blocks are
  \begin{align}
    s_{AJ} =& (p_A+q_b)^2, & s_{BJ} =& (p_B+q_a)^2 .
  \end{align}
In a symmetric situation, where both the jet and the impact factors provide a hard scale, a natural choice for the scales is similar to that in the total cross section
\begin{align}
  s_0 =& |\ka|\,|\kjet|, & s_0' =& |\kjet|\,|\kb|.
\label{sosoprime}
\end{align}
These choices can now be related to the relative rapidity between the jet and the external particles. To set the ground for the NLO discussion of the next section we introduce an additional integration over the rapidity $\eta$ of the central system:
\begin{multline}
\frac{d\sigma}{\ddkjet dy_J} =
\int \ddqa \int \ddqb \int d\eta\\
\times\left[\int\frac{\ddka}{2\pi\ka^2} \, \Phi_A(\ka) \, 
\int_{\delta-i\infty}^{\delta+i\infty}\frac{d\omega}{2\pi i} e^{\omega(y_A-\eta)} f_\omega(\ka,\qa)\right]\, \mathcal{V}(\qa,\qb,\eta;\kjet,y_J)\\
\times \left[\int\frac{\ddkb}{2\pi\kb^2} \, 
\Phi_B(\kb) \int_{\delta-i\infty}^{\delta+i\infty}\frac{d\omega'}{2\pi i} e^{\omega'(\eta-y_B)}f_{\omega'}(-\qb,-\kb)\right] 
\label{eq:masterformula1}
\end{multline}
with the LO emission vertex being
\begin{equation}
\mathcal{V}(\qa,\qb,\eta;\kjet,y_J) = \mathcal{K}_r^{(B)}\left(\qa,-\qb\right) \, \del{\qa+\qb-\kjet}\,\delta(\eta-y_J). 
\label{eq:jetvertexloy}
\end{equation}
Eqs. \eqref{eq:masterformula1} and \eqref{eq:jetvertexloy} will be the starting point for the NLO jet production in the symmetric configurations. 

Let us now switch to the asymmetric case. In general we can write $q_a$ and $q_b$ as 
\begin{align}
  q_a=&\alpha_a p_A+\beta_a p_B + q_{a \perp} & q_b=&\alpha_b p_A+\beta_b p_B + q_{b \perp} .
\end{align}
The strong ordering in the rapidity of emissions translates into the conditions $\alpha_a\gg\alpha_b$ and $\beta_b\gg\beta_a$. This, together with momentum 
conservation $q_a+q_b=k_J$, leads us to $\alpha_J=\alpha_a+\alpha_b\approx
\alpha_a$, $\beta_J=\beta_a+\beta_b\approx\beta_b$ and
\begin{align}
  s_{AJ} =& \beta_J s, & s_{BJ}=&\alpha_J s.
\label{eq:sajsbjfirst}
\end{align}
While the longitudinal momentum of $q_a (q_b)$ is a linear combination of 
$p_A$ and $p_B$ we see that only its component along $p_A (p_B)$ matters. 

If the colliding external particles provide no perturbative scale as it is 
the case in hadron--hadron collisions, then the jet is the only hard scale 
in the process and we have to deal with an asymmetric situation. Thus the 
scales $s_0$ and $s_0'$ should be chosen as $\kjet^2$ alone. At LO accuracy $s_0$ is arbitrary and we are indeed free to make this choice.
Then the arguments of the gluon Green's functions can be written as
\begin{align}
  \label{eq:sajs0lo}
  \frac{s_{AJ}}{s_0}=&\frac{1}{\alpha_a}, &
  \frac{s_{BJ}}{s_0}=&\frac{1}{\beta_b}. 
\end{align}
The description in terms of these longitudinal components is particularly 
useful if one is interested in jet production in a hadronic environment. 
Here one can introduce the concept of an {\it unintegrated gluon 
density} in the hadron. This represents the probability of resolving a gluon 
carrying a longitudinal momentum fraction $x$ from the incoming hadron, and 
with a certain transverse momentum $k_T$. 
With the help of Eq.~\eqref{eq:sajs0lo} a LO unintegrated gluon distribution $g$ can be defined from Eq.~\eqref{eq:masterformula0} as
\begin{equation}
  \label{eq:updflo}
  g(x,\kpure) = \int\frac{\ddqpure}{2\pi \qpure^2}\,\Phi_{P}(\qpure)\,\int_{\delta-i\infty}^{\delta+i\infty}\frac{d\omega}{2\pi i}\, x^{-\omega} f_\omega(\qpure,\kpure).
\end{equation}

Then we can rewrite Eq.~\eqref{eq:masterformula0} as
\begin{multline}
\label{eq:masterformula2}
\frac{d\sigma}{\ddkjet dy_J} = \int \ddqa\int dx_1 \int \ddqb\int  dx_2\;\\
\times g(x_1,\qa)g(x_2,\qb)\mathcal{V}(\qa,x_1,\qb,x_2;\kjet,y_J),
\end{multline}
with the LO jet vertex for the asymmetric situation being
\begin{multline}
\label{eq:jetvertexlo}
\mathcal{V}(\qa,x_1,\qb,x_2;\kjet,y_J)=\mathcal{K}_r^{(B)}\left(\qa,-\qb\right)\\
\times \del{\qa+\qb-\kjet}\,\delta\left(x_1-\sqrt{\frac{\kjet^2}{s}}e^{y_J}\right)\delta\left(x_2-\sqrt{\frac{\kjet^2}{s}}e^{-y_J}\right).
\end{multline}
Having presented our framework for the LO case, in both $\gamma^*\gamma^*$ and 
hadron--hadron collisions, we now proceed to explain 
in detail what corrections are needed to define our cross sections at NLO. 
Special attention should be put into the treatment of those scales with do not 
enter the LO discussion, but are crucial at higher orders. 

\section{Inclusive jet production at NLO}

A similar approach to that shown in section \ref{sec:jetproductionlo} remains valid when jet 
production is considered at NLO. The crucial step in this direction is 
to modify  the LO jet vertex of Eq.~\eqref{eq:jetvertexloy} and 
Eq.~\eqref{eq:jetvertexlo} to include new configurations present 
at NLO. We show how this is done in the following first subsection. 
In the second subsection we implement this vertex in the symmetric 
$\gamma^* \gamma^*$ case, and we repeat the steps from Eq.~\eqref{eq:total} to 
Eq.~\eqref{eq:sajsbjfirst}, carefully describing the choice of energy scale at 
each of the subchannels. In the third subsection hadron--hadron scattering is 
taken into consideration, and we extend the concept of unintegrated gluon 
density of Eq.~\eqref{eq:updflo} to NLO accuracy. Most importantly, it is 
shown that a correct choice of intermediate energy scales in this case 
implies a modification of the impact factors, the jet vertex, and the evolution kernel.


\subsection{The NLO jet vertex}

For those 
parts of the NLO kernel responsible for one gluon production we proceed in exactly the 
same way as at LO. The treatment of those terms related to two particle production is more complicated since  
for them it is necessary to introduce a jet algorithm. In general terms, if the two 
emissions generated by the kernel are nearby in phase space they will be considered as 
one single jet, otherwise one of them will be identified as the jet whereas the other 
will be absorbed as an untagged inclusive contribution. Hadronization effects in the 
final state are neglected and we simply define a cone of radius $R_0$ in the 
rapidity--azimuthal angle space such that two particles form a single jet if 
$R_{12} \equiv \sqrt{(\phi_1-\phi_2)^2+(y_1-y_2)^2} < R_0$. As long as only 
two emissions are involved this is equivalent to the $k_T$--clustering algorithm. 

To introduce the jet definition in the $2 \rightarrow 2$ components of the 
kernel it is convenient to start by considering the gluon and quark matrix 
elements together:
\begin{multline}
\label{eq:k2q2g}
\left(\mathcal{K}_{\rm QMRK} + \mathcal{K}_{Q\bar{Q}} \right)(\qa,-\qb)  =
\int\dktwo \int dy_2 \\
\times \Big(\agsquare{\qa}{\qb}{\kone}{\ktwo}\theta(s_\Lambda-s_{12}) +\aqsquare{\qa}{\qb}{\kone}{\ktwo}\Big),
\end{multline}
with ${\cal A}_{2P}$ being the two particle production amplitudes of which
 only the gluonic one also contributes to MRK. This is why a step function is 
needed to separate it from MRK. Momentum conservation implies that 
$\kone = \qa + \qb - \ktwo$.

Expression~\eqref{eq:k2q2g} is not complete  as it stands since we should 
also include the MRK contribution as it was previously done in 
Eq.~\eqref{eq:krrgg}: 
\begin{align}
&\left( {\cal K}_{GG} + {\cal K}_{Q\bar Q} \right) (\qa,-\qb) \equiv 
\int\dktwo\int dy_2\,\bsquare{\qa}{\qb}{\kone}{\ktwo} \non
 =& \int\dktwo\int dy_2\,
\Bigg\{\agsquare{\qa}{\qb}{\kone}{\ktwo}\theta(s_\Lambda-s_{12})\non
&-{\cal K}^{(B)}(\qa,\qa-\kone) \, {\cal K}^{(B)}(\qa-\kone,-\qb)\;\frac{1}{2}\,\theta\left(\ln\frac{s_{\Lambda}}{\ktwo^2}-y_2\right)\theta\left(y_2-\ln\frac{\kone^2}{s_{\Lambda}}\right)\non
& +\aqsquare{\qa}{\qb}{\kone}{\ktwo}\Bigg\}. 
\label{eq:defbsquare}
\end{align}
We are now ready to introduce the jet definition for the double emissions. The 
NLO versions of Eq.~\eqref{eq:jetvertexloy} and Eq.~\eqref{eq:jetvertexlo} then read, respectively,
\begin{align}
\mathcal{V}(\qa,\qb,\eta;\kjet,y_J)= & \left(\mathcal{K}_r^{(B)} +\mathcal{K}_r^{(v)}\right)(\qa,-\qb) \Big|_{(a)}^{[y]}\non
  &\hspace{-2.5cm}+ \int\dktwo\;dy_2\bsquare{\qa}{\qb}{\kjet-\ktwo}{\ktwo}\theta(R_0-R_{12})\Big|_{(b)}^{[y]}\non
  &\hspace{-2.5cm}+ 2\int\dktwo\;dy_2\bsquare{\qa}{\qb}{\kjet}{\ktwo}\theta(R_{J2}-R_0)\Big|_{(c)}^{[y]},\label{eq:jetvertexnloy}\\
\mathcal{V}(\qa,x_1,\qb,x_2;\kjet,y_J)= & \left(\mathcal{K}_r^{(B)} +\mathcal{K}_r^{(v)}\right)(\qa,-\qb) \Big|_{(a)}^{[x]}\non
  &\hspace{-2.5cm}+ \int\dktwo\;dy_2\bsquare{\qa}{\qb}{\kjet-\ktwo}{\ktwo}\theta(R_0-R_{12})\Big|_{(b)}^{[x]}\non
  &\hspace{-2.5cm}+ 2\int\dktwo\;dy_2\bsquare{\qa}{\qb}{\kjet}{\ktwo}\theta(R_{J2}-R_0)\Big|_{(c)}^{[x]}.\label{eq:jetvertexnlox}
\end{align}
In these two expressions we have introduced the notation 
\begin{align}
  &\Big|_{(a,b)}^{[y]}&&\hspace{-.9cm}=\del{\qa+\qb-\kjet} \delta (\eta - y^{(a,b)}),  \\
  &\Big|_{(c)}^{[y]}&&\hspace{-.9cm}=\del{\qa+\qb-\kjet-\ktwo} \delta \left(\eta-y^{(c)}\right), \\
   &\Big|_{(a,b)}^{[x]}&&\hspace{-.9cm}=\del{\qa+\qb-\kjet} \delta \left(x_1 - x_1^{(a,b)}\right)\delta \left(x_2 - x_2^{(a,b)}\right),  \\
  &\Big|_{(c)}^{[x]}&&\hspace{-.9cm}=\del{\qa+\qb-\kjet-\ktwo} \delta \left(x_1 - x_1^{(c)}\right)   \delta \left(x_2 - x_2^{(c)}\right) .
\end{align}
The various jet configurations demand different $y$ and $x$ configurations. These are related to the properties of the produced jet in different ways 
depending on the origin of the jet: if only one gluon was produced in MRK this 
corresponds to the configuration (a) in the table below, if two particles in 
QMRK form a jet then we have the case (b), and finally case (c) if the jet is 
produced out of one of the partons in QMRK. The factor of 2 in the last term 
of Eq.~\eqref{eq:jetvertexnloy} and Eq.~\eqref{eq:jetvertexnlox} accounts for the possibility that either emitted 
particle can form the jet. Just by kinematics we get the explicit expressions for the different $x$ configurations listed in the following table:
\begin{center}
\begin{tabular}[h!]{c|c|cc}
  JET  & $y$ configurations & \multicolumn{2}{c}{$x$ configurations}
\\\hline
a) \raisebox{-1ex}{\includegraphics[height=.7cm]{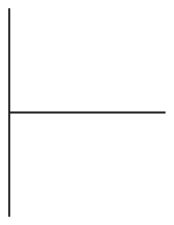}}  & 
   $y^{(a)}=y_J$ & $x_1^{(a)}=\frac{|\kjet|}{\sqrt{s}}e^{y_J}$ & 
   $x_2^{(a)}=\frac{|\kjet|}{\sqrt{s}}e^{-y_J}$ \\
b) \raisebox{-1.5ex}{\includegraphics[height=.7cm]{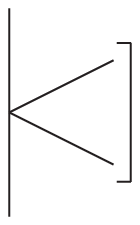}} & 
   $y^{(b)}=y_J$ & $x_1^{(b)}=\frac{\sqrt{\Sigma}}{\sqrt{s}}e^{y_J}$ & 
   $x_2^{(b)}=\frac{\sqrt{\Sigma}}{\sqrt{s}}e^{-y_J}$\\
c) \raisebox{-2ex}{\includegraphics[height=.7cm]{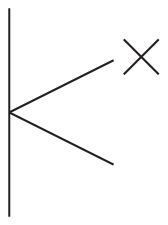}} & 
   $y^{(c)}=\frac{1}{2}\ln\frac{x_1^{(c)}}{x_2^{(c)}}$ & 
  {\small $ x_1^{(c)}=\frac{|\kjet|}{\sqrt{s}}e^{y_J}+\frac{|\ktwo|}{\sqrt{s}}e^{y_2}$} &   
  {\small $ x_2^{(c)}=\frac{|\kjet|}{\sqrt{s}}e^{-y_J}+\frac{|\ktwo|}{\sqrt{s}}e^{-y_2}$}
\end{tabular}
\end{center}
The variable $\Sigma$ is defined below in Eq.~\eqref{eq:defsigma}. Due to 
the analogue treatment of the emission vertex either expressed in terms of 
rapidities or longitudinal momentum fractions in the remaining of this 
section we will imply the same analysis for both. In particular, we will not 
explicitly mention these arguments when we come to 
Eqs.~(\ref{eq:jetvertexnlo2}, \ref{eq:freeofsing}). 

The introduction of the jet definition divides the phase space into 
different sectors. It is now needed to show that the final result is 
indeed free of any infrared divergences. In the following we proceed to 
independently calculate several contributions to the kernel to be able, in 
this way, to study its singularity structure.

The NLO virtual correction to the one--gluon emission kernel, 
${\cal K}^{(v)}$, was originally calculated in 
Ref.~\cite{Fadin:1993wh,Fadin:1994fj,Fadin:1996yv}. Its expression reads
\begin{multline}
  \mathcal{K}_r^{(v)} \left(\qa,-\qb\right)= 
\frac{\bar{g}_\mu^4 \mu^{-2\epsilon}}{\pi^{1+\epsilon}\Gamma(1-\epsilon)}\frac{4}{\delt^2}\Bigg\{ 2\left(\frac{\delt^2}{\mu^2}\right)^\epsilon \left(-\frac{1}{\epsilon^2}+\frac{\pi^2}{2}-2 \, \epsilon \, \zeta(3)\right) \\
+\frac{\beta_0}{N_c}\frac{1}{\epsilon}+\,\frac{3\delt^2}{\qa^2-\qb^2}\ln\left(\frac{\qa^2}{\qb^2}\right)
-\ln^2\left(\frac{\qa^2}{\qb^2}\right) \\
+\,\left(1-\frac{n_f}{N_c}\right)\bigg[
\frac{\delt^2}{\qa^2-\qb^2}\left(1-\frac{\delt^2(\qa^2+\qb^2-4 \, \qa\qb)}{3(\qa^2-\qb^2)^2}\right)\ln\left(\frac{\qa^2}{\qb^2}\right)\\
-\,\frac{\delt^2}{6\qa^2\qb^2}(\qa-\qb)^2+\frac{\delt^4 \, (\qa^2+\qb^2)}{6\qa^2\qb^2(\qa^2-\qb^2)^2}(\qa^2+\qb^2-4 \, \qa\qb)\bigg]\Bigg\},
\label{eq:kernelv}
\end{multline}
with $\beta_0 = (11N_c-2n_f)/3$, $\zeta(n)= \sum_{k=1}^\infty k^{-n}$ and 
$\delt = \qa + \qb$. ${\bar g}_\mu$ can be expressed in terms of 
the  renormalized coupling constant $g_\mu$ in the $\overline{\rm MS}$ renormalization scheme by the relation $\bar{g}_\mu^2 = g_\mu^2 \, N_c \, \Gamma(1-\epsilon) \, (4\pi)^{-2-\epsilon}$. Note that the expression for the virtual contribution given in \cite{Ostrovsky:1999kj} lacks the log squared.

Those pieces related to two--gluon production in QMRK can be rewritten 
in terms of their corresponding matrix elements as
\begin{multline}
  \mathcal{K}_{\rm QMRK}(\qa,-\qb) = \int\dktwo\int dy_2 \agsquare{\qa}{\qb}{\kone}{\ktwo} \theta(s_\Lambda-s_{12})\\
 = \frac{g_\mu^2 \mu^{-2\epsilon}N_c^2}{\pi(2\pi)^{D+1}\qa^2\qb^2}\int\dktwoeps\int dy_2\; A_{\text{gluons}}\,\theta(s_\Lambda-s_{12}),
\label{eq:kernelagluons}
\end{multline}
and those related to quark--antiquark production are
\begin{align}
\mathcal{K}_{Q\bar{Q}}(\qa,-\qb) =& \int\dktwo\int dy_2 \aqsquare{\qa}{\qb}{\kone}{\ktwo} \non
=& \frac{g_\mu^2 \mu^{-2\epsilon}N_c^2}{\pi(2\pi)^{D+1}\qa^2\qb^2}\int\dktwoeps\int dy_2\; A_{\text{quarks}}\label{eq:kernelaquarks}.
\end{align}

\begin{figure}
\begin{center}
\includegraphics{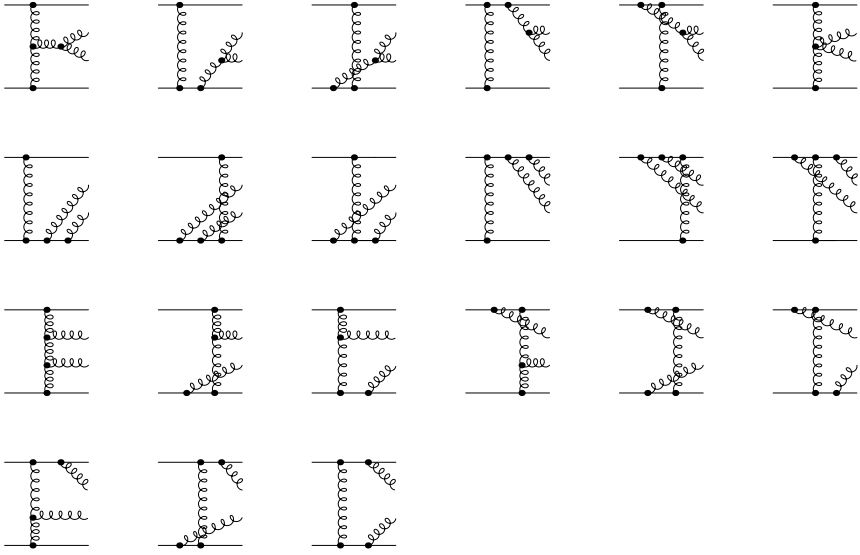}
\end{center}
\caption{Feynman diagrams for the process $q+q\to q+\;\,g+g\;\,+q$. For the second to fourth row the diagrams with crossed gluons have to be added.}
\label{fig:ggproduction}
\end{figure}

\begin{figure}
\begin{center}
\includegraphics{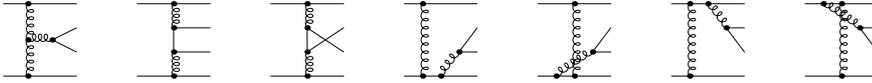}
\end{center}
\caption{Feynman diagrams for the process $q+q\to q+\;\,q+\bar q\;\,+q$. }
\label{fig:qqproduction}
\end{figure}

For our purposes we need the corresponding amplitudes including those parts which vanish in the fully inclusive case. Therefore we have calculated these amplitudes at which the ingoing legs are \ofs .
To obtain them in a gauge invariant way it is necessary to consider an appropriate embedding as a quark--quark scattering where  the gluon or quark--antiquark pair is produced additionally. The according Feynman diagrams are depicted in Figs.~\ref{fig:ggproduction} and ~\ref{fig:qqproduction}, and using the Mandelstam invariants $\shat$, $\that$, and $\uhat$, and we extracted the following results 
\begin{align}
  A_{\text{gluons}}=& \;\qa^2\qb^2 \Bigg\{-\frac{1}{\that\uhat}+\frac{1}{4\that\uhat}\frac{\qa^2\qb^2}{\kone^2\ktwo^2}-\frac{1}{4}\left(\frac{1-x}{x}\frac{1}{\ktwo^2\that}+\frac{x}{1-x}\frac{1}{\kone^2\uhat}\right)+\frac{1}{4\kone^2\ktwo^2}\non
&+\frac{1}{\Sigma}\Bigg[-\frac{1}{\shat}\left(2+\left(\frac{1}{\that}-\frac{1}{\uhat}\right)\left(\frac{1-x}{x}\kone^2-\frac{x}{1-x}\ktwo^2\right)\right)+\frac{1}{4}\left(\frac{\Sigma}{\shat}+1\right)\non
&\phantom{+\frac{1}{\Sigma}\Bigg[}\times\left(\frac{1-x}{x}\frac{1}{\ktwo^2}+\frac{x}{1-x}\frac{1}{\kone^2}\right) -\frac{\qb^2}{4\shat}\left(\frac{1}{(1-x)\that}+\frac{1}{x\uhat}\right)\non
&\phantom{+\frac{1}{\Sigma}\Bigg[}-\frac{\qa^2}{4\shat}\left(\left[1+\frac{x}{1-x}\frac{\ktwo^2}{\kone^2}\right]\frac{1}{\that}+\left[1+\frac{1-x}{x}\frac{\kone^2}{\ktwo^2}\right]\frac{1}{\uhat}\right)\Bigg]\Bigg\}\non
&\hspace{-1cm}+\frac{D-2}{4}\Bigg\{\left(\frac{(\kone-\qa)^2(\ktwo-\qa)^2-\kone^2\ktwo^2}{\that\uhat}\right)^2\non
&\hspace{-.5cm}-\frac{1}{4}\left(\frac{(\ktwo-\qa)^2-\frac{x}{1-x}\ktwo^2}{\uhat}+\frac{E}{\shat}\right)
\left(\frac{(\kone-\qa)^2-\frac{1-x}{x}\kone^2}{\that}-\frac{E}{\shat}\right)\Bigg\},
\label{eq:Agluons}
\end{align}
\begin{align}
A_{\text{quarks}} =& \frac{n_f}{4N_c}\Bigg\{\frac{\qa^2\qb^2}{\shat\Sigma}\left(2+\left(\frac{1}{\that}-\frac{1}{\uhat}\right)\left(\frac{1-x}{x}\kone^2-\frac{x}{1-x}\ktwo^2\right)\right)\non
&-\left(\frac{(\kone-\qa)^2(\ktwo-\qa)^2-\kone^2\ktwo^2}{\that\uhat}\right)^2\non
&+\frac{1}{2}\left(\frac{(\ktwo-\qa)^2-\frac{x}{1-x}\ktwo^2}{\uhat}+\frac{E}{\shat}\right)
\left(\frac{(\kone-\qa)^2-\frac{1-x}{x}\kone^2}{\that}-\frac{E}{\shat}\right)\Bigg\}\non
&+\frac{n_f}{4N_c^3}\Bigg\{\left(\frac{(\kone-\qa)^2(\ktwo-\qa)^2-\kone^2\ktwo^2}{\that\uhat}\right)^2-\frac{\qa^2\qb^2}{\that\uhat}\Bigg\} .
\end{align}
These expressions are in agreement with the corresponding ones obtained in 
Ref.~\cite{Ostrovsky:1999kj}. The following notation has been used:
\begin{align}
  x =& \frac{|\kone|}{|\kone|+|\ktwo|e^{\Delta y}}, \\
 \lambd =& (1-x)\kone-x\ktwo, \\
\Sigma =& \shat+\delt^2 \, = \, \frac{\lambd^2}{x(1-x)} + \delt^2 , \label{eq:defsigma}\\
 E =&\phantom{-} 2(2x-1)\qa^2+4\lambd\qa +\frac{1-2x}{x(1-x)}\lambd^2 \non
=&- 2 x (1-x)\left((2x-1)\delt^2+2\lambd\delt\right)\frac{\qa^2}{x(1-x)\delt^2+\lambd^2}
\label{eq:e}.
\end{align}
We now study those terms which contribute to generate soft and collinear divergences after integration over the two--particle phase space. They should be able to cancel the $\epsilon$ poles of the virtual contributions in Eq.~\eqref{eq:kernelv}, {\it i.e.}
\begin{equation}
  \mathcal{K}^{(v)}_{\rm singular} \left(\qa,\qb\right)=
\frac{\bar{g}_\mu^4 \mu^{-2\epsilon}}{\pi^{1+\epsilon}\Gamma(1-\epsilon)}\frac{4}{\delt^2}\Bigg\{\left(\frac{\delt^2}{\mu^2}\right)^\epsilon \left(-\frac{2}{\epsilon^2}\right) +\frac{\beta_0}{N_c}\frac{1}{\epsilon}\Bigg\}.
\label{eq:kernelvsingular}
\end{equation}
Here we identify those pieces responsible for the generation of these poles.

One of the divergent regions is defined by the two emissions with momenta 
$k_1=\alpha_1 p_A+\beta_1 p_B +k_{1\perp}$ and $k_2=\alpha_2 p_A+\beta_2 p_B +k_{2\perp}$ becoming collinear. This means that, for a real parameter $\lambda$, $k_1 \simeq \lambda \, k_2$, {\it i.e.} $k_{1\perp} \simeq \lambda \, k_{2\perp}$, $\alpha_1 \simeq \lambda \, \alpha_2$ and thus $\alpha_2 k_{1\perp} - \alpha_1 k_{2\perp} \simeq 0$. Since $x=\frac{\alpha_1}{\alpha_1+\alpha_2}$ this is equivalent to the condition $\lambd \simeq 0$. In the collinear region $\shat=\frac{\lambd^2}{x(1-x)}$ tends to zero and the dominant contributions which are 
purely collinear are 
\begin{align}
  A_{\text{gluons}}^{\text{singular}}\Big|_{\rm collinear} =& 
- \frac{\qa^2\qb^2}{\Sigma}\frac{2}{\shat} +\frac{D-2}{16}\frac{E^2}{\shat^2} \equiv A_{(1)}+A_{(2)}\label{eq:agluonscollinear},\\
  A_{\text{quarks}}^{\text{singular}}\Big|_{\rm collinear} =& \frac{n_f}{2N_c}\frac{\qa^2\qb^2}{\shat\Sigma}-\frac{n_f}{8N_c}\frac{E^2}{\shat^2} \label{eq:aquarkscoll}.
\end{align}
The quark--antiquark production does not generate divergences when 
$\kone$ or $\ktwo$ become soft, therefore we have that the only 
purely soft divergence is
\begin{equation}
  A_{\text{gluons}}^{\text{singular}}\Big|_{\rm soft} = \qa^2\qb^2\Bigg(\frac{1}{4\that\uhat}\frac{\qa^2\qb^2}{\kone^2\ktwo^2}+\frac{1}{4\kone^2\ktwo^2}\Bigg) \equiv A_{(3)}+A_{(4)}
~\rightarrow~ 2A_{(4)},
\label{eq:agluonssoft}
\end{equation}
where we have used the property that, in the soft limit, the $\that\uhat$ 
product tends to $\qa^2\qb^2$. We will see that these terms will be 
responsible for simple poles in $\epsilon$. The double poles will be generated 
by the regions with simultaneous soft and collinear divergences. They are 
only present in the gluon--gluon production case and can be written as
\begin{align}
  A_{\text{gluons}}^{\text{singular}}\Big|_{\rm soft \& collinear} =& 
\frac{\qa^2\qb^2}{4\shat}\left[\frac{1-x}{x}\frac{1}{\ktwo^2}
+\frac{x}{1-x}\frac{1}{\kone^2}\right] \non
&\hspace{-3cm}-\frac{\qa^2\qb^2}{4\shat\Sigma}
\Bigg[\phantom{+}\qb^2\left(\frac{1}{(1-x)\that}+\frac{1}{x\uhat}\right)\non
&\hspace{-3cm}\phantom{-\frac{\qa^2\qb^2}{4\shat\Sigma}\Bigg[}
+\qa^2\left(\left[1+\frac{x}{1-x}\frac{\ktwo^2}{\kone^2}\right]\frac{1}{\that}
+\left[1+\frac{1-x}{x}\frac{\kone^2}{\ktwo^2}\right]
\frac{1}{\uhat}\right)\Bigg].\non
=&A_{(5)}+A_{(6)}.\label{eq:agluonssoftcoll}
\end{align}
Focusing on the divergent structure it turns out that in the soft and 
collinear region the first line of Eq.~\eqref{eq:agluonssoftcoll}, $A_{(5)}$,  
has exactly the same limit as the second line, $A_{(6)}$. This is very 
convenient since we can then simply write
\begin{equation}
  A_{\text{gluons}}^{\text{singular}}\Big|_{\rm soft \& collinear} 
\;\rightarrow \;
\frac{\qa^2\qb^2}{2\shat}\left(\frac{1-x}{x}\frac{1}{\ktwo^2}+\frac{x}{1-x}\frac{1}{\kone^2}\right)=2A_{(5)}.\label{eq:agluonssoftcoll2}
\end{equation}
The MRK contribution of Eq.~\eqref{eq:defbsquare} has the form 
$A_{\rm MRK}= - 4 A_{(4)}$ and when added to all the other singular terms we 
get the expression
\begin{multline}
\int\dktwo\int dy_2\,\bssquare{\qa}{\qb}{\ktwo}{\kone} \equiv\\
\frac{g_\mu^2 \mu^{-2\epsilon}N_c^2}{\pi(2\pi)^{D+1}\qa^2\qb^2}
\int\dktwoeps \int dy_2 \, \Big\{A^{\rm singular}_{\text{gluons}} \, 
\theta(s_\Lambda-s_{12})+ A^{\rm singular}_{\text{quarks}}\Big\}
\label{eq:defbssquare},
\end{multline}
with
\begin{multline}
A^{\rm singular}_{\text{gluons}} \, 
\theta(s_\Lambda-s_{12})+ A^{\rm singular}_{\text{quarks}} = \\
\left\{\underbrace{- \frac{\qa^2\qb^2}{\Sigma}\frac{2}{\shat}}_{\rm Gluon|_{coll_1}} +\underbrace{\frac{D-2}{16}\frac{E^2}{\shat^2}}_{\rm Gluon|_{coll_2}}-\underbrace{\frac{\qa^2\qb^2}{2\kone^2\ktwo^2}}_{\rm Gluon|_{soft}}+\underbrace{\frac{\qa^2\qb^2}{2\shat}\left(\frac{1-x}{x}\frac{1}{\ktwo^2}+\frac{x}{1-x}\frac{1}{\kone^2}
\right)}_{\rm Gluon|_{\rm soft \& coll}}\right\}  \\
\times\theta(s_\Lambda-s_{12})\;+ \underbrace{\frac{n_f}{2N_c}\frac{\qa^2\qb^2}{\shat\Sigma}}_{\rm Quark|_{coll_1}}-\underbrace{\frac{n_f}{8N_c^3}\frac{E^2}{\shat^2}}_{\rm Quark|_{coll_2}}.
\label{eq:differentterms}
\end{multline}
We have labeled the different terms to study how each of them produces the 
$\epsilon$ poles. We will do this in section \ref{sec:cancellation}.

With the singularity structure well identified we now return to 
Eqs.~(\ref{eq:jetvertexnloy}, \ref{eq:jetvertexnlox}) and show that they are free of any divergences.
Only if the divergent terms belong to the same 
configuration, this cancellation can be shown analytically. With this in mind we 
add the singular parts of the two particle production of 
Eq.~\eqref{eq:defbssquare} in the configuration $(a)$ 
multiplied by $0=1-\theta(R_0-R_{12})-\theta(R_{12}-R_0)$:
\begin{align}
\mathcal{V} =& \bigg[\left(\mathcal{K}_r^{(B)}+\mathcal{K}_r^{(v)}\right)(\qa,-\qb) \non
&\hspace{1cm}+\int\dktwo\;dy_2\bssquare{\qa}{\qb}{\kjet-\ktwo}{\ktwo}\bigg] \Big|_{(a)}\non
&+\int\dktwo\;dy_2\bigg[\phantom{-}\bsquare{\qa}{\qb}{\kjet-\ktwo}{\ktwo}\Big|_{(b)}\non
&\phantom{+\int\dktwo\;dy_2\bigg[}-\bssquare{\qa}{\qb}{\kjet-\ktwo}{\ktwo}\Big|_{(a)}\bigg]\theta(R_0-R_{12})\non
&+\bigg[2\int\dktwo\;dy_2\bsquare{\qa}{\qb}{\kjet}{\ktwo}\theta(R_{J2}-R_0)\Big|_{(c)}\non
&\phantom{+\bigg[}-\int\dktwo\;dy_2\bssquare{\qa}{\qb}{\kjet-\ktwo}{\ktwo}\theta(R_{12}-R_0)\Big|_{(a)}\bigg].
\label{eq:jetvertexnlo2}
\end{align}

The cancellation of divergences within the first two lines is now the same 
as in the calculation of the full NLO kernel. In section \ref{sec:cancellation} we will show how the first two lines of 
Eq.~\eqref{eq:jetvertexnlo2} are free of any singularities in the form of 
$\epsilon$ poles. In doing so we will go into the details of the r\^ole of $s_\Lambda$.
The third and fourth lines are also explicitly free 
of divergences since these have been subtracted out. The sixth line has a 
$\kone \leftrightarrow \ktwo$ symmetry which allows us to write
\begin{align}
\mathcal{V} =& \bigg[\left(\mathcal{K}_r^{(B)}+\mathcal{K}_r^{(v)}\right)(\qa,-\qb)\non
&\hspace{1cm} +\int\dktwo\;dy_2\bssquare{\qa}{\qb}{\kjet-\ktwo}{\ktwo}\bigg] \Big|_{(a)}\non
&+\int\dktwo\;dy_2\bigg[\phantom{-}\bsquare{\qa}{\qb}{\kjet-\ktwo}{\ktwo}\Big|_{(b)}\non
&\phantom{+\int\dktwo\;dy_2\bigg[}-\bssquare{\qa}{\qb}{\kjet-\ktwo}{\ktwo}\Big|_{(a)}\bigg]\theta(R_0-R_{12})\non
&+2\int\dktwo\;dy_2\bigg[\bsquare{\qa}{\qb}{\kjet}{\ktwo}\theta(R_{J2}-R_0)\Big|_{(c)}\non
&\hspace{.5cm}-\bssquare{\qa}{\qb}{\kjet-\ktwo}{\ktwo}\theta(R_{12}-R_0)\theta(|\kone|-|\ktwo|)\Big|_{(a)}\bigg].
\label{eq:freeofsing}
\end{align}
We can now see that the remaining possible divergent regions of the last line 
are regulated by the cone radius $R_0$. 

It is worth noting that, apart from an overall 
${\bar \alpha}_s^2 (\mu^2)$ factor, the NLO terms in the last four lines in 
Eq.~\eqref{eq:freeofsing} do not carry any renormalization scale dependence since 
they are finite when $\epsilon$ is set to zero. The situation is different for the 
first two lines since $\mathcal{V}$ contains a logarithm of $\mu^2$ in the form
\begin{equation}
  \label{eq:runningcoupling}
  \mathcal{V}= \mathcal{V}^{(B)}\left(1-\frac{\alpha_s(\mu^2)}{4\pi}\frac{\beta_0}{N_c}\ln\frac{\kjet^2}{\mu^2}\right) + \Delta\mathcal{V}.
\end{equation}
where $\Delta\mathcal{V}$ contains the third to sixth lines and the $\mu$--independent part of the first two lines of \eqref{eq:freeofsing}. It is then natural to absorb this term in a redefinition of the running of the 
coupling and replace $\alpha_s(\mu^2)$ by $\alpha_s(\kjet^2)$. For a explicit derivation 
of this term we refer the reader to section \ref{sec:cancellation}.

Therefore we now have a finite expression for the jet vertex suitable for numerical 
integration. 
This numerical analysis will be performed elsewhere since here we are mainly 
concerned with the formal introduction of the jet definition 
and the correct separation of the different contributions to the kernel. 

What remains to be proven is the cancellation of divergences between 
Eq.~\eqref{eq:kernelvsingular} and Eq.~\eqref{eq:defbssquare}. This will be 
performed in section \ref{sec:cancellation}. Before doing so, in the next two subsections, we indicate 
how to introduce our vertex in the definition of the differential cross 
section. Special care must be taken in the treatment of the energy scale 
in the reggeized gluon propagators since in the symmetric case it is directly  
related to the rapidity difference between subsequent emissions, as we will  
show in the next subsection, but in the asymmetric case of hadron--hadron 
collisions it depends on the longitudinal momentum fractions 
of the $t$--channel Reggeons. 

\subsection{Production of jets in $\gamma^*\gamma^*$ scattering}

We now have all the ingredients required to describe the inclusive single jet 
production in a symmetric process at NLO. To be definite, we consider $\gamma^*\gamma^*$ scattering with the 
virtualities of the two photons being large and of the same order. All we need is to 
take Eq.~\eqref{eq:masterformula0} for the differential cross section as a 
function of the transverse momentum and rapidity of the jet. The vertex 
${\cal V}$ to be used is that of Eq.~\eqref{eq:freeofsing} in the 
representation based on rapidity variables of Eq.~\eqref{eq:jetvertexnloy}. 
The rapidities of the emitted particles are the natural variables to 
characterize the partonic evolution and $s$--channel production since we 
assume that all transverse momenta are of the same order. 

Let us note that the rapidity difference between two emissions can be 
written as
\begin{equation}
y_i - y_{i+1} = \ln{\frac{s_{i,i+1}}{\sqrt{\ki^2 \kip^2}}}
\end{equation}
which supports the choice $s_{R;i,i+1}=\sqrt{\ki^2\kip^2}$ in 
Eq.~\eqref{eq:news17nlo}. This is also technically more convenient since it 
simplifies the final expression for the cross section in Eq.~\eqref{MRKfinal}.

\begin{center}
\begin{figure}[htbp]
  \centering
  \includegraphics[width=8cm]{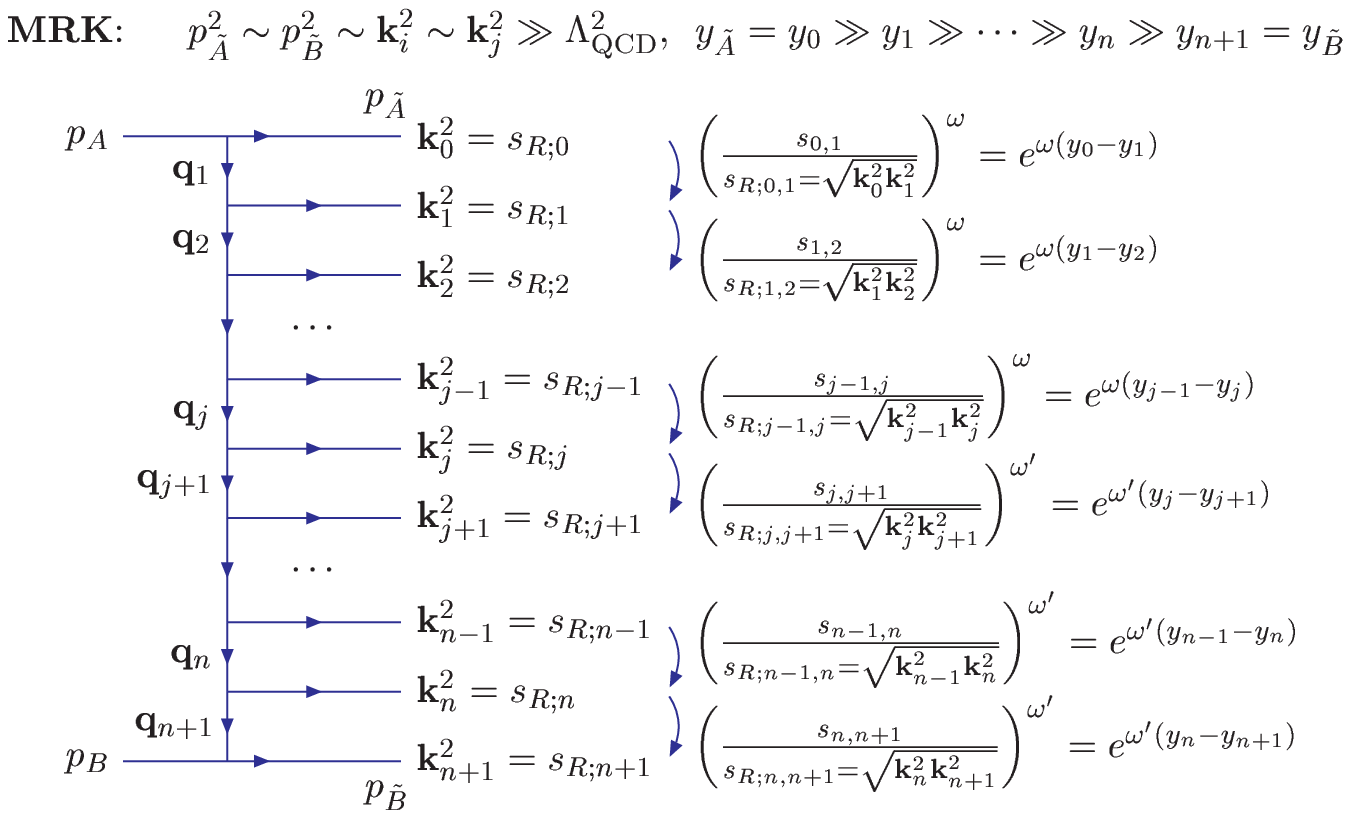}
  \caption{Momenta for $2 \rightarrow 2 + (n-1) + {\rm jet}$ amplitude 
in the symmetric configuration with MRK. The produced jet has rapidity 
$y_J=y_j$ and transverse momentum $\kjet=\kj$.}
  \label{SymmetricMRK}
\end{figure}
\end{center}

In Fig.~\ref{SymmetricMRK} we illustrate the different scales 
participating in the scattering and the variables of the evolution. We give the conditions for MRK: all transverse momenta are of similar size and 
much larger than the confining scale, the rapidities are strongly ordered in 
the evolution from one external particle to the other. At each stage of the 
evolution the propagation of the Reggeized gluons, which generates  
rapidity gaps, takes place between two real emissions. There are many 
configurations contributing to the differential cross section, each of them 
with a different weight. Eq.~\eqref{eq:masterformula0} represents the sum of
these production processes.

\subsection{The unintegrated gluon density and jet production in 
hadron--hadron collisions}

In this subsection we now turn to the case of hadron collisions where 
MRK has to be necessarily modified to include some evolution in the 
transverse momenta, since the momentum of the jet will be much 
larger than the typical transverse scale associated to the hadron.

In the LO case we have already explained that, in order to move from the symmetric 
case to the asymmetric one, it is needed to change the energy scale from the choice in
Eq.~\eqref{sosoprime} to the one in Eq.~\eqref{eq:sajs0lo}. This is equivalent to 
changing the description of the evolution in terms of rapidity 
differences between emissions to longitudinal momentum fractions of the
Reggeized gluons in the $t$--channel. 
Whereas in LO this change of scales has no consequences, in NLO accuracy it
leads to modifications, not only of the jet emission vertex, but
also of the evolution kernels above and below the jet vertex. 
\begin{figure}
  \centering
  \includegraphics[width=7cm]{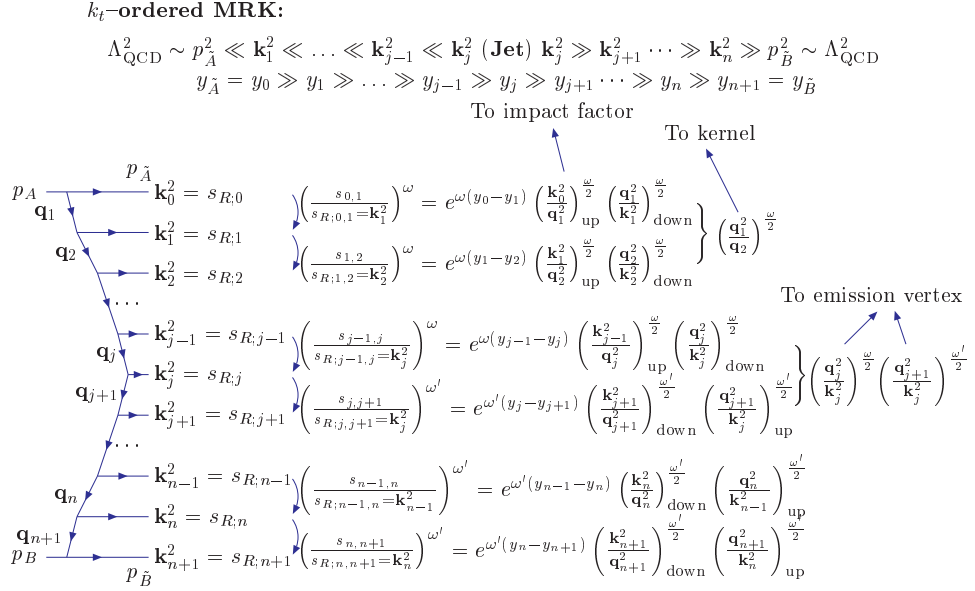}
  \caption{Momenta for $2 \rightarrow 2 + (n-1) + {\rm jet}$ amplitude 
in the asymmetric configuration with $k_t$--ordered MRK.}
  \label{AsymmetricMRK}
\end{figure}
These new definitions will allow the cross section still to be written in a factorizable way and the 
evolution of the gluon Green's function still to be described by an integral 
equation. 

To understand this in detail we start by writing the solution 
to the NLO BFKL equation iteratively, i.e.
\begin{equation}
\int \ddka f_\omega(\ka,\qa) = 
\frac{1}{\omega} \sum_{j=1}^\infty 
\left[\prod_{i=1}^{j-1}\int \ddqi\frac{1}{\omega}\mathcal{K}(\qi,\qip)\right],
\label{eq:ggfiterando}
\end{equation}
where $\qone = \ka$ and $\qj = \qa $.
As both sides of the evolution are similar, we now focus on one side of the 
evolution towards the hard scale and use 
Fig.~\ref{AsymmetricMRK} as a graphical reference. Starting 
with the symmetric case, the differential cross section for jet 
production contains the following evolution between particle $A$ and 
the jet:
\begin{multline}
  \frac{d\sigma}{\ddkjet dy_J} = \int \ddqa \int \ddka 
\frac{\Phi_A(\ka)}{2\pi\ka^2} \\
\times\int\frac{d\omega}{2\pi i} 
f_\omega(\ka,\qa)\left(\frac{s_{AJ}}{\sqrt{\ka^2\kjet^2}}\right)^\omega 
{\cal V}(\qa,\qb;\kjet,y_J)\dots
\end{multline}

In the asymmetric situation where $\kjet^2\gg \ka^2$ the scale 
$\sqrt{\ka^2\kjet^2}$ should be replaced by $\kjet^2$. In order to do so 
we rewrite the term related to the choice of energy scale. To be consistent 
with Fig.~\ref{AsymmetricMRK} we take $\kj = \kjet$, 
$\kzero =  -\ka = -\qone$ and $\qj = \qa$. To start, it is convenient to introduce a chain of scale changes in every kernel:
\begin{equation}
 \left(\frac{s_{AJ}}{\sqrt{\ka^2\kjet^2}}\right)^\omega =
\left[\prod_{i=1}^{j}\left(\frac{\ki^2}{\kim^2}\right)^{\frac{\omega}{2}}\right]  \left(\frac{s_{AJ}}{\kjet^2}\right)^\omega, 
\end{equation}
which can alternatively be written in terms of the $t$--channel momenta as
\begin{equation}
 \left(\frac{s_{AJ}}{\sqrt{\ka^2\kjet^2}}\right)^\omega = \left[\prod_{i=1}^{j-1}\left(\frac{\qip^2}{\qi^2}\right)^{\frac{\omega}{2}}\right]\left(\frac{\kjet^2}{\qa^2}\right)^{\frac{\omega}{2}}\left(\frac{s_{AJ}}{\kjet^2}\right)^\omega.
\end{equation}
For completeness we note that we are indeed changing the variable of evolution 
from a difference in rapidity:
\begin{equation}
\frac{s_{AJ}}{\sqrt{\ka^2\kjet^2}} = e^{y_{\tilde A}-y_J} 
\end{equation}
to the inverse of the longitudinal momentum fraction, {\it i.e.}
\begin{equation}
\frac{s_{AJ}}{\kjet^2} = \frac{1}{\alpha_J}.
\label{jetinverselong}
\end{equation}

This shift in scales translates into the following expression for the cross 
section:
\begin{multline}
\frac{d\sigma}{\ddkjet dy_J} =
\int\frac{d\omega}{2\pi i \, \omega}\sum_{j=1}^\infty 
\left[\prod_{i=1}^{j}\int \ddqi\right] \frac{\Phi_A(\qone)}{2\pi\qone^2}\\
\times\left[\prod_{i=1}^{j-1}\left(\frac{\qip^2}{\qi^2}\right)^{\frac{\omega}{2}}\frac{1}{\omega}\mathcal{K}(\qi,\qip)\right]\left(\frac{\kjet^2}{\qa^2}\right)^{\frac{\omega}{2}}
{\cal V}(\qa,\qb;\kjet,y_J)\left(\frac{s_{AJ}}{\kjet^2}\right)^\omega\ldots
\end{multline}
As we mentioned above these changes can be absorbed at NLO in the kernels and 
impact factors, we just need to perturbatively expand the integrand. The 
impact factors get one single contribution, as can be seen in 
Fig.~\ref{AsymmetricMRK}, and they explicitly change as
\begin{equation}
\label{newimpactfactor}
 \widetilde{\Phi}(\ka)= \Phi(\ka) -\frac{1}{2}{\ka^2}\int \ddqpure 
\frac{\Phi^{(B)}(\qpure)}{\qpure^2}\mathcal{K}^{(B)}(\qpure,\ka)
\ln\frac{\qpure^2}{\ka^2}.
\end{equation}
The kernels in the evolution receive a double contribution from the different 
energy scale choices of both the incoming and outgoing Reggeons (see 
Fig.~\ref{AsymmetricMRK}). This amounts to the following correction:
\begin{equation}
\label{newkernel}
  \widetilde{\mathcal{K}}(\qone,\qtwo) = \mathcal{K}(\qone,\qtwo)
-\frac{1}{2}\int \ddqpure \, \mathcal{K}^{(B)}(\qone,\qpure) 
\, \mathcal{K}^{(B)}(\qpure,\qtwo)\ln\frac{\qpure^2}{\qtwo^2}.
\end{equation}
There is a different type of term in the case of the emission vertex 
where the jet is defined. This correction has also two contributions 
originating from the two different evolution chains from the hadrons $A$ and 
$B$. Its expression is
\begin{multline}
\label{newemissionvertex}
  \widetilde{\cal V}(\qa,\qb) = {\cal V}(\qa,\qb)
-\frac{1}{2}\int \ddqpure \,  
\mathcal{K}^{(B)}(\qa,\qpure) {\cal V}^{(B)}(\qpure,\qb)
\ln\frac{\qpure^2}{(\qpure-\qb)^2}\\
-\frac{1}{2}\int \ddqpure \, {\cal V}^{(B)}(\qa,\qpure) \,
\mathcal{K}^{(B)}(\qpure,\qb)\ln\frac{\qpure^2}{(\qa-\qpure)^2}.
\end{multline}

These are all the modifications we need to write down our differential 
cross section for the asymmetric case. The final expression is
\begin{multline}
  \frac{d\sigma}{\ddkjet dy_J} = \int \ddqa\int \ddka 
\frac{\widetilde{\Phi}_A(\ka)}{2\pi\ka^2}\\
\times \int\frac{d\omega}{2\pi i}
\tilde{f}_\omega(\ka,\qa)\left(\frac{s_{AJ}}{\kjet^2}\right)^\omega 
\widetilde{\cal V}(\qa,\qb;\kjet,y_J)\ldots
\end{multline}
As in the LO case, we can use Eq.~\eqref{jetinverselong} to define the 
NLO unintegrated gluon density as
\begin{equation}
  g(x,\kpure) = \int \ddqpure\frac{\widetilde{\Phi}_P(\qpure)}{2\pi\qpure^2}\int\frac{d\omega}{2\pi i}\tilde{f}_\omega(\kpure,\qpure)\, x^{-\omega}.
\end{equation}
The gluon Green's function ${\tilde f}_\omega$ is the solution to a new 
BFKL equation with the modified kernel of Eq.~\eqref{newkernel} which includes the energy shift at 
NLO, i.e.
\begin{equation}
  \omega \tilde{f}_\omega(\ka,\qa) = \del{\ka-\qa} +
\int \ddqpure \, \widetilde{\mathcal{K}}(\ka,\qpure) \,
\tilde{f}_\omega(\qpure,\qa).
\end{equation}
In this way the unintegrated gluon distribution follows the evolution equation 
\begin{equation}
  \frac{\partial g(x,\qa)}{\partial\ln 1/x} 
= \int \ddqpure \, \widetilde{\mathcal{K}} (\qa,\qpure) \, g(x,\qpure).
\end{equation}
Finally, taking into account the evolution from the other hadron, the 
differential cross section reads
 \begin{equation}
  \frac{d\sigma}{\ddkjet dy_J} = \int \ddqa \int \ddqb 
\, g(x_a,\qa) \, g(x_b,\qb) \, \widetilde{\cal V}(\qa,\qb;\kjet,y_J),
\label{ppfinal} 
\end{equation}
with the emission vertex taken from Eq.~\eqref{newemissionvertex}.

We would like to indicate that with the prescription derived in this 
subsection we managed to express the new kernels, emission vertex and 
impact factors as functions of their incoming momenta only. It is also 
worth mentioning that the proton impact factor contains 
non--perturbative physics which can only be modeled by, e.g.
\begin{equation}
\Phi_P(\qpure) \;\sim\; (1-x)^{p_1} x^{- p_2} 
\left(\frac{\qpure^2}{\qpure^2+Q_0^2}\right)^{p_3},
\label{eq:protonIF}
\end{equation}
where $p_i$ are positive free parameters, with $Q_0^2$ representing a momentum
scale of the order of the confinement scale. 
Though Regge theory suggests a $p_2\approx 0.08$ \cite{Donnachie:1992ny} the common collinear parton distributions available work with a rather valence like value of $p_2\approx -0.09$ \cite{Martin:2001es}. But in any case the initial $x$ dependence in this expression would be of non--perturbative origin.  

Let us also point out that the prescription to modify the kernel as in 
Eq.~\eqref{newkernel} was originally suggested in Ref.~\cite{Fadin:1998py} in the context of deep inelastic scattering.  
This new kernel can be considered as the first term in 
an all orders perturbative expansion due to the change of scale. When all 
terms are included, the kernel acquires improved convergence properties and 
matches collinear evolution as shown in section~\ref{sec:bfklresummation}.

\section[Cancellation of divergences]{Cancellation of divergences and a closer look at the separation 
between MRK and QMRK}
\label{sec:cancellation}

During the calculation of a NLO BFKL cross section, both at a fully inclusive 
level and at a more exclusive one, there is a need to separate 
the contributions from MRK and QMRK. In order to do so we followed 
Ref.~\cite{Fadin:1998sh} and introduced the parameter $s_\Lambda$ in 
Eq.~\eqref{eq:kqmrk} and Eq.~\eqref{eq:phiqmrk}. In principle, at NLO 
accuracy, our final results should not depend on this extra scale. 
In fact, as we have remarked earlier in our discussion of the total
cross section (after Eq.~\eqref{eq:krrgg}), 
we could have taken the limit $s_{\Lambda} \to \infty$: the logarithms 
of $s_{\Lambda}$ cancel, and the corrections to the finite pieces die 
away as ${\cal O}(s_\Lambda^{-1})$.
In the context of the inclusive cross section, however, we prefer to treat    
$s_{\Lambda}$ as a physical parameter: it separates MRK from QMRK and, hence, 
cannot be arbitrarily large. We will therefore retain the dependence upon 
$s_{\Lambda}$: in the remainder of this section we demonstrate that, 
in our inclusive cross section, all logarithmic terms cancel 
(analogous to Eq.~\eqref{eq:krrgg}), and we 
 leave the study of the corrections of the order 
$\order{s_\Lambda^{-1}}$ for a numerical analysis. It will also be 
interesting to see how this dependence on $s_\Lambda$ could be 
related to the rapidity veto introduced in Ref.~\cite{Schmidt:1999mz,Forshaw:1999xm,Andersson:1995ju,Chachamis:2004ab}. 

Let us consider the $s_\Lambda$ dependent terms in 
Eq.~\eqref{eq:defbssquare} which are only present in the gluon piece:
\begin{multline}
\left(\frac{g_\mu^2 \mu^{-2\epsilon}N_c^2}{\pi(2\pi)^{D+1}} \right)^{-1}
\int\dktwo\int dy_2\,\bssquare{\qa}{\qb}{\ktwo}{\kone} \Big|_{s_\Lambda}\\
~\equiv~
\int\dktwoeps \int dy_2 \, 
\frac{A^{\rm singular}_{\text{gluons}}}{\qa^2\qb^2}\, \theta(s_\Lambda-s_{12})
~=~ \sum_{i=I}^{IV} \mathcal{S}_i,
\label{eq:allJs}
\end{multline}
where we have used the numbering $(I,II,III,IV)$ corresponding to, 
respectively, $({\rm Gluon|_{coll_1}},{\rm Gluon|_{coll_2}},
{\rm Gluon|_{soft}},{\rm Gluon|_{soft \& coll}})$ in 
Eq.~\eqref{eq:differentterms}.

To calculate each of the $\mathcal{S}_i$ terms we start by transforming the rapidity 
integral into an integral over $x$ in the form 
$\int d\Delta y=\int\frac{dx}{x(1-x)}$. We consider $s_{\Lambda}$ much larger 
than any of the typical transverse momenta. In the limit of large $s_\Lambda$, 
the theta function $\theta(s_\Lambda-\shat)$ amounts to the  
limits $\frac{\kone^2}{s_\Lambda}+\orderslambda{-2}$ and 
$1-\frac{\ktwo^2}{s_\Lambda}+\orderslambda{-2}$ for the $x$ integral. 

We firstly consider $\mathcal{S}_{III}$ which is 
\begin{multline}
-\int\dktwoeps\int_{\frac{\kone^2}{s_\Lambda}}^{1-\frac{\ktwo^2}{s_\Lambda}}\frac{dx}{x(1-x)}\frac{1}{2\,\kone^2\ktwo^2} = \frac{-\pi}{(4\pi)^\epsilon}\frac{1}{\delt^2}\frac{\Gamma(1-\epsilon)\Gamma(\epsilon)^2}{\Gamma(2\epsilon)} \\
\times\left(\ln\frac{s_\Lambda}{\delt^2}+\psi(1-\epsilon)-\psi(\epsilon)+\psi(2\epsilon)-\psi(1)\right)\left(\frac{\delt^2}{\mu^2}\right)^{\epsilon}+\orderslambda{-1}.
\label{eq:SIII}
\end{multline}
We are only interested in the logarithmic dependence on $s_\Lambda$ and hence 
we do not need to calculate $\orderslambda{-1}$ or $s_\Lambda$ independent 
factors.

The next $s_\Lambda$ contribution we calculate is $\mathcal{S}_{IV}$ which reads
\begin{align}
&\int\dktwoeps\int_{\frac{(\delt-\ktwo)^2}{s_\Lambda}}^{1-\frac{\ktwo^2}{s_\Lambda}}\frac{dx}{x(1-x)}
\left(\frac{(1-x)^2}{\ktwo^2(\ktwo-(1-x)\delt)^2}+\frac{x^2}{\ktwo^2(\ktwo-x\delt)^2}\right)\non
=& \int\dktwoeps\Bigg[\frac{2}{(\delt-\ktwo)^2\ktwo^2}\ln\frac{s_\Lambda}{\ktwo^2}+\frac{2\,(\delt-\ktwo)\ktwo}{(\delt-\ktwo)^2\ktwo^2\sqrt{\ktwo^2\delt^2-(\delt\ktwo)^2}}\non
&\times\left(
 \arctan\frac{\delt(\delt-\ktwo)}{\sqrt{\ktwo^2\delt^2-(\delt\ktwo)^2}}
+\arctan\frac{\delt\ktwo}{\sqrt{\ktwo^2\delt^2-(\delt\ktwo)^2}}\right)\Bigg]+\orderslambda{-1}.
\end{align}
The part with logarithmic $s_\Lambda$ dependence can be calculated analytically:
\begin{multline}
\int\dktwoeps\frac{1}{(\delt-\ktwo)^2\ktwo^2}\ln\frac{s_\Lambda}{\ktwo^2} = 
\frac{\pi}{(4\pi)^\epsilon}\frac{1}{\delt^2}\frac{\Gamma(1-\epsilon)\Gamma(\epsilon)^2}{\Gamma(2\epsilon)}\\
\times \left(\ln\frac{s_\Lambda}{\delt^2}+\psi(1-\epsilon)-\psi(\epsilon)+\psi(2\epsilon)-\psi(1)\right)\left(\frac{\delt^2}{\mu^2}\right)^{\epsilon}.
\end{multline}
It is then clear that this logarithmic $s_\Lambda$ contribution cancels 
against that of $\mathcal{S}_{III}$ in Eq.~\eqref{eq:SIII}.

Let us proceed now to show that the contribution of $\mathcal{S}_{I}$ is directly of $\orderslambda{-1}$ and does not contribute with any logarithm of $s_\Lambda$. In the 
relevant integral we introduce the change of variable 
$\ktwo\to\lambd=(1-x)\delt-\ktwo$ and obtain
\begin{multline}
\int\dlambdeps\int_{\frac{\lambd^2}{s_\Lambda}}^{1-\frac{\lambd^2}{s_\Lambda}}\frac{dx}{x(1-x)}
\left(\frac{x^2(1-x)^2}{\lambd^2(\lambd^2+x(1-x)\delt^2)}\right)=\\
 \int\dlambdeps
\left(\frac{1}{\delt^2\lambd^2 }-\frac{2\ln\left(1+\frac{\delt^2+\sqrt{\delt^2(\delt^2+4\lambd^2)}}{2\lambd^2}\right)}{\delt^2\sqrt{\delt^2(\delt^2+4\lambd^2)}}\right)+\orderslambda{-1}.
\end{multline}
We do not write here the lengthier, but similar expression which corresponds 
to $\mathcal{S}_{II}$ and also only contributes to $\orderslambda{-1}$.

With this we have shown that the sum of different terms
 in Eq.~\eqref{eq:allJs} 
is free of logarithmic dependences on $s_\Lambda$ proving, in this way, 
that the remaining $\orderslambda{-1}$ corrections vanish at large values 
of $s_\Lambda$. In particular, it is possible to take the 
$s_\Lambda \rightarrow \infty$ limit in order to completely eliminate the 
dependence on this scale. This is convenient in the fully inclusive 
case where it is very useful to write a Mellin transform in the $k_T$ 
dependence of the NLO BFKL kernel.

If we perform this $s_\Lambda \rightarrow \infty$ limit, then  
$\mathcal{S}_{III}$ and $\mathcal{S}_{IV}$ can be combined and their sum is
\begin{align}
&\mathcal{S}_{III}+\mathcal{S}_{IV} \non
=& \int_0^1\frac{dx}{x(1-x)}\int\dktwoeps\left[\frac{1}{2\shat}\left(\frac{1-x}{x\ktwo^2}+\frac{x}{(1-x)\kone^2}\right)-\frac{1}{2\kone^2\ktwo^2}\right]\non
=& \int_0^1\frac{dx}{2x(1-x)}\int\dktwoeps\Bigg[\frac{(1-x)^2}{\ktwo^2(\ktwo-(1-x)\delt)^2}\non
&\hphantom{\int_0^1\frac{dx}{2x(1-x)}\int\dktwoeps\Bigg[}+\frac{x^2}{\kone^2(\kone-x\delt)^2}-\frac{1}{\ktwo^2(\delt-\ktwo)^2}\Bigg]\non
=&\frac{1}{\delt^2}\frac{\pi}{(4\pi)^\epsilon}\frac{\Gamma(1-\epsilon)\Gamma^2(1+\epsilon)}{\epsilon \, \Gamma(1+2\epsilon)}\left(\frac{1}{\epsilon}+2\psi(1)-2\psi(1+2\epsilon)\right)\left(\frac{\delt^2}{\mu^2}\right)^\epsilon.\label{eq:sthreeplusfour}
\end{align}
Regarding $\mathcal{S}_{I}$, one obtains from  the integration
\begin{align}
\mathcal{S}_{I}=&-2\int_{0}^{1}\frac{dx}{x(1-x)}\int\dlambdeps\left[\frac{x^2(1-x)^2}{\lambd^2(\lambd^2+x(1-x)\delt^2)}\right]\non
=& -2 \int_{0}^{1}\frac{dx}{x(1-x)}\left[\frac{\pi}{(4\pi)^\epsilon}\frac{x(1-x)}{\delt^2}\frac{\Gamma(1-\epsilon)\Gamma(\epsilon)}{\Gamma(1+\epsilon)}\left(\frac{x(1-x)\delt^2}{\mu^2}\right)^\epsilon\right] \non
=& -\frac{2}{\delt^2}\frac{\pi}{(4\pi)^\epsilon}\frac{\Gamma(1-\epsilon)\Gamma(1+\epsilon)^2}{\epsilon\Gamma(2+2\epsilon)}\left(\frac{\delt^2}{\mu^2}\right)^\epsilon.
\end{align}
The contribution from $\mathcal{S}_{II}$ is more complicated and the relevant integral 
can be obtained in the following way:
\begin{align}
\int\dlambdeps\frac{E^2}{8\qa^2\qb^2\shat^2}
=& \int\dlambdeps\frac{x^2(1-x)^2E^2}{8\qa^2\qb^2\lambd^4} \non
&\hspace{-4cm}= \int\dlambdeps \bigg[\frac{x^2(1-x)^2(2x-1)^2\delt^2}{2\qb^2\lambd^2 (x(1-x)\delt^2+\lambd^2)}\non
&\hspace{-3.5cm}-\frac{x^3(1-x)^3(2x-1)^2\delt^2\qa^2}{\qb^2\lambd^4(x(1-x)\delt^2+\lambd^2)}+\frac{x^4(1-x)^4(2x-1)^2\delt^4\qa^2}{2\qb^2\lambd^4(x(1-x)\delt^2+\lambd^2)^2}\non
&\hspace{-3.5cm}-\frac{4x^3(1-x)^3 (\delt\lambd)(\lambd\qa)}{\qb^2\lambd^2(x(1-x)\delt^2+\lambd^2)}+\frac{2x^4(1-x)^4(\delt\lambd)^2\qa^2}{\qb^2\lambd^2(x(1-x)(\delt^2+\lambd^2)^2}\bigg]\non
&\hspace{-4cm}= \frac{\pi}{(4\pi)^\epsilon}\frac{\Gamma(2-\epsilon)\Gamma(\epsilon)}{\Gamma(1+\epsilon)}\frac{\left(x(1-x)\delt^2\right)^{\epsilon-1}}{\mu^{2\epsilon}}\bigg[\frac{1}{1-\epsilon}\frac{x^2(1-x)^2(2x-1)^2\delt^2}{\qb^2}\non
&\hspace{-3.5cm}+\frac{1}{1-\epsilon}\frac{x^2(1-x)^2(2x-1)^2\qa^2}{\qb^2}-\frac{2-\epsilon}{1-\epsilon}\frac{x^2(1-x)^2(2x-1)^2\qa^2}{2\qb^2}\non
&\hspace{-3.5cm}-\frac{2}{1-\epsilon^2}\frac{x^3(1-x)^3\delt\qa}{\qb^2}+\frac{1}{1+\epsilon}\frac{x^3(1-x)^3\qa^2}{\qb^2}\bigg]
\end{align}
We now need to integrate it over $x$ to obtain:
\begin{equation}
  \mathcal{S}_{II} = \frac{1}{\delt^2}\frac{\pi}{(4\pi)^\epsilon}\frac{\pi(1+\epsilon)\Gamma(2+\epsilon)}{\sin(\pi\epsilon)\Gamma(4+2\epsilon)}\left(\frac{\delt^2}{\mu^2}\right)^\epsilon .
\end{equation}
 This result gives the same poles in $\epsilon$ as the result given in \cite{Ostrovsky:1999kj}, but differs for the finite contribution.
To obtain all the $\epsilon$ poles we now also include the quark contributions 
present in Eq.~\eqref{eq:defbssquare}. We denote them as
\begin{equation}
\int\dktwoeps \int dy_2 \, \frac{A^{\rm singular}_{\text{quarks}}}{\qa^2\qb^2} =
\sum_{i=V}^{VI} \mathcal{S}_i,
\label{eq:onlyquarksinBs}
\end{equation}
where the correspondence with Eq.~\eqref{eq:differentterms} is 
$(V,VI) \rightarrow ({\rm Quark|_{coll_1}},{\rm Quark|_{coll_2}})$. 
Adding everything up, the sum of all the terms reads
\begin{equation}
  \sum_{i=I}^{VI}\mathcal{S}_i = \frac{1}{\delt^2}\frac{\pi\Gamma(1-\epsilon)}{(4\pi)^\epsilon}\left(\frac{\delt^2}{\mu^2}\right)^\epsilon\left[\frac{1}{\epsilon^2}-\frac{\beta_0}{2N_c}\frac{1}{\epsilon}+\frac{67}{18}-\frac{5n_f}{9N_c}-\frac{5\pi^2}{6}+\order{\epsilon}\right].
\end{equation}
The final expression for Eq.~\eqref{eq:defbssquare} is thus given by
\begin{multline}
  \int\dktwo\int dy_2 \bssquare{\qa}{\qb}{\kjet-\ktwo}{\ktwo} = \\
\frac{\bar{g}_\mu^4 \mu^{-2\epsilon}}{\pi^{1+\epsilon}\Gamma(1-\epsilon)}\frac{4}{\kjet^2}\left(\frac{\kjet^2}{\mu^2}\right)^\epsilon\left[\frac{2}{\epsilon^2}-\frac{\beta_0}{N_c}\frac{1}{\epsilon}+\frac{67}{9}-\frac{10n_f}{9N_c}-\frac{5\pi^2}{3}+\order{\epsilon}\right] \label{eq:resultbsquare}.
\end{multline}
When we combine this result with the singular terms of Eq.~\eqref{eq:kernelv} 
then we explicitly prove the cancellation of any singularity in our 
subtraction procedure to introduce the jet definition. The finite remainder 
reads
\begin{equation}
\frac{\bar\alpha_s^2(\mu^2)}{\pi}
\frac{1}{\kjet^2}
\left[-\frac{\beta_0}{4 N_c}\ln{\frac{\kjet^2}{\mu^2}}+\frac{1}{12}\left(
4 - 2 \pi^2 + 5 \frac{\beta_0}{N_c}\right)\right].
\end{equation}
We have already discussed the logarithmic term due to the running of the 
coupling in Eq.~\eqref{eq:runningcoupling}. The non--logarithmic part is 
similar to that present in other calculations involving soft gluon 
resummations~\cite{Catani:1990rr,Dokshitzer:1995ev} where terms of the form 
\begin{equation}
{\bar \alpha}_s \left(1+ {\cal S} \,{\bar \alpha}_s\right)
\end{equation}
appear and offer the possibility to change from the $\overline{\rm MS}$ 
renormalization scheme to the so--called {\sl gluon--bremsstrahlung} (GB) 
scheme by shifting the position of the Landau pole, {\it i.e.}
\begin{equation}
\Lambda_{\rm GB} = \Lambda_{\rm \overline{\rm MS}} 
\exp{\left({\cal S}\frac{2 N_c}{\beta_0}\right)}.
\end{equation}
The factor $\mathcal{S}$ differs from ours in the $\pi^2$ term:
\begin{equation}
\label{eq:defS}
{\cal S} = \frac{1}{12}\left(4 - \pi^2 + 5 \frac{\beta_0}{N_c}\right).
\end{equation}
The origin of this discrepancy lies in the fact that we used the simplest 
form of subtraction procedure. In appendix \ref{sec:alternative}  we suggest a different 
subtraction term which is more complicated in the sense that it 
subtracts a larger portion of the matrix element in addition to the infrared 
divergent pieces. When this is done and we put together the divergent pieces 
of Eq.~\eqref{eq:kernelv} and the second line of 
Eq.~\eqref{eq:resultbsquarenew}, then we recover the same ${\cal S}$ term.

\clearpage{\pagestyle{empty}\cleardoublepage}
\chapter{Angular decorrelation}
\label{sec:decorrelation}
 
After the construction of a jet vertex in the central region at NLO, we dedicate this chapter to another aspect of jet physics, namely its angular correlation. 

In the first section, we start with a short review of the LO description of Mueller-Navelet jets and introduce some notation, while in the second subsection the extension to NLO is presented. We then study the phenomenology and compare our results with experiment. In this context, we discuss the need for a partial resummation of even higher order contributions and show its consequences. Thereafter, we turn to the case of electron-proton scattering. The angular decorrelation between the electron and the most forward jet is governed by the same physics as in the Mueller-Navelet case. To make use of our experiences from proton-proton collisions for a study of angular correlation in deep inelastic scattering (DIS) is hence a natural extension which will be presented in the second section of this chapter.


\section{Angular correlation of Mueller-Navelet jets}

A discussion of  the Mueller-Navelet jet  has been already given in the context of the power like rise of the cross section in section~\ref{sec:phenomenology}. 
Our focus, now, lies on the angular correlation of the jets as a more exclusive observable. Before we go into the details, it is worthwhile to point out that the Mueller-Navelet jets lie at the interface of collinear factorization and BFKL dynamics. The partons emitted from the hadrons carry large longitudinal momentum fractions. After scattering off each other, they produce the Mueller-Navelet jets. Because of the large transverse momentum of the jets, the partons are hard and obey the collinear factorization. In particular, its scale dependence is given by the DGLAP evolution equations. Between the jets, on the other hand, we require a large rapidity gap which is described by BFKL dynamics. The hadronic cross section, hence, factorizes into two conventional collinear parton distribution functions convoluted with the partonic cross section, to be described by the BFKL equation. With respect to the partonic  cross section, the incoming partons, consequently, are considered to be \ons\ and collinear to the incident hadrons.

For the angular correlation, theoretical predictions from LO BFKL exist \cite{DelDuca:1993mn,Stirling:1994zs}, and even improvements due the running of the coupling and proper treatment of the kinematics have been implemented \cite{Orr:1997im,Kwiecinski:2001nh}. A first step to a NLO description has been made in Ref.~\cite{Vera:2006un,Vera:2006xa} on which our work -- starting with the second subsection --  builds up. But for pedagogical reasons we start with the LO description.

\subsection{Mueller-Navelet jets at LO}

Our starting point is the total cross section of two particles as given in Eq.~\eqref{eq:total}. For the case of Mueller-Navelet jets these particles are partons out of the protons. 
We initially  consider the partonic gluon-gluon scattering:
\begin{equation}
  \label{eq:gluongluonscattering}
\hat\sigma(\shat) = \int\frac{\ddqone}{2\pi\qtwo^2}\int\frac{\ddqtwo}{2\pi\qtwo^2}\int\frac{d\omega}{2\pi i} \, 
\Phi_{\rm gluon,1}(\qone) \, \Phi_{\rm gluon,2}(\qtwo) f_\omega(\qone,\qtwo)e^{\omega\Delta y}.
\end{equation}
In this symmetric situation we have chosen $s_0$ such that it leads to a description in terms of the rapidity $\Delta y=\ln\frac{\shat}{|\qone|\,|\qtwo|}$ between the two scattered gluons. To be observed by the detector as a jet, a gluon has to have a minimal transverse momentum. We incorporate this resolution scale $p^2$ by a Heaviside function defining the jet vertex:
\begin{equation}
  \label{eq:mnjetvertex}
  \Phi_{\rm gluon,1}(\qone)\quad\rightarrow\quad\Phi_{\rm jet,1}(\qone)=\frac{\Phi_{\rm gluon,1}(\qone)}{\sqrt{2}\pi^2\,\asbar\,\qone^2}\theta(\qone^2-p_1^2) .
\end{equation}
Simultaneously we rearranged the constants in a convenient way and, again, make use of the convention $\asbar=\alpha_s N_c/\pi$.
At LO we are completely free in our choice of  $s_0$. The introduction of the resolution scale for the jets suggests a more convenient rapidity variable $Y=\ln\frac{\shat}{p_1 p_2}$. We will comment on this change in more detail at the next section dealing with the NLO description. 

According to the normalization of the BFKL equation used in this work, the gluon impact factor at LO in our case reads
\begin{equation}
  \Phi_{\rm gluon,1}(\qone) = \frac{\pi\asbar}{\sqrt{2}} .
\end{equation}
Due to high-energy factorization the differences between quark-quark, quark-gluon, and gluon-gluon scattering concern only a constant factor in front of the impact factors, namely
\begin{equation}
  \Phi_{\rm quark,1}(\qone) = \frac{C_F}{C_A}\Phi_{\rm gluon,1}(\qone),
\end{equation}
where $C_A=N_c$ and $C_F=(N_c^2-1)/(2N_c)$ are the Casimir invariants of the adjoint and fundamental representation, respectively. 
Hence, we can incorporate the combined quark and gluon contributions to the hadronic cross section by the effective parton density \cite{Combridge:1983jn}
\begin{equation}
  \label{eq:feff}
  f_{\rm eff}(x,\mu_F^2) = G(x,\mu_F^2)+\frac{4}{9}\sum_f\left[Q_f(x,\mu_F^2)+\bar{Q}_f(x,\mu_F^2)\right],
\end{equation}
where the sum runs over all quark flavors, and $\mu_F$ denotes the factorization scale. For the proton-proton scattering the Mueller-Navelet cross section then reads
\begin{equation}
  \frac{d\sigma(s)}{dx_1\, dx_2} = f_{\rm eff}(x_1,\mu_F^2)f_{\rm eff}(x_2,\mu_F^2)\hat\sigma(x_1x_2s).
\end{equation}

\begin{figure}[htbp]
  \centering
  \includegraphics[width=7cm]{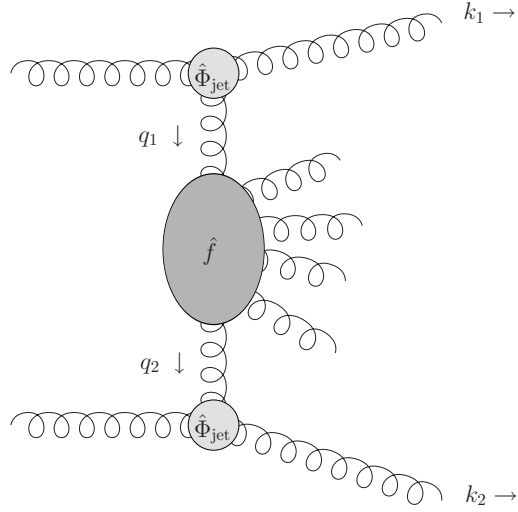}
  \caption{Kinematics of the partonic cross section in the Mueller-Navelet case}
  \label{fig:mnkinematics}
\end{figure}

For the following steps we focus on the partonic cross section, and the convolution with the parton densities will be performed at the end. The kinematics are depicted in Fig.~\ref{fig:mnkinematics}, and we can write the partonic cross section \eqref{eq:gluongluonscattering} as
\begin{align}
  \label{eq:oppartonic}
  \hat\sigma(\shat) =& \frac{\pi^2\asbar^2}{2}\int\ddkone\int\ddktwo\int\frac{d\omega}{2\pi i} \bra{\kone} \opjetone\opf_\omega\opjettwo\ket{\ktwo}e^{\omega Y}\\
=& \frac{\pi^2\asbar^2}{2}\sum_{n,n'=-\infty}^\infty\int\ddkone\int\ddktwo\int\frac{d\omega}{2\pi i}\int\ddqone\int\ddqtwo \int d\nu\int d\nu' \non
 & \bra{\kone} \opjetone\ket{\qone}\braket{\qone}{\nu,n}\bra{n,\nu}\opf_\omega\ket{\nu',n'}\braket{n',\nu'}{\qtwo}\bra{\qtwo}\opjettwo\ket{\ktwo} e^{\omega Y}.\label{eq:oppartonic2}
\end{align}
Here we made use of the operator representation introduced in section~\ref{sec:bfklmellin}.
The integrals over transverse momenta have to be taken over the whole two dimensional space while the $\nu$ integrations go from $-\infty$ to $\infty$. The contour of the $\omega$ integration  is to the right of all $\omega$-plane singularities of the integrand.

Due to the cylindric symmetry around the beam axis, the cross section can only depend on the difference between the azimuthal angles of $k_1$ and $k_2$, but not on their absolute values.
Nevertheless, it is convenient for our derivation to fix an arbitrary axis in the azimuthal plane and to denote the azimuthal angles of $k_1$ and $k_2$ with $\alpha_1$ and $\alpha_2$, respectively. 
In the end, we are interested in the angle $\phi=\alpha_1-\alpha_2-\pi$ between both jets. Therefore, we transform the integral measure in the following way and later integrate one of the angles ($\alpha_2$) out (see Eq.~\eqref{eq:oppartoniccompact}) to give a factor of $2\pi$.
\begin{align}
  \ddkone\,\ddktwo\,\ddqone\,\ddqtwo &= 
\frac{d\alpha_1\;d \kone^2}{2}\frac{d\alpha_2\;d \ktwo^2}{2}\;\ddqone\;\ddqtwo\non
&= d\phi\; d\alpha_2\;\frac{ d \kone^2 \;\ddqone}{2}\;\frac{ d \ktwo^2 \;\ddqtwo}{2} .
\end{align}

The integrations belonging to the second jet vertex can be easily performed. The result leads to the following definition of the coefficient $c_2(\nu')$: 
\begin{align}
&  \frac{1}{2}\int d \ktwo^2 \int\ddqtwo \; \braket{n',\nu'}{\qtwo}\bra{\qtwo}\opjettwo\ket{\ktwo} \non
=& \frac{1}{2}\int d \ktwo^2 \int\ddqtwo \; \frac{1}{\pi\sqrt{2}}\minusinup{\qtwo^2}e^{-in'\theta_2} \Phi_{\rm jet,2}(\ktwo) \del{\qtwo-\ktwo}\non
 =&\frac{1}{2}\int d \ktwo^2 \frac{1}{\pi\sqrt{2}}\minusinup{\ktwo^2}e^{-in'\alpha_2}\frac{\theta(\ktwo^2-p_2^2)}{\ktwo^2}\non
 =& \frac{1}{\sqrt{2}}\frac{1}{\frac{1}{2}+i\nu'}\minusinup{p_2^2}\frac{e^{-in'\alpha_2}}{2\pi} 
=: c_2(\nu')\frac{e^{-in'\alpha_2}}{2\pi} \label{eq:defc2}.
\intertext{When performing the same calculation for the first jet vertex, we have to note that, due to conventions, $q_1$ is an incoming momentum with respect to the jet vertex while $q_2$ is an outgoing one. This means that $\qone$ has to equal $-\kone$, and hence we get for the first jet vertex}
&  \frac{1}{2}\int d \kone^2 \int\ddqone \; \bra{\kone} \opjetone\ket{\qone}\braket{\qone}{\nu,n}\non
 =& \frac{1}{\sqrt{2}}\frac{1}{\frac{1}{2}-i\nu}\plusinu{p_1^2}\frac{e^{in(\alpha_1-\pi)}}{2\pi}
 =: c_1(\nu)\frac{e^{in(\alpha_1-\pi)}}{2\pi}\label{eq:defc1}.
\end{align}

Recalling Eq.~\eqref{eq:fnnulo2} we have
\begin{align}
  \int\frac{d\omega}{2\pi i}\,\bra{n,\nu}\opf_\omega\ket{\nu',n'} e^{\omega Y} =& \int\frac{d\omega}{2\pi i}\,\frac{1}{\omega-\asbar\chi_0\left(|n|,\frac{1}{2}+i\nu\right)}\delta(\nu-\nu')\delta_{nn'} e^{\omega Y}\non
=& e^{\asbar\chi_0\left(|n|,\frac{1}{2}+i\nu\right) Y}\;\delta(\nu-\nu')\delta_{nn'}  .
\label{eq:fnnulo0}
\end{align}

Inserting these results in Eq.~\eqref{eq:oppartonic} leads to the following compact expression
\begin{align}
  \label{eq:oppartoniccompact}
  \hat\sigma(\shat) =& \frac{\pi^2\asbar^2}{2}\sum_{n,n'=-\infty}^\infty\int d\phi \int d\alpha_2 \int d\nu\int  d\nu' \left[c_1(\nu)\frac{e^{in(\alpha_1-\pi)}}{2\pi}\right]\non
&\times e^{\asbar\chi_0\left(|n|,\frac{1}{2}+i\nu\right) Y} \delta(\nu-\nu')\delta_{nn'} \left[c_2(\nu')\frac{e^{-in'\alpha_2}}{2\pi}\right] \non
=&\frac{\pi^2\asbar^2}{2}\sum_{n=-\infty}^\infty\int d\phi\int d\nu\; c_1(\nu)c_2(\nu)e^{\asbar\chi_0\left(|n|,\frac{1}{2}+i\nu\right) Y} \frac{e^{in\phi}}{2\pi}\non
=& \int d\phi\;\;\frac{\pi^2\asbar^2}{4\sqrt{p_1^2p_2^2}}\sum_{n=-\infty}^\infty e^{in\phi}
\int \frac{d\nu}{2\pi}\frac{1}{\frac{1}{4}+\nu^2}\left(\frac{p_1^2}{p_2^2}\right)^{i\nu}e^{\asbar\chi_0\left(|n|,\frac{1}{2}+i\nu\right) Y},
\end{align}
from which we can read off the angular distribution
\begin{equation}
  \label{eq:dsigmadphi}
 \frac{\hat\sigma(Y)}{d\phi} =\frac{\pi^2\asbar^2}{4\sqrt{p_1^2p_2^2}}\sum_{n=-\infty}^\infty e^{in\phi}\,\mathcal{C}^{\rm LL}_n( Y),
\end{equation}
with
\begin{equation}
  \mathcal{C}^{\rm LL}_n( Y) = \int \frac{d\nu}{2\pi}\frac{1}{\frac{1}{4}+\nu^2}\left(\frac{p_1^2}{p_2^2}\right)^{i\nu}e^{\asbar\chi_0\left(|n|,\frac{1}{2}+i\nu\right) Y} \label{eq:cll}.
\end{equation}

This decomposition is very well suited to study the angular correlation. Besides the differential cross section itself, the mean values $<\cos (n\phi)>$ can be easily obtained from Eq.~\eqref{eq:cll} due the orthogonality of the functions $\exp(i n\phi)$:
\begin{equation}
  \label{eq:cosmeanvalue}
  < \cos (n\phi) > = \frac{\mathcal{C}_n(Y)}{\mathcal{C}_0(Y)} .
\end{equation}

\subsection{Mueller-Navelet Jets at NLO}
\label{sec:mnatnlo}

In general, a complete NLO calculation modifies the jet vertices $\Phi_{\rm jet}$ and the Green's function $f_\omega$ as well. 
The decorrelation between the jets is assumed to be mainly caused by the Green's function even if the NLO jet vertex contains one additional emission. In particular the rapidity dependence is exclusively driven by the Green's function. 
Therefore, we focus on the effect of the NLO Green's function while keeping the jet vertices at LO.

Nevertheless, we have to state that the NLO Mueller-Navelet jet vertex in principle has been calculated \cite{Bartels:2001ge,Bartels:2002yj}. Since it is the link between the collinear factorization from the hadron side and the BFKL dynamics between the jets, the NLO calculation reveals divergent contributions which have to be combined with the NLO collinear parton density and others which overlap with the BFKL evolution. It has been shown in Refs.~\cite{Bartels:2001ge,Bartels:2002yj} that after all the factorization remains valid with a finite result for the jet vertex. However, the final results are not yet in a form appropriate for a direct implementation.

Already in the previous chapter we stressed that the choice of $s_0$ at NLO is no longer completely free. In the LO discussion we made use of the possibility to perform the $q_{1/2}$ and $k_{1/2}$ integrations without touching the Green's function. This was possible since we switched from the physical rapidity $\Delta y$ to an effective rapidity variable $Y$.
A consistent NLO calculation demands to compensate for this energy scale change
\begin{equation}
  \label{eq:scalechange}
  \left(\frac{\shat}{s_0}\right)^\omega = \left(\frac{x_1 x_2 s}{\sqrt{\qone^2\qtwo^2}}\right)^\omega=e^{\omega \Delta y}
  \quad \longrightarrow \quad
  \left(\frac{\shat}{s_0'}\right)^\omega= \left(\frac{x_1 x_2 s}{\sqrt{p_1^2p_2^2}}\right)^\omega=e^{\omega Y}
\end{equation}
by a change of the impact factors
\begin{equation}
  \label{eq:impactfactorchange}
  \Phi^{(\Delta y)}(\qone) \longrightarrow
  \Phi^{(Y)}(\qone) = \Phi^{(\Delta y)}(\qone)-\frac{1}{2}\int \ddqprime \Phi^B( \qprime)\mathcal{K}^B(\qprime,\qone)\ln\frac{\qprime^2}{p_1^2}.
\end{equation}
The superscript $(\Delta y)$ marks the correct expression for the energy scale $s_0$ and $(Y)$ that for the energy scale $s_0'$. The Born expressions are labeled with a $B$. Therefore, this scale change is a NLO effect to the impact factors which we intend to postpone to a future study. As a matter of fact, this choice of energy scale separates the integration over the transverse momenta from that over the longitudinal momentum fractions and from the rapidity itself. Due to this separation the integration over the longitudinal momentum fractions factorizes and  just leads to an overall factor.

To calculate $\bra{n,\nu}\opf_\omega\ket{\nu',n'}$, we first have to determine the matrix elements $\bra{n,\nu}\opk\ket{\nu',n'}$ with the help of Eq.~\eqref{eq:KactingKotikov}:
\begin{subequations}
\label{eq:actionofnloktoto}
\begin{align}
&  \bra{n,\nu}\opk\ket{\nu',n'} = \int\ddqone\int\ddqtwo\;\braket{n,\nu}{\qone}\bra{\qone}\opk\ket{\qtwo}\braket{\qtwo}{n'\nu'}\non
=&\frac{1}{2\pi^2}\int\ddqone\; \asbar(\qone^2)\left[\chi_0\left(|n'|,\frac{1}{2}+i\nu'\right)+\frac{\asbar(\qone^2)}{4}\delta\left(|n'|,\frac{1}{2}+i\nu'\right)\right]\non
&\quad\times\left(\qone^2\right)^{i(\nu'-\nu)-1}e^{i(n'-n)\theta_1} .
\end{align}
We can perform the $\qone$ integration if we take into account the explicit form of the running of the coupling $\asbar(\qone^2)=\asbar(\mu^2)-\asbar^2(\mu^2)\frac{\beta_0}{4N_c}\ln\frac{\qone^2}{\mu^2}$. The logarithm of $\qone^2$ can be expressed as a derivative with respect to either $\nu$ or $\nu'$.  We choose a version, symmetrized  in $\nu$ and $\nu'$:
\begin{align}
&  \bra{n,\nu}\opk\ket{\nu',n'} \non
=& \frac{\asbar(\mu^2)}{2\pi^2}\Bigg[\chi_0\left(|n'|,\frac{1}{2}+i\nu'\right)\non
&\hphantom{\frac{\asbar(\mu^2)}{2\pi^2}\Bigg[}-\frac{\asbar(\mu^2)\beta_0}{4N_c}\chi_0\left(|n'|,\frac{1}{2}+i\nu'\right)\left\{\frac{1}{2}\left(-i\frac{\partial}{\partial\nu'}+i\frac{\partial}{\partial\nu}\right)-\ln\mu^2\right\}\non
&\hphantom{\frac{\asbar(\mu^2)}{2\pi^2}\Bigg[}+\frac{\asbar(\mu^2)}{4}\delta\left(|n'|,\frac{1}{2}+i\nu'\right)\Bigg]\int\ddqtwo\left(\qone^2\right)^{i(\nu'-\nu)-1}e^{i(n'-n)\theta_1} \non
=& \asbar\Bigg[\chi_0\left(|n'|,\frac{1}{2}+i\nu'\right)+\asbar\chi_1\left(|n'|,\frac{1}{2}+i\nu'\right)\non
&\hphantom{\asbar\Bigg[}-\frac{\asbar\beta_0}{8N_c}\chi_0\left(|n'|,\frac{1}{2}+i\nu'\right)\left\{-i\frac{\partial}{\partial\nu'}+i\frac{\partial}{\partial\nu}-2\ln\mu^2\right\}\non
&\hphantom{\asbar\Bigg[}+i\frac{\asbar\beta_0}{8N_c}\frac{\chi_0\left(|n'|,\frac{1}{2}+i\nu'\right)}{\partial\nu'}\Bigg]\delta_{n,n'}\delta(\nu-\nu')
\label{eq:actionofnloknearly} .
\end{align}
\end{subequations}

For the last transformation we used the explicit form of the function $\delta(n,\gamma)$ given in Eq.~\eqref{eq:definitiondelta} and, from now on, $\asbar=\asbar(\mu^2)$. The derivative operators inside the curly brackets now act on the Dirac delta distribution which -- at the end --  has to be resolved by integration by parts. The $\nu$ derivative, with flipped sign, will then act on $\nu$ dependent terms outside this kernel matrix element. The same holds for the $\nu'$ derivative, but it will also act on the leading order kernel attached to this derivative operator. That term cancels the derivative of $\chi_0$ from the NLO kernel. This is very convenient since it is the only term on the r.h.s. of  Eq.~\eqref{eq:KactingKotikov} which is not a real valued function for real $\nu$ or, in other terms, which is not symmetric under the exchange $\gamma\leftrightarrow 1-\gamma$. Since in the full expression of the cross section the $\nu$ dependent part is to the left of the Green's function and the $\nu'$ dependent part to the right, we already evaluate the Dirac and Kronecker deltas by explicitly indicating that one derivative is acting to the left and the other one to the right. Therefore, we can write the diagonal matrix elements of Eq.~\eqref{eq:actionofnloknearly} without the Dirac and Kronecker deltas as
\begin{equation}
\asbar\mathfrak{X}:=\asbar\left[\check\chi_0-\frac{\asbar\beta_0}{8N_c}\left\{-i\overleftarrow{\partial_{\nu}}\check\chi_0+i\check\chi_0\overrightarrow{\partial_{\nu}}-2\ln\mu^2\right\}+\asbar\check\chi_1\right]
  \label{eq:actionofnlok} ,
\end{equation}
where we introduced the short hand notations $\check\chi_{0/1}=\chi_{0/1}(|n|,1/2+i\nu)$. 

As shown in chapter~\ref{sec:bfkl}, the operator of the gluon Green's function is given as the inverse of another operator $\opf_\omega=(\omega-\opk)^{-1}$ (see Eq.~\eqref{eq:opbfkl}). To calculate the matrix element $\bra{n,\nu}\opf_\omega\ket{\nu',n'}$, we there simply used Eq.~\eqref{eq:fnnulo} to write 
\begin{equation}
  \opf_\omega=(\omega-\opk)^{-1}\quad\Rightarrow\quad\bra{n,\nu}\opf_\omega\ket{\nu',n'} = \frac{1}{\omega-\bra{n,\nu}\opk\ket{\nu',n'}} .
\end{equation}
One might pose the question whether an operator in the denominator of a fraction is well defined.
From the mathematical point of view, this is not trivial but well defined since that operator inversion actually is given by the Neumann series
\begin{equation}
  \opf_\omega=(\omega-\opk)^{-1}=\frac{1}{\omega}\sum_{k=0}^\infty\left(\frac{\opk}{\omega}\right)^k .
\end{equation}
Since the kernel acts on its eigenfunctions, the eigenvalues stand in for the kernel, and the series becomes a simple geometric series. That series can again be written as one simple fraction.  Due to the running of the coupling, the NLO kernel is not really diagonal with respect to the LO eigenfunctions such that the ``eigenvalues'' are still operators. Therefore, we have to study the action of the $\nu$-derivatives in the series representation of $\bra{n,\nu}\opf_\omega\ket{\nu',n'}$. At this stage we can write the partonic cross section as
\begin{align}
  \hat{\sigma}(\shat) =& \frac{\pi^2\asbar^2}{2}\sum_{n=-\infty}^\infty\int d\phi\int\frac{d\omega}{2\pi i}\int d\nu \;
c_1(\nu)\frac{1}{\omega}\sum_{k=0}^\infty\left(\frac{\asbar\mathfrak{X}}{\omega}\right)^k c_2(\nu)\frac{e^{i n\phi}}{2\pi}e^{\omega Y},\label{eq:fnloacting}
\end{align}
where $\mathfrak{X}$ is defined in Eq.~\eqref{eq:actionofnlok}. The derivative operators within $\mathfrak{X}$ now act on other $\mathfrak{X}$'s and on the jet vertex coefficient functions $c_1(\nu)$ and $c_2(\nu)$. 
For every operator $\overrightarrow{\partial_\nu}$ we find an operator $\overleftarrow{\partial_\nu}$ at the mirrored position which acts on the exactly mirrored bunch of $\mathfrak{X}$'s.
Since the operators, acting to the right, come with a different sign compared to those, acting to the left, these contributions cancel. The only exception emerges when the derivatives act on the coefficient functions $c_{1/2}(\nu)$. Therefore, the operator $\overrightarrow{\partial_\nu}$ effectively can be replaced by $c'_{2}(\nu)/c_{2}(\nu)$ if the according operator $\overleftarrow{\partial_\nu}$ at the mirrored position is replaced by $c'_{1}(\nu)/c_{1}(\nu)$ at the same time. In this way we can evaluate all derivatives such that every $\mathfrak{X}$ is replaced by a simple function:
\begin{equation}
  \mathfrak{X}\quad\longrightarrow\quad
\check\chi_0-\frac{\asbar\beta_0}{8N_c}\left\{-i\left[\frac{\partial}{\partial\nu}\ln\frac{c_1(\nu)}{c_2(\nu)}\right]-2\ln\mu^2\right\}+\asbar\check\chi_1 \label{eq:fnloacting2}.
\end{equation}
For the case of our jet vertices defined in Eqs.~(\ref{eq:defc2}, \ref{eq:defc1}) the logarithmic derivative in Eq.~\eqref{eq:fnloacting2} reads
\begin{equation}
-i  \frac{\partial}{\partial\nu}\ln\frac{c_1(\nu)}{c_2(\nu)} = 
\ln(p_1^2p_2^2)+\frac{1}{\frac{1}{4}+\nu^2} \label{eq:impcontributionmn}.
\end{equation}

Now the operator series in Eq.~\eqref{eq:fnloacting} reduces to a simple geometric series, and it is straight forward to perform the Mellin transformation to calculate the NLO coefficient
\begin{equation}
  \label{eq:cnll}
  \mathcal{C}_n^{\rm NLL}(Y) =
\infint \frac{d\nu}{2\pi}\left(\frac{p_1^2}{p_2^2}\right)^{i\nu}
\frac{e^{\asbar Y\left(\check\chi_0+\asbar\left[\check\chi_1-\frac{\beta_0}{8N_c}\check\chi_0\left(2\ln\frac{p_1 p_2}{\mu^2}+\frac{1}{\frac{1}{4}+\nu^2}\right)\right]\right)}}{\frac{1}{4}+\nu^2} .
\end{equation}
In the derivation of Eq.~\eqref{eq:actionofnloktoto} we expressed the coupling constant at the scale $\qone^2$ in terms of one at a general renormalization scale $\mu$. Since this was an intermediate step, it is natural to redo it for the final result. Thus we can reexpress the running coupling term in Eq.~\eqref{eq:cnll} in the following way
\begin{equation}
  \label{eq:redorunning}
  \asbar-\asbar^2\frac{\beta_0}{4N_c}\ln\frac{p_1 p_2}{\mu^2} \quad\longrightarrow\quad \asbar(p_1p_2) ,
\end{equation}
and obtain as coefficient
\begin{equation}
  \label{eq:cnll2}
\mathcal{C}_n^{\rm NLL}(Y) =\infint \frac{d\nu}{2\pi}\left(\frac{p_1^2}{p_2^2}\right)^{i\nu}
\frac{e^{\asbar(p_1 p_2) Y\left(\check\chi_0+\asbar(p_1 p_2)\left[\check\chi_1-\frac{\beta_0}{8N_c}\frac{\check\chi_0}{\frac{1}{4}+\nu^2}\right]\right)}}{\frac{1}{4}+\nu^2} .
\end{equation}

In section~\ref{sec:bfklresummation} we have shown that it is necessary to resum terms of even higher order than NLO to obtain a reliable kernel. The resummed kernel, including the impact factor contribution $-\frac{\beta_0}{8N_c}\frac{\check\chi_0}{\frac{1}{4}+\nu^2}$, then replaces the kernel in the exponent of Eq.~\eqref{eq:cnll2} and will be referred to it as 
\begin{equation}
  \label{eq:cresummed}
  \mathcal{C}_n^{\rm resum}(Y) =
\infint \frac{d\nu}{2\pi}\left(\frac{p_1^2}{p_2^2}\right)^{i\nu}
\frac{e^{\omega^{\rm resum}(n,\nu)\,Y}}{\frac{1}{4}+\nu^2} .
\end{equation}
The technical details, how to include this additional contribution to the kernel in the resummation procedure, can be found in appendix~\ref{sec:impfactor}.

It is not possible to perform the $\nu$-integration in Eq.~\eqref{eq:cresummed} analytically. In principle there are two ways two evaluate it: either by a numerical integration or by an analytical approximation. In our studies we perform the integration numerically although this is quite time consuming as the resummed kernel is given by an implicit equation. 

The saddle point method is the common way to approximate this kind of integrals analytically. It relies on the assumptions that the exponent $\omega(n,\nu)$ is well approximated by a Taylor expansion around its maximum at $\nu=0$
\begin{equation}
  \omega(n,\nu)\approx \omega(n,0)+\frac{\nu}{2}\frac{d^2\omega(n,\nu)}{d\nu^2}\big|_{\nu=0} ,
\end{equation}
and the multiplicative function in front of the exponential is `slowly varying' around this maximum. If we write a general kernel as $\omega(n,\nu)=\asbar\chi(n,1/2+i\nu)$ and make use of these approximations, the saddle point approximation to Eq.~\eqref{eq:cresummed} reads
\begin{equation}
  \mathcal{C}_n(Y) \approx \frac{2}{\pi}\sqrt{\frac{2\pi}{\asbar\chi''\left(n,\frac{1}{2}\right)Y}} \; \; e^{\asbar \chi \left(n,\frac{1}{2}\right)Y},
\end{equation}
where $\chi''(n,\gamma)=d^2\chi(n,\gamma)/d\gamma^2$.

\begin{figure}[htbp]
  \centering
  \includegraphics[width=10cm]{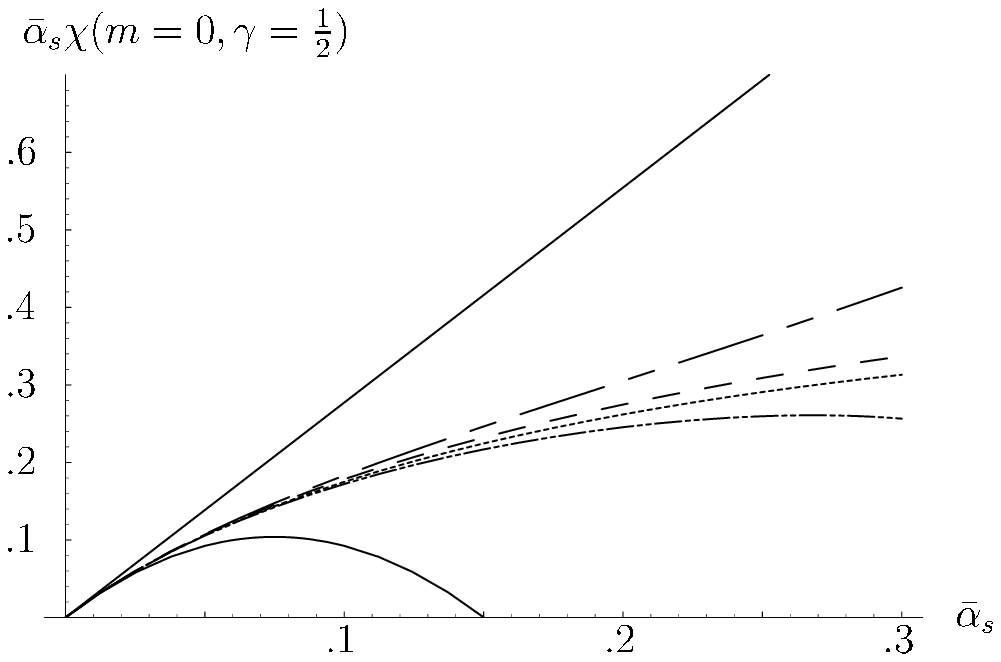}
  \includegraphics[width=10cm]{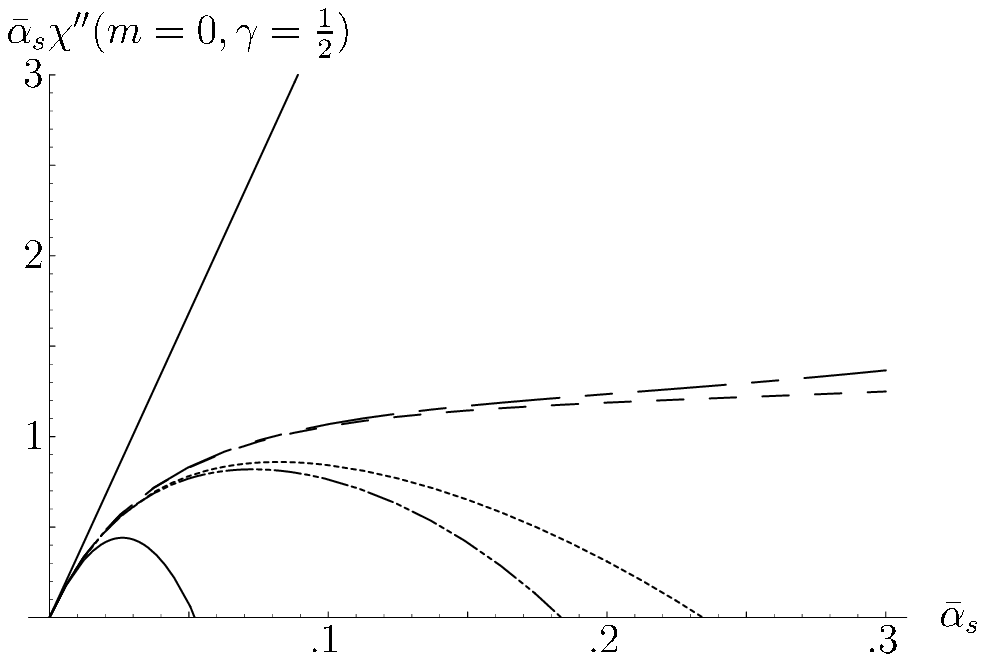}
   \caption{The kernel and its second derivative in dependence of the coupling $\asbar$. The solid lines show the results for the LO (straight line) and NLO (bended line) kernel without resummation. The results for the different resummation schemes lie between these both. In both diagrams they are from top to bottom: scheme 4, scheme 3, scheme 1, scheme 2.}
  \label{fig:saddlepoint}
\end{figure}

\begin{figure}[htbp]
  \centering
  \includegraphics[width=10cm]{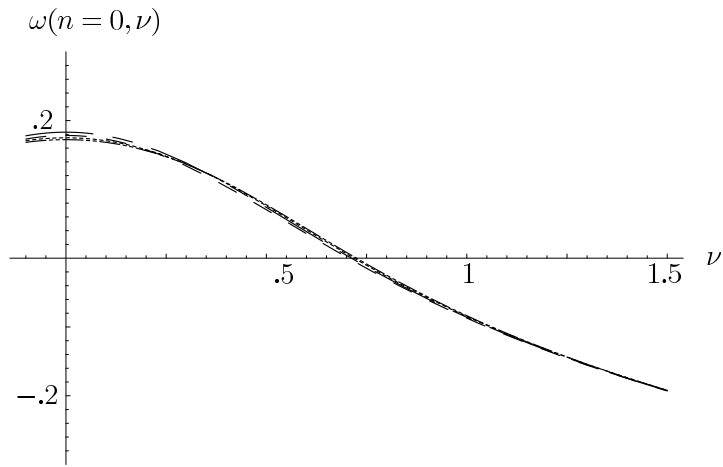}
  \caption{The kernel for conformal spin 0 is shown for  $\asbar=.1$. The different lines correspond to the four resummation schemes, plotted in the same style as  in Fig.~\ref{fig:saddlepoint}.}
  \label{fig:saddlepoint3}
\end{figure}

\begin{figure}[htbp]
  \centering
  \includegraphics[width=10cm]{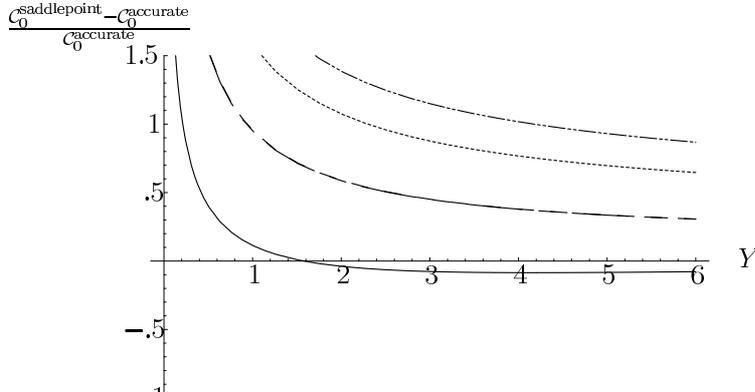}
  \caption{The rapidity dependence of the ratio between  the saddle point approximation and the accurate numeric evaluation for the coefficient $\mathcal{C}_0$ is shown for $\asbar=.1$. The solid line displays the LO calculation. The resummation schemes from top to bottom are: scheme 2, scheme 1, scheme 3, scheme 4. The lines for scheme 3 and 4 lie on top of each other.}
  \label{fig:saddlepoint23}
\end{figure}

For the LO kernel this approximation works quite well, but for the pure NLO kernel it breaks down since $\chi''(n,1/2)<0$ in that case. Even after the resummation procedure the saddle point method is not sufficient for the resummed kernel. To illustrate this, we show in Fig.~\ref{fig:saddlepoint} the resummed kernel and its second derivative with respect to $\gamma=1/2+i\nu$. Since the resummed kernel is implicitly $\omega$ dependent, we have
\begin{equation}
  \chi''\Big(n,\frac{1}{2},\omega\Big)=\frac{d^2\chi(n,\gamma,\omega)}{d\gamma^2}\Big|_{\gamma=\frac{1}{2}} = \left[1-\asbar\frac{\partial\chi(n,\gamma,\omega)}{\partial\omega}\right]^{-1}\frac{\partial^2\chi(n,\gamma,\omega)}{\partial\gamma^2}\Big|_{\gamma=\frac{1}{2}} .
\end{equation}
Fig.~\ref{fig:saddlepoint} illustrates that the kernel at the saddle point for the resummed version is positive also for larger values of $\asbar$ while the NLO kernel without resummation turns and becomes negative. The different resummation schemes give compatible results for the kernel. In contrast, the second derivative shows a stronger dependence on the resummation scheme. In scheme 1 and 2 the maximum even turns into a minimum for larger $\asbar$ as it is the case for the NLO kernel without resummation  at $\asbar\approx .05$. Already at $\asbar=.1$ the second derivatives deviate from each other significantly at the saddle point, but the complete kernel is very similar for the different resummation schemes, as it can be seen in Fig.~\ref{fig:saddlepoint3}. Therefore, we conclude that it is mandatory to perform the $\nu$-integration numerically to obtain reliable results. As Fig.~\ref{fig:saddlepoint23} shows, even at large rapidities the saddle point method does not approximate the accurate result reasonably.

\subsection{Phenomenology of Mueller-Navelet jets}

\begin{figure}[htbp]
  \centering
  \includegraphics[width=6cm]{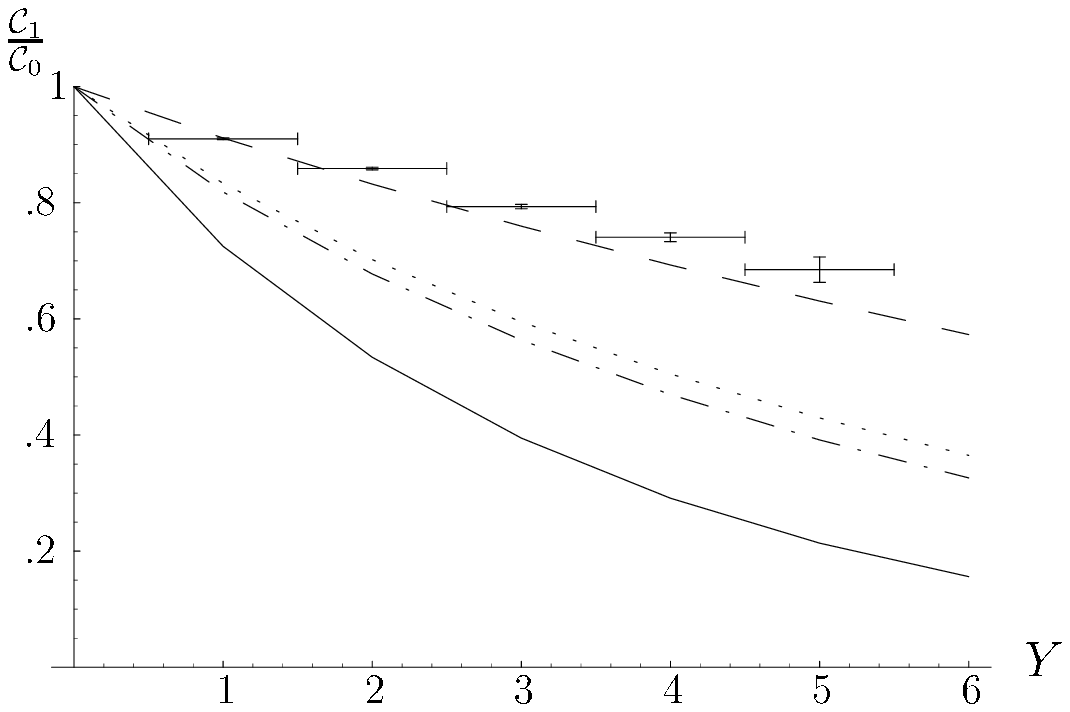}
  \includegraphics[width=6cm]{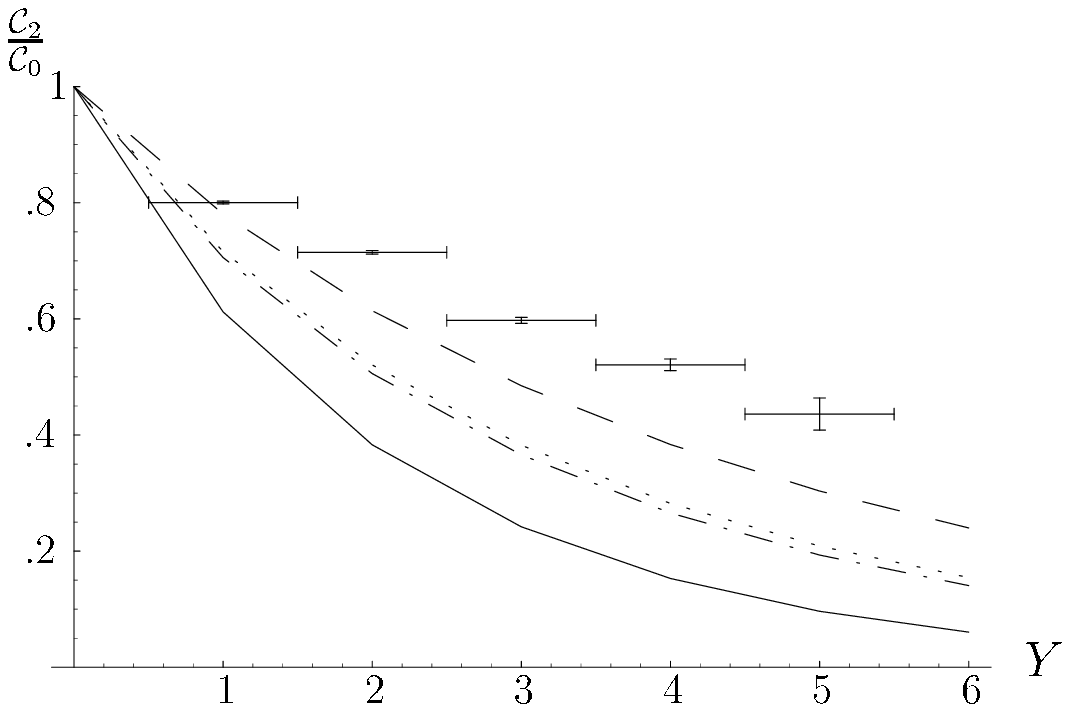}
  \includegraphics[width=6cm]{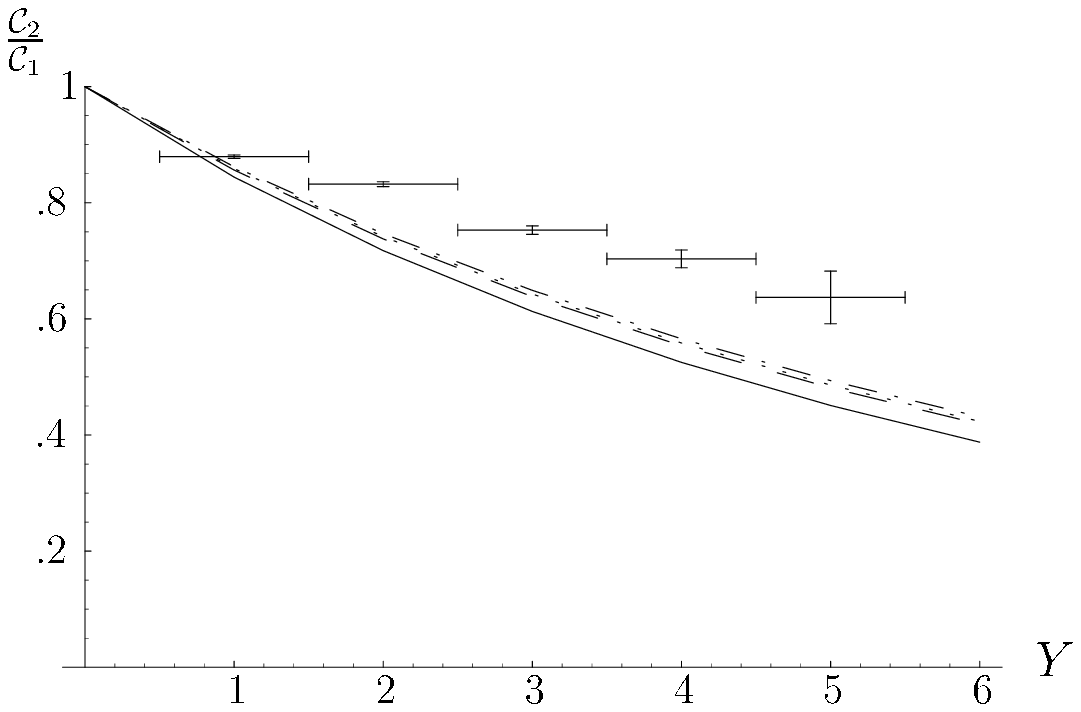}
   \caption{$<\cos\phi>$, $<\cos 2\phi>$, and $\frac{<\cos2\phi>}{<\cos\phi>}$ at a $p\bar{p}$ collider with a center of mass energy $\sqrt{s}$=1.8TeV at leading (solid) and next to leading order (dashed). The results of the resummation following scheme 1 (dotted) and scheme 3 (dash-dotted) are given as well. Tevatron data points are taken from Ref.~\cite{Kim:1996dg}.}
  \label{fig:tevatron}
\end{figure}

Ten years ago the {D$\emptyset$} collaboration at the Tevatron measured the azimuthal decorrelation between Mueller-Navelet jets \cite{Abachi:1996et}. At that time merely the LO BFKL equation was available which failed to describe the data by estimating too much decorrelation, while an exact fixed NLO $(\alpha_s^3)$ Monte Carlo calculation by the program JETRAD \cite{Giele:1993dj} underestimated the decorrelation. In contrast, the Monte Carlo program HERWIG \cite{Marchesini:1991ch}  was in perfect agreement with the data.
In Fig.~\ref{fig:tevatron} we show the data on $<\cos\phi>$ and $<\cos 2\phi>$ and compare them with our resummed NLO prediction developed in the previous section using Eq.~\eqref{eq:cosmeanvalue}, which expresses the angular mean values in terms of the coefficients $\mathcal{C}_n$. For comparison we give the LO, and not-resummed-NLO BFKL results as well. 

It can be seen that the NLO corrections to the BFKL calculation change the results significantly leading to a slower decorrelation with increasing rapidity. As stated in the previous section, the choice of the rapidity variable $Y=\ln\frac{\shat}{p_1p_2}$ turns the convolution with the effective parton distributions into a simple global factor which cancels whenever we study ratios  of cross sections or coefficients $\mathcal{C}_n$. For such observables, the hadronic level calculation does therefore not differ from the partonic one. Thereby, uncertainties of parton distribution functions do not spoil our calculation.

It is worth remarking that of all eigenvalues associated to the NLO 
BFKL kernel the one corresponding to the $n = 0$ conformal spin is the one with 
the poorest convergence in the transition from LO. This becomes evident if we consider ratios of coefficients
\begin{equation}
  \frac{\mathcal{C}_m}{\mathcal{C}_n} = \frac{<\cos(m\phi)>}{<\cos(n\phi)>},
\label{eq:rmn}
\end{equation}
as we did for the case $m=2$, $n=1$ in the third plot of Fig.~\ref{fig:tevatron}. This observable reveals that the not-resummed-NLO calculation just accidentally fits the data for $<\cos\phi>$.

\begin{figure}[htbp]
  \centering
  \includegraphics[width=6cm]{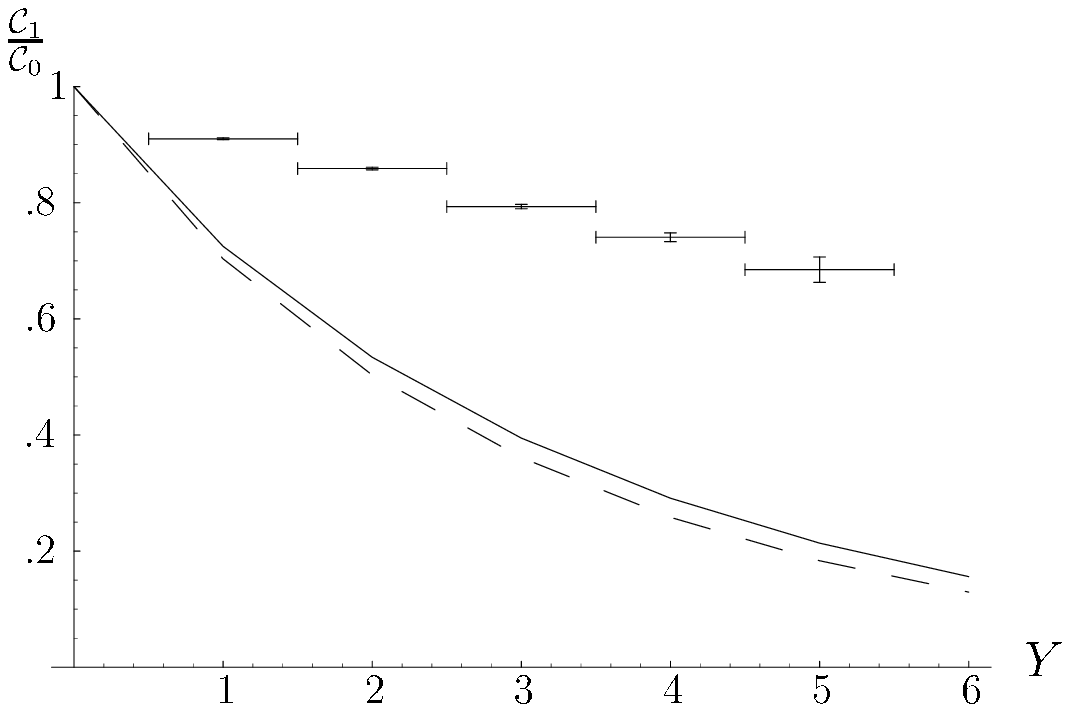}\hspace{.5cm}
  \includegraphics[width=6cm]{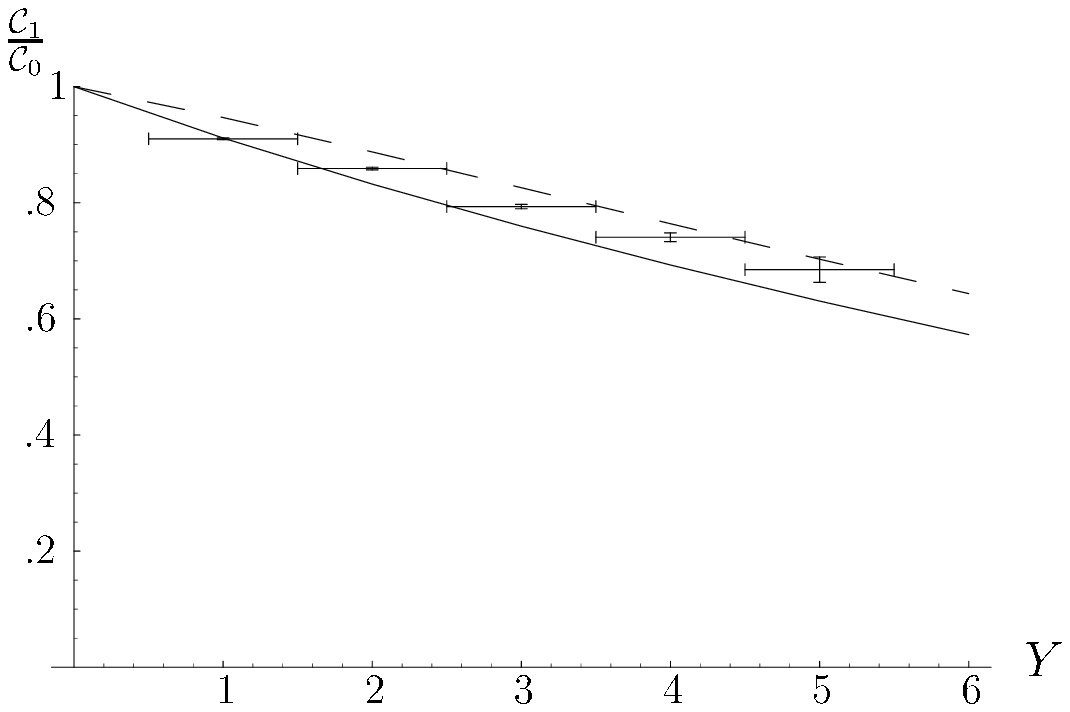}\\
  \includegraphics[width=6cm]{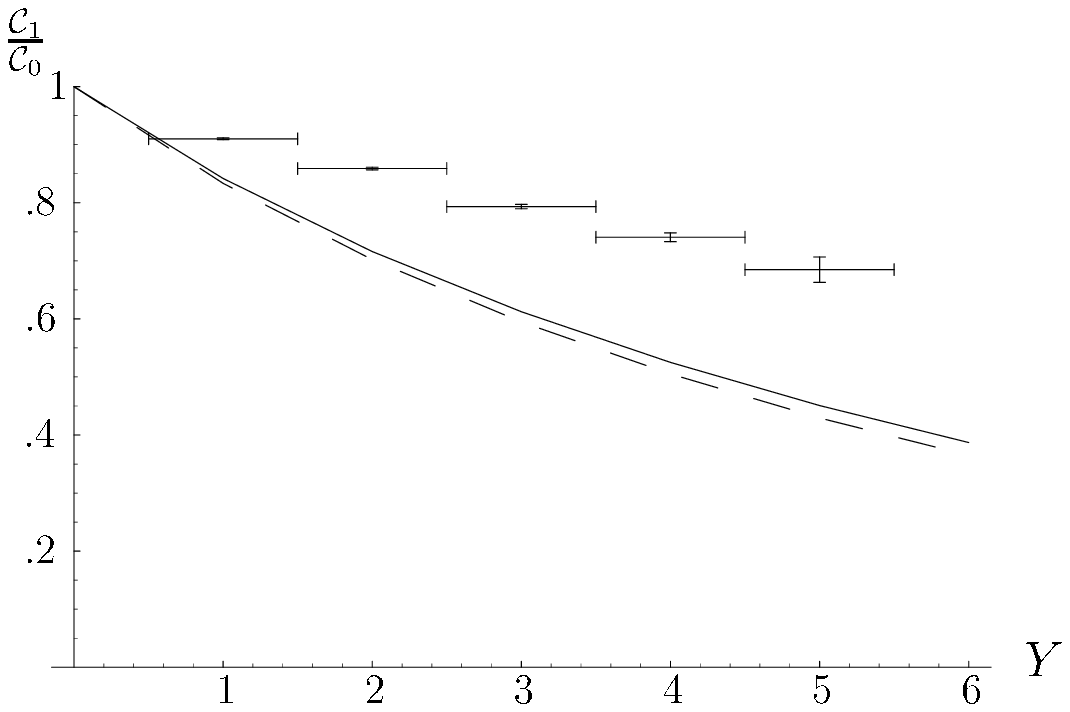}\hspace{.5cm}
  \includegraphics[width=6cm]{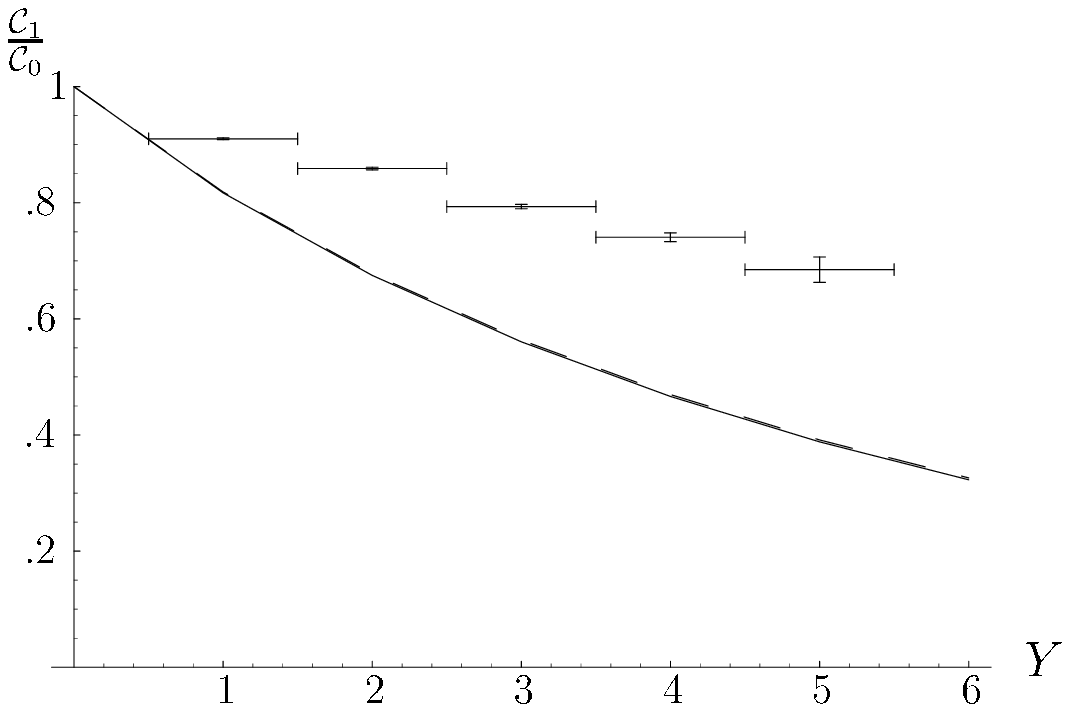}
   \caption{$<\cos\phi>$ at a $p\bar{p}$ collider with a center of mass energy $\sqrt{s}$=1.8TeV. Results for $\overline{\rm MS}$-renormalization are shown as a solid line, GB-renormalization is given as a dashed line. The different plots show in the first row LO and NLO, while the second row contains the resummation schemes 1 and 3. Tevatron data points are taken from Ref.~\cite{Kim:1996dg}.}
  \label{fig:tevatronresummed}
\end{figure}

To further motivate the necessity of a resummed kernel, we now discuss the sensitivity against a change of the renormalization scheme.
In Fig.~\ref{fig:tevatronresummed} we compare the results for $\overline{\rm MS}$ renormalization scheme with those for the gluon-bremsstrahlung (GB) renormalization scheme \cite{Catani:1990rr,Dokshitzer:1995ev}. The relevance of this more physical renormalization scheme has been discussed in a variety of works \cite{Vera:2005jt,Kotikov:2000pm,Salam:1999cn,Cacciari:2001cw,Banfi:2004yd} dealing with soft gluon emission. Instead of the LO calculation being more dependent on the renormalization scheme than the NLO one, nearly the opposite observation has to be made. In contrast, the resummed calculations are nearly independent of the renormalization scheme, with scheme 3 being the most robust. This underlines the necessity to resum such formally sub-leading terms although a quick glance at just the observable $<\cos\phi>$ might suggest that the pure NLO kernel is sufficient.

\begin{figure}[htbp]
  \centering
  \includegraphics[width=9cm]{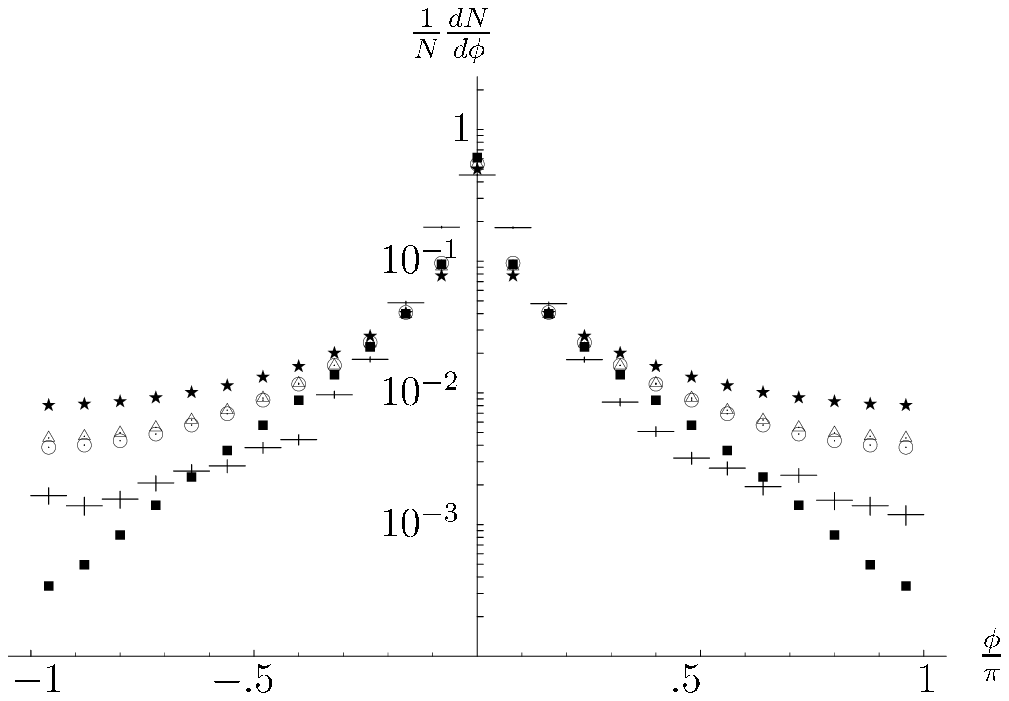}
  \includegraphics[width=9cm]{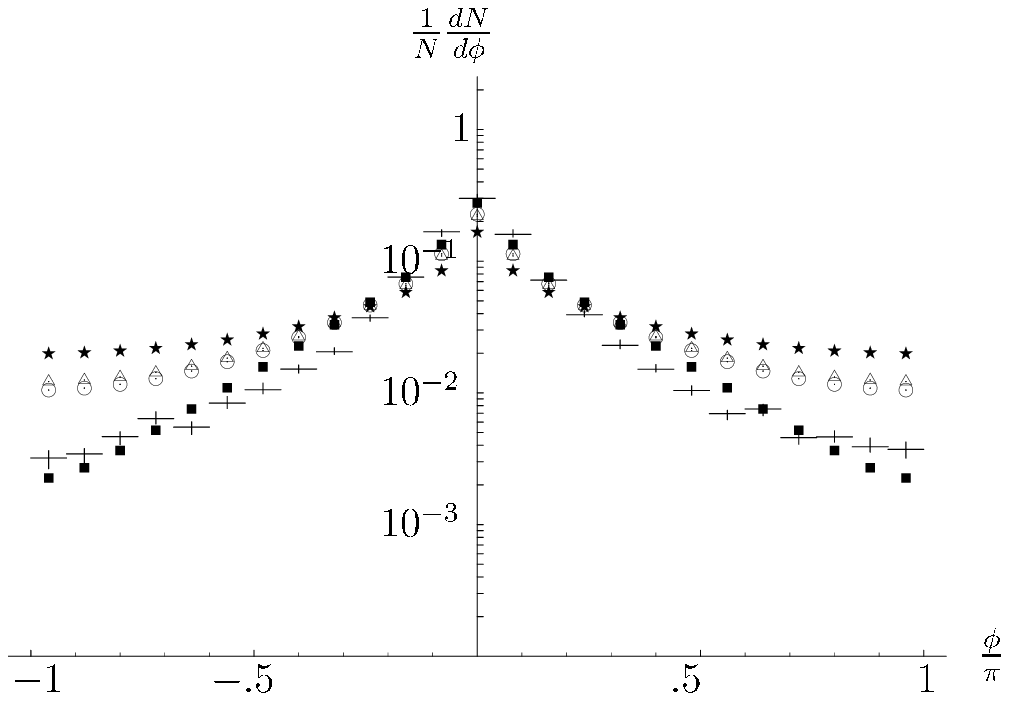}
  \includegraphics[width=9cm]{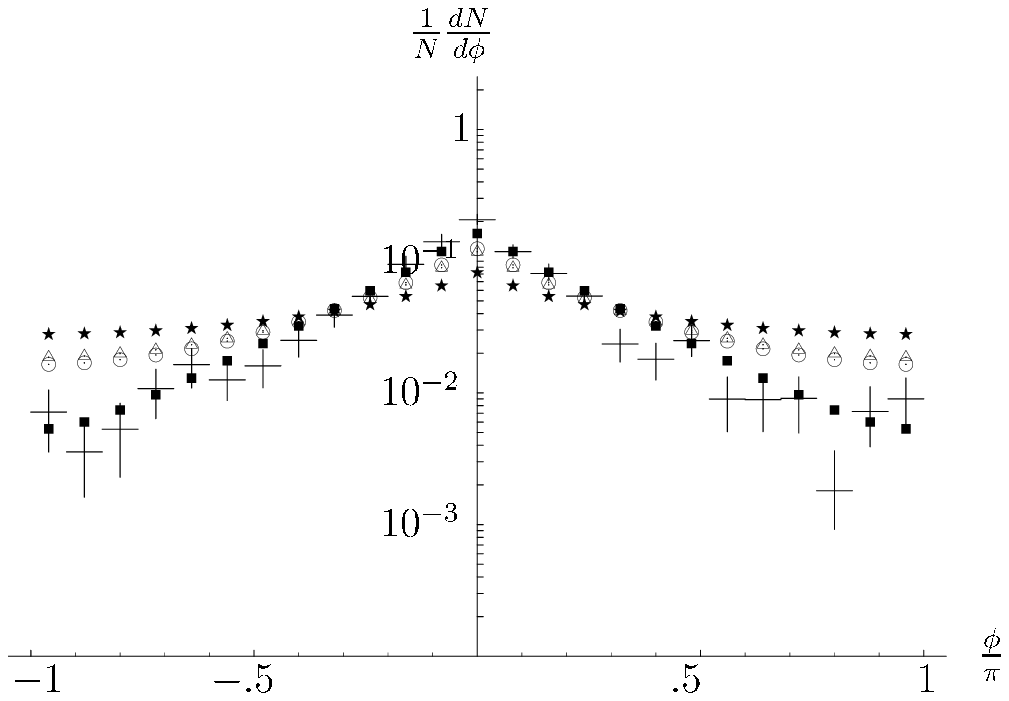}
   \caption{$\frac{1}{N}\frac{dN}{d\phi}$ at a $p\bar{p}$ collider with a center of mass energy $\sqrt{s}$=1.8TeV at leading (stars) and next to leading order (squares). The results of the resummation following scheme 1 (circles) and scheme 3 (triangles) are given as well. Plots are shown for $Y=1$, $Y=3$, and $Y=5$.  Tevatron data points with error bars are taken from Ref.~\cite{Abachi:1996et}.}
  \label{fig:tevatrondsigma}
\end{figure}

In addition to the deduced observable $\mathcal{C}_{m}/\mathcal{C}_{n}$ defined in Eq. \eqref{eq:rmn}, we study directly the differential angular distribution as given in Eq.~\eqref{eq:dsigmadphi}. 
The {D$\emptyset$} collaboration published their measurement of the normalized angular distribution for different rapidities \cite{Abachi:1996et}.
In Fig.~\ref{fig:tevatrondsigma} we compare this measurement with the predictions obtained in our approach from  LO, NLO, and resummed BFKL kernel. This comparison throws more light on the question whether a resummed NLO kernel is needed. While the weights $<\cos (n\phi)>$ seemed to favor the pure NLO calculation, the shape of the differential distribution puts this impression  into the right perspective. Although neither the resummed nor the pure NLO kernel matches the data, the resummed kernel agrees in shape and is closer to the data than the LO estimate. Nevertheless, it still overestimates the decorrelation, and a pure calculation of $\chi^2/n.d.f.$ still favors the NLO calculation.

A very strong hint, that the resummation is necessary, is given by the instability of the pure NLO kernel against a change of the renormalization scheme. Switching to GB-renormalization the angular distribution even becomes negative for large $\phi$. This unphysical behavior clearly is not acceptable. 

To comment on the differences between scheme 1 and 3, we have to anticipate some details from appendix \ref{sec:resummation} where we explicate why scheme 3 is  more robust than scheme 1. 

\begin{figure}[htbp]
  \centering
  \begin{tabular}{ccc}
  \includegraphics[width=4.4cm]{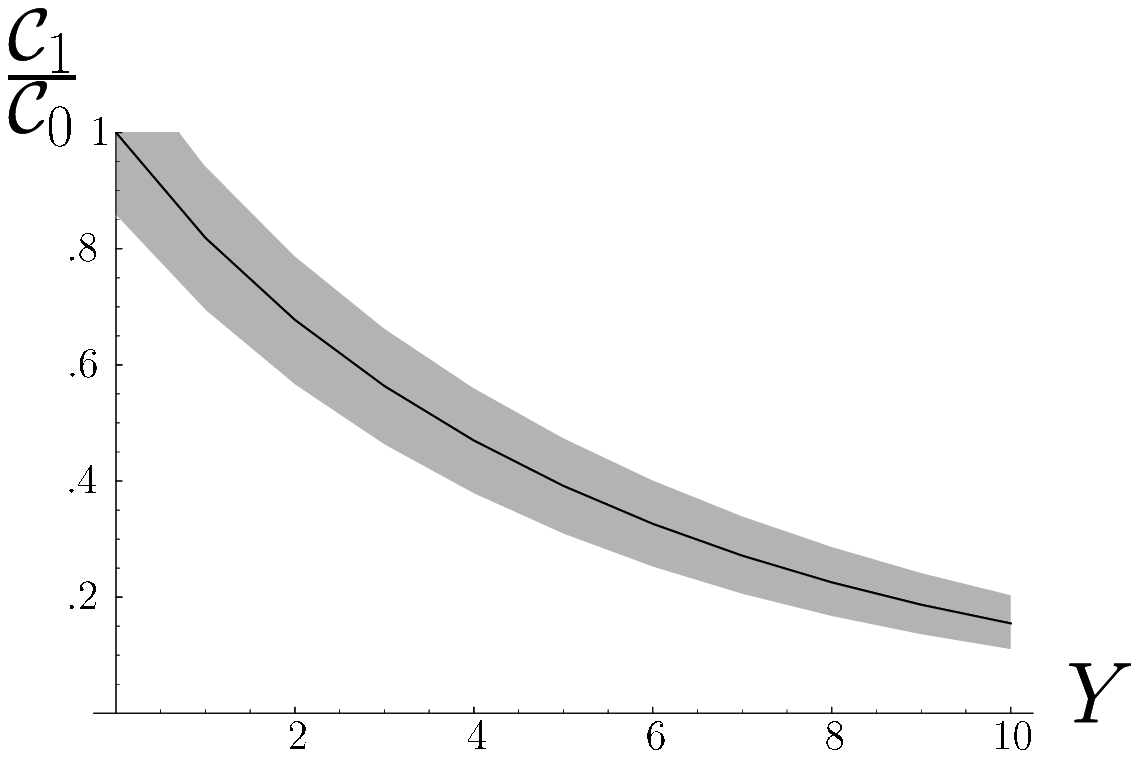} &  \hspace{-1cm} 
  \includegraphics[width=4.4cm]{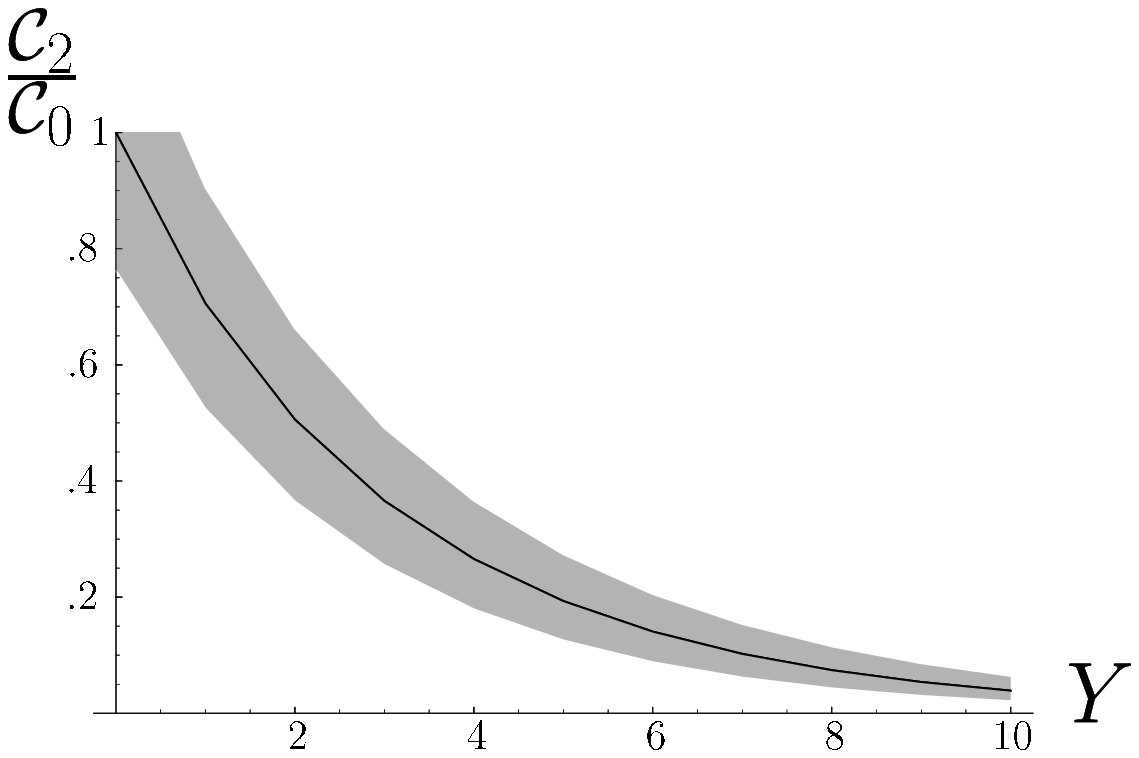} &  \hspace{-1cm} 
  \includegraphics[width=4.4cm]{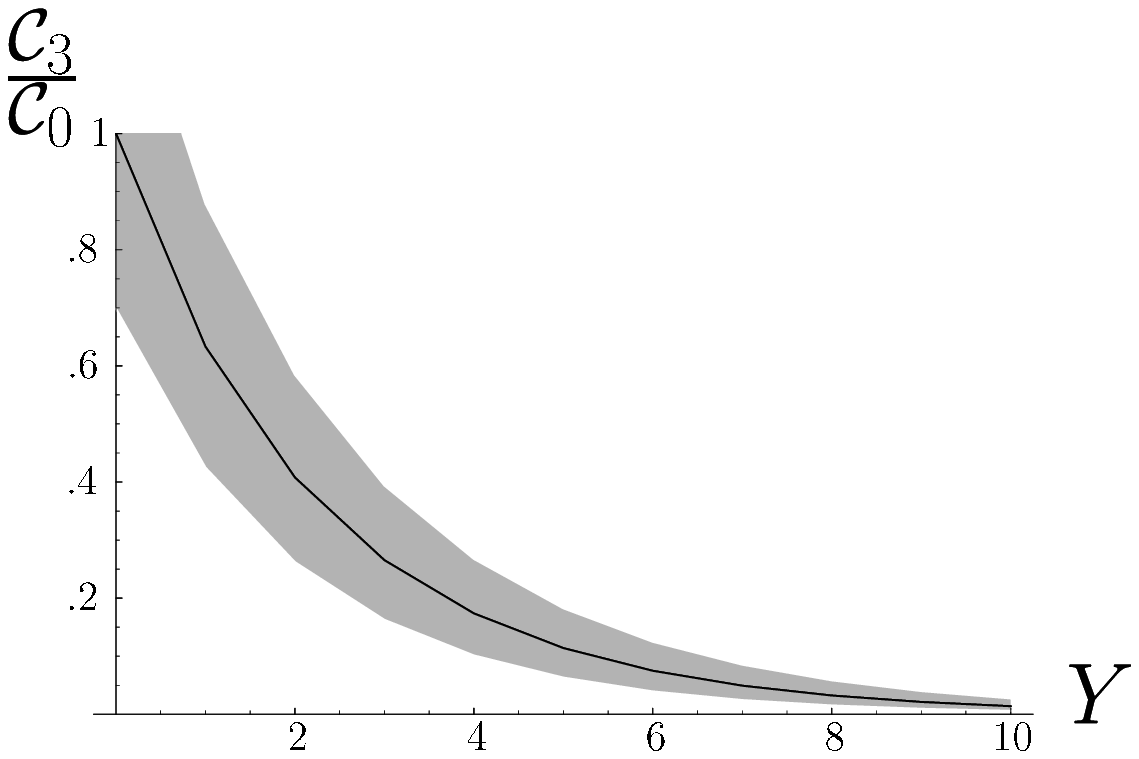} \\
  &  \hspace{-1cm} 
  \includegraphics[width=4.4cm]{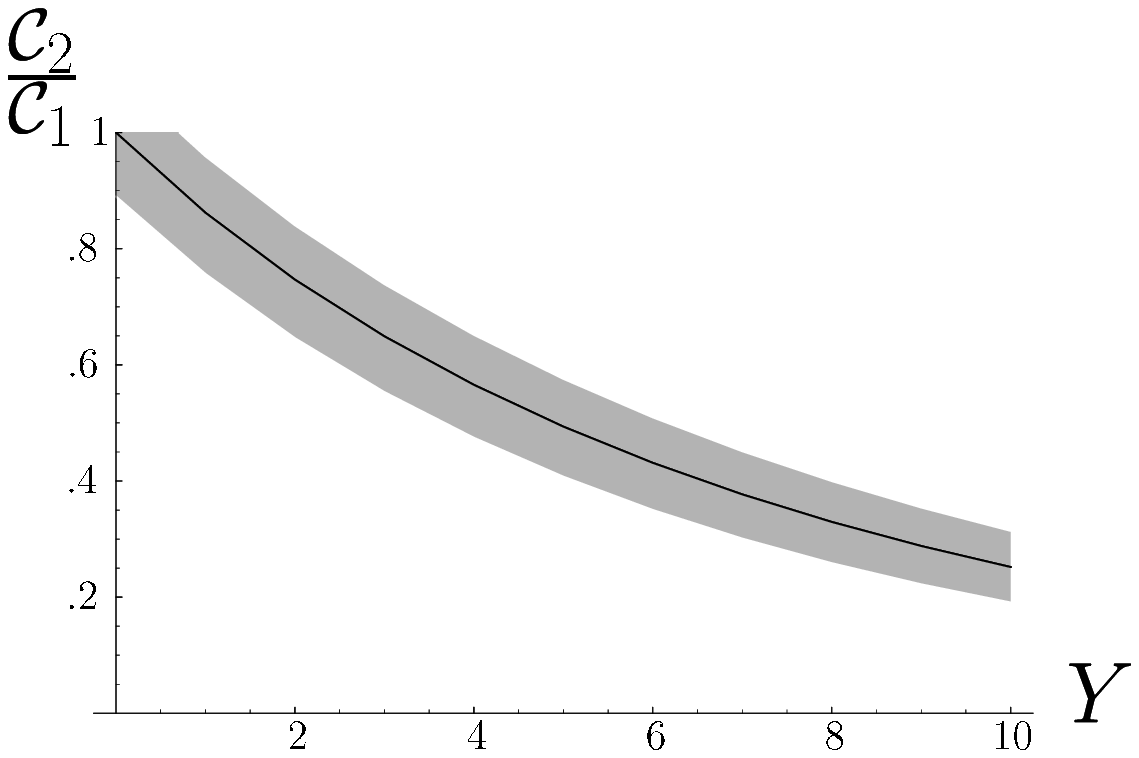}&  \hspace{-1cm} 
  \includegraphics[width=4.4cm]{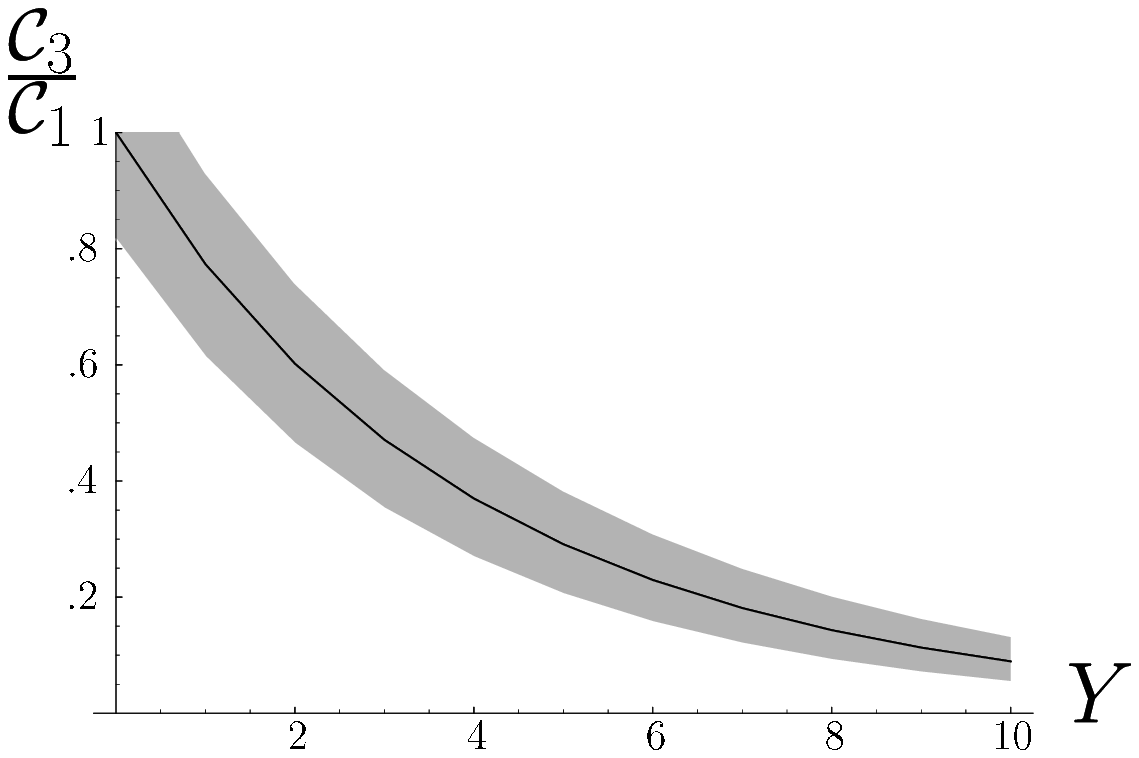} \\
  && \hspace{-1cm} 
  \includegraphics[width=4.4cm]{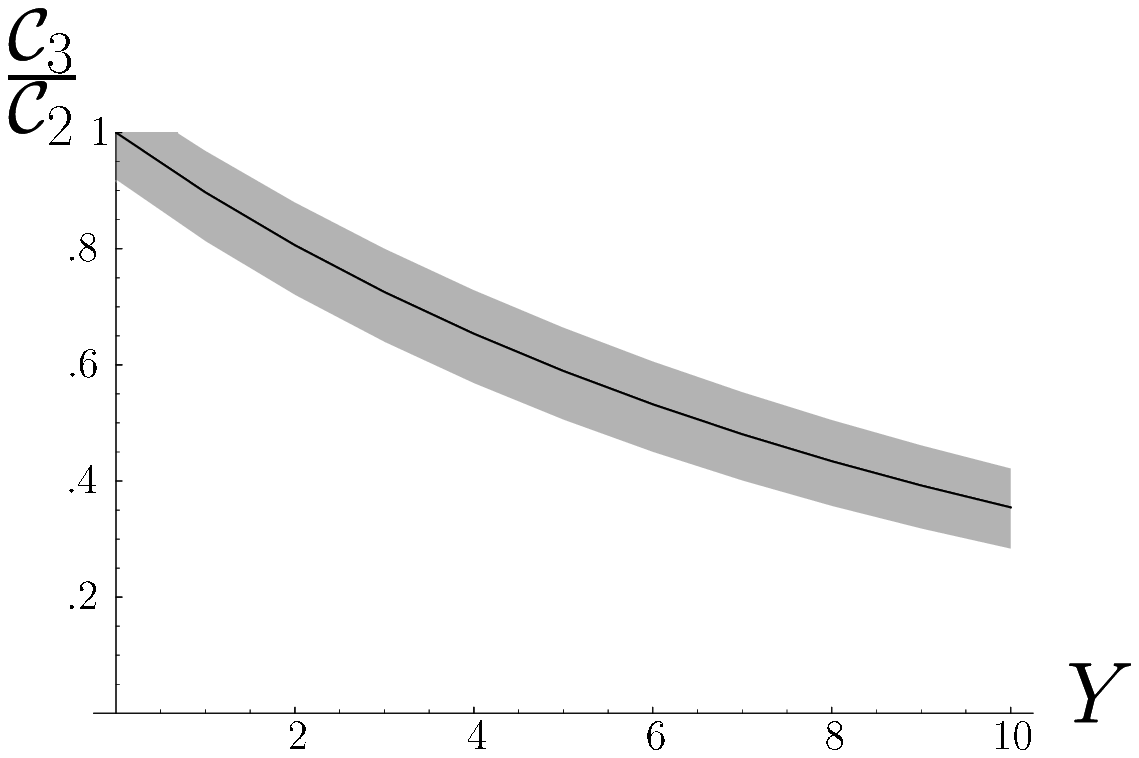} 
  \end{tabular}
  \caption{Different ratios of coefficients $\mathcal{C}_n$ obtained by resummation scheme 3. The gray band reflects the uncertainty in $s_0$ and in the renormalization scale $\mu$.}
  \label{fig:lhc}
\end{figure}

\begin{figure}[htbp]
  \centering
  \includegraphics[width=9cm]{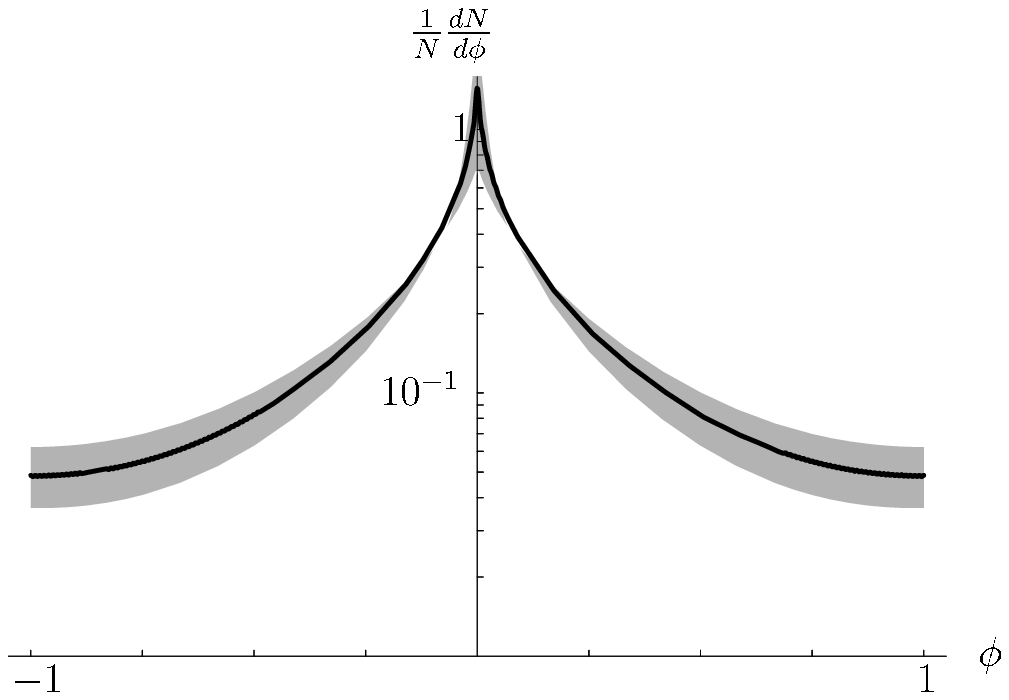}
  \includegraphics[width=9cm]{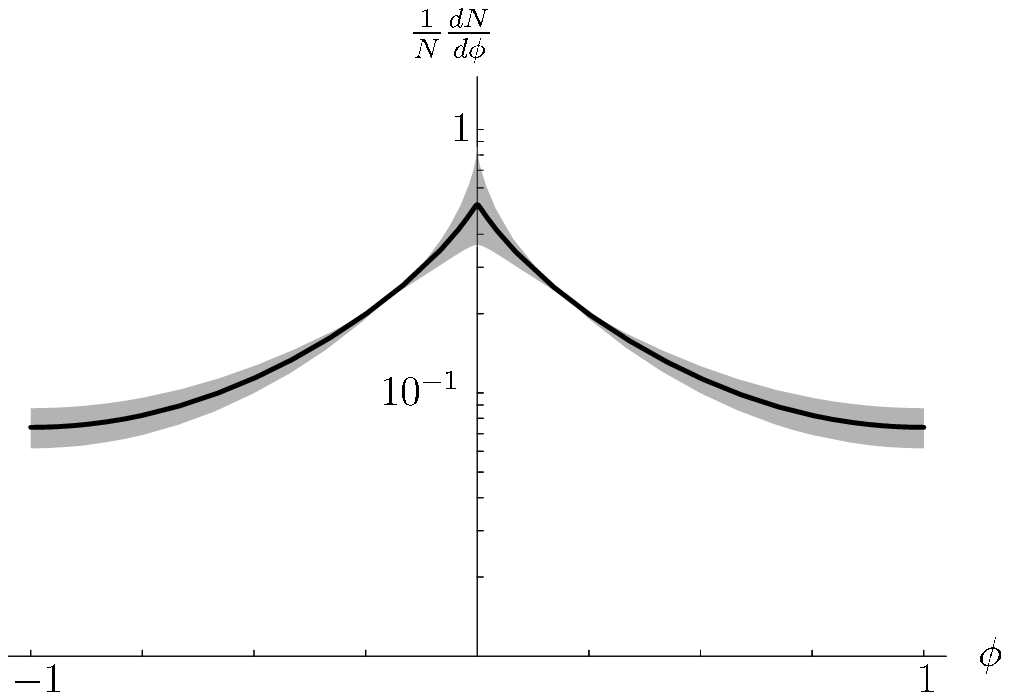}
  \includegraphics[width=9cm]{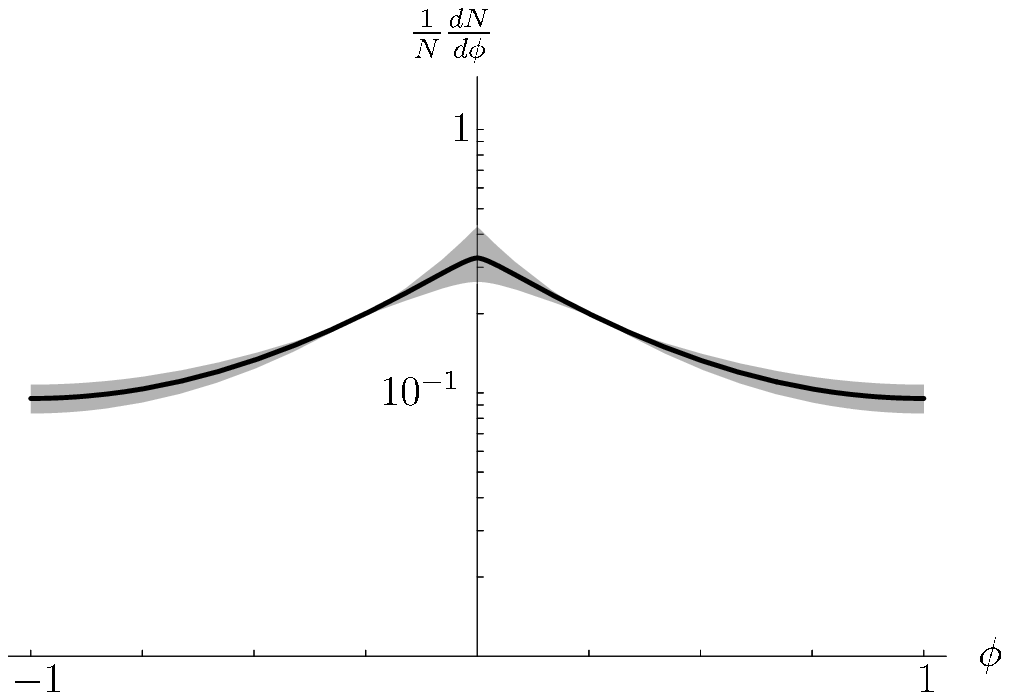}
   \caption{$\frac{1}{N}\frac{dN}{d\phi}$ obtained by resummation scheme 3 for rapidities $Y=3,5,7$ from top to bottom. The gray band reflects the uncertainty in $s_0$ and in the renormalization scale $\mu$.}
  \label{fig:lhcdsigma}
\end{figure}

Before we turn to the study of electron-proton collisions in the next section, we like to state that we strongly suggest to measure the angular correlation of Mueller-Navelet jets at the forthcoming Large Hadron Collider (LHC) as well. Besides the possibility of an independent check of the data published by the {D$\emptyset$} collaboration, measurements for even larger rapidity are desirable. This region of forward physics is interesting in many aspects, as e.g. to study diffraction or saturation. Since the available rapidity range is  restricted rather by the geometry of the detector than by the energy of the colliding particles, the experimentalists spare no effort to extend the capability of the detectors to the very forward region. Therefore, we provide our numerical calculations of ratios $\mathcal{C}_{m}/\mathcal{C}_{n}$ for a broader range of rapidity as a prediction for LHC in Fig.~\ref{fig:lhc}.  For the differential cross section itself we provide predictions in Fig.~\ref{fig:lhcdsigma}, starting with rapidity $Y=3$. Since BFKL dynamics require large rapidities, predictions based on the BFKL equation for even smaller rapidities are not reasonable. 

Although we soundly justified to ignore the NLO contribution to the jet vertex, we studied the possible effects of corrections to it. We investigated the impact of the running of the coupling and of that part of the NLO Mueller-Navelet jet vertex originating from the splitting functions. These contributions can be read off easily from the results of Refs.~\cite{Bartels:2001ge,Bartels:2002yj}. It turns out, that the effect on the overall normalization can become large, as it has been shown in Ref.~\cite{Kepka:2006xe} as well, but the change of the ratios that we consider is only of the order of a few percent.

Furthermore, are aware of the fact, that the change from physical rapidity $\Delta y$ to an effective rapidity $Y$ is note compensated (See Eqs.~(\ref{eq:scalechange}, \ref{eq:impactfactorchange})). We take the resulting uncertainty into account by varying our choice of $s_0=p_1 p_2$ by a factor of 2. We varied the renormalization scale $\mu$ by a factor of two as well, and give the uncertainty arising from these two sources by gray bands in  Figs.~\ref{fig:lhc} and \ref{fig:lhcdsigma}. 

\section{Forward jets at an $ep$ collider}

At an electron proton collision one can study the angular correlation between the electron and  a forward jet form the proton. A LO study can be found in Ref.~\cite{Bartels:1996wx}. Here we apply our NLO formalism to this subject. While the jet vertex of the forward jet, in principle, remains the same as in the Mueller-Navelet case, we have to deal with a new vertex for the leptonic part, namely, for the coupling of the electron to the gluon ladder. Since the electron does not interact strongly, the coupling is mediated by an additionally produced quark--antiquark pair which couples to the electron via a virtual photon. Again we restrict ourselves to LO jet vertices, especially since the challenging calculation of the NLO photon impact factor which is needed for the leptonic part is still work in progress \cite{Bartels:2000gt,Bartels:2001mv,Bartels:2002uz,Bartels:2004bi,Fadin:2001ap,Fadin:2002tu}. 

\begin{figure}[htbp]
  \centering
  \includegraphics[width=8cm]{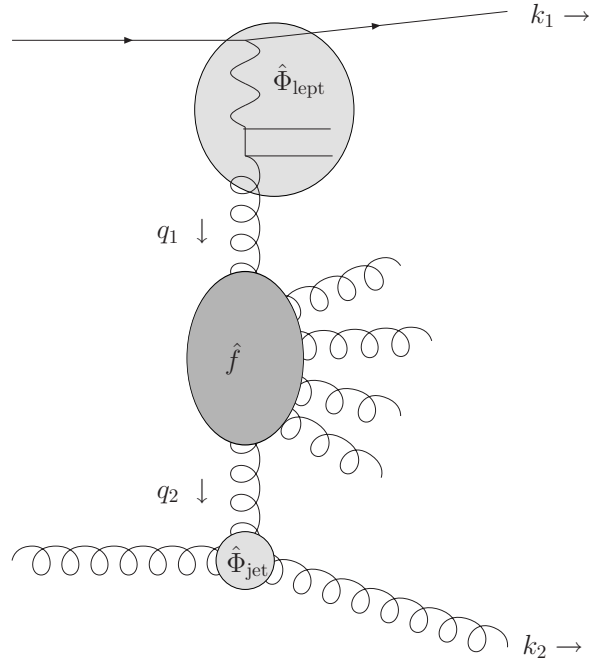}
  \caption{Kinematics of the partonic cross section in the $ep$ case}
  \label{fig:herakinematics}
\end{figure}

The kinematical situation is depicted in Fig~\ref{fig:herakinematics} where we indicate that, for the leptonic vertex, we take the quark--antiquark pair as being inclusive. We focus on the outgoing electron with momentum $k_1$ and the gluon with momentum $q_1$ which couples to the gluon Green's function. Again, we denote the azimuthal angle of $k_1$ as $\alpha_1$ and the azimuthal angle of $q_1$ as $\theta_1$.

It is convenient to invoke the common language of deep inelastic scattering, i.e. denoting the proton momentum as $P$ and the momentum of the virtual photon as $q_\gamma$, we introduce the virtuality $Q=-q_\gamma^2$, the Bjorken scaling variable $x_{\rm Bj}=\frac{Q^2}{2Pq_\gamma}$ and the inelasticity $y=\frac{Pq_\gamma}{P(q_\gamma+k_1)}$. The longitudinal momentum fraction of the forward jet will be denoted as $x_{\rm FJ}$. Using the relation $\kone^2=(1-y)Q^2$ and the specific structure of the leptonic vertex, we can transform the integration over $\kone$ and write
\begin{align}
  \label{eq:oppartonicep}
  \hat\sigma(\shat) =& \frac{\pi^2\asbar^2}{2}\int\ddkone\int\ddktwo\int\frac{d\omega}{2\pi i} \bra{\kone} \oplept\opf_\omega\opjettwo\ket{\ktwo}e^{\omega Y}\\
=& \frac{\pi^2\asbar^2}{2}\sum_{n,n'=-\infty}^\infty\int d\alpha_1\int dy \int\ddktwo\int\frac{d\omega}{2\pi i}\int\ddqone\int\ddqtwo \int  d\nu\int d\nu' \non
 & \bra{y,\alpha_1} \oplept\ket{\qone}\braket{\qone}{\nu,n}\bra{n,\nu}\opf_\omega\ket{\nu',n'}\braket{n',\nu'}{\qtwo}\bra{\qtwo}\opjettwo\ket{\ktwo} e^{\omega Y}\label{eq:oppartonicep2},
\end{align}
where the rapidity is defined as $Y=\ln x_{\rm FJ}/x_{\rm Bj}$.

\subsection{The leptonic vertex and a modified jet vertex}

It has been shown in Ref.~\cite{Bartels:1996wx} (taking into account the different conventions) that the leptonic vertex reads
\begin{align}
& \bra{y,\alpha_1} \oplept\ket{\qone} \non
=& \int dQ^2 \frac{4\alpha^2}{\pi^2N_c\;y\,Q^2}\sum_q e_q^2\int_0^1\int_0^1\frac{ d\xi\, d\zeta}{\xi(1-\xi)Q^2+\zeta(1-\zeta)\qone^2}\non
&\times\Bigg\{\left(\frac{1}{2}-\xi(1-\xi)-\zeta(1-\zeta)+2\xi(1-\xi)\zeta(1-\zeta)\right)y^2\non
&\phantom{\times\Bigg\{}+\Big(1-2\xi(1-\xi)-2\zeta(1-\zeta)+12\xi(1-\xi)\zeta(1-\zeta)\Big)(1-y)\non
&\phantom{\times\Bigg\{}-4\xi(1-\xi)\zeta(1-\zeta)(1-y)\cos\big(2(\theta_1-\alpha_1)\big)\Bigg\}\non
=: & \int dQ^2\left[2a_1^{(0)}(\qone^2,y,Q^2) + 2a_1^{(2)}(\qone^2,y,Q^2)\cos\big(2(\theta_1-\alpha_1)\big)\right],
\end{align}
where $\alpha$ denotes the electromagnetic fine structure constant and $\sum_q e_q^2$ is the sum over the electric charges of the quark of the quark--antiquark pairs which are produced. As a next step we like to compute the projection onto the basis $\ket{\nu,n}$. We have
\begin{align}
&  \int\ddqone \bra{y,\alpha_1} \oplept\ket{\qone}\braket{\qone}{\nu,n}\non
=&\int dQ^2\int\frac{ d\theta_1\,d\qone^2}{2}\left[2a_1^{(0)}(\qone^2,y,Q^2)+2a_1^{(2)}(\qone^2,y,Q^2)\cos\big(2(\theta_1-\alpha_1)\big)\right]\non
&\hphantom{\int dQ^2\frac{ d\theta_1\,d\qone^2}{2}}\times\frac{1}{\pi\sqrt{2}}\plusinu{\qone^2}e^{in\theta_1}\non
=& \int dQ^2\left[2A_1^{(0)}\left(\nu,y,Q^2\right)+A_1^{(2)}\left(\nu,y,Q^2\right)\left(\delta_{n,-2}e^{-2i\alpha_1}+\delta_{n,2}e^{2i\alpha_1}\right)\right],
\end{align}
with
\begin{equation}
  A_1^{(0/2)}\left(\nu,y,Q^2\right) = \frac{1}{\sqrt{2}}\int d\qone^2\; a_1^{(0/2)}(\qone^2,y,Q^2)\plusinu{\qone^2} .
\end{equation}
To calculate these coefficients, we need integrals of the following type:
\begin{multline}
  \int_0^1 d\xi\,(\xi(1-\xi))^{t_\xi}\int_0^1d\zeta\,(\zeta(1-\zeta))^{t_\zeta}\int_0^\infty d\qone^2\frac{\plusinu{\qone^2}}{\xi(1-\xi)Q^2+\zeta(1-\zeta)\qone^2}\\
=\plusinu{Q^2}\frac{\pi}{\cosh (\pi\nu)} B\left(\frac{1}{2}+t_\xi+i\nu,\frac{1}{2}+t_\xi+i\nu\right)\\
\times B\left(\frac{1}{2}+t_\zeta-i\nu,\frac{1}{2}+t_\zeta-i\nu\right),\label{eq:leptimpproj}
\end{multline}
with $B$ the Euler beta function. Using this formula with $t_{\xi/\zeta}\in\{0,1\}$ we obtain
\begin{subequations}
\begin{align}
  A_1^{(0)}\left(\nu,y,Q^2\right) =& \phantom{\times}\frac{\alpha^2\sqrt{2}}{y\,N_c}\plusinu{Q^2}\sum_q e_q^2\frac{1}{16\nu(\nu^2+1)}\frac{\tanh(\pi\nu)}{\cosh(\pi\nu)}\non
&\times\left(\frac{4\nu^2+9}{2}y^2+(12\nu^2+11)(1-y)\right)\label{eq:defA0},\\ 
  A_1^{(2)}\left(\nu,y,Q^2\right) =& \phantom{\times}\frac{\alpha^2\sqrt{2}}{y\,N_c}\plusinu{Q^2}\sum_q e_q^2\frac{1}{16\nu(\nu^2+1)}\frac{\tanh(\pi\nu)}{\cosh(\pi\nu)}\non
&\times\left(-(4\nu^2+1)(1-y)\right)\label{eq:defA2} .
\end{align}
\end{subequations}

Unfortunately,  the angular correlation between the electron and a forward jet has not been measured so far. Nevertheless, to provide theoretical predictions ready for  a comparison with experiment, we try to implement the same kinematical cuts and constraints as the ones that will be used from experimental side. Therefore, we consider a vertex for the forward jet, which is slightly modified, although we could -- in principle -- proceed with the jet vertex of Eq.~\eqref{eq:mnjetvertex} presented in the Mueller-Navelet scenario. 
The ZEUS collaboration (see e.g. Ref.~\cite{Chekanov:2005yb}) and the H1 collaboration (see e.g. Ref.~\cite{Aktas:2005up}) impose  also  an upper cut  for the transverse momentum of the forward jet in terms of the photon virtuality $Q^2$ to ensure that both ends of the gluon ladder have a similar scale:
\begin{subequations}
\begin{align}
 & {\rm ZEUS:}  & \hspace{-2cm}\frac{1}{2}<\frac{\ktwo^2}{Q^2}<2\\
 & {\rm H1:}  & \hspace{-2cm}\frac{1}{2}<\frac{\ktwo^2}{Q^2}<5 .
\end{align}
\end{subequations}
This requirement suppresses DGLAP evolution without effecting the BFKL dynamics. The necessary modification of Eq.\eqref{eq:defc2} is straight forward and yields for the ZEUS condition
\begin{equation}
  c_2(\nu') = \frac{1}{\sqrt{2}}\frac{1}{\frac{1}{2}+i\nu'}\minusinup{\frac{Q^2}{2}}\left[1-\plusinup{\frac{1}{4}}\right]\label{eq:defc2mod}.
\end{equation}
For the H1 condition the $1/4$ has to be replaced be a $1/10$. For simplicity, we keep the ZEUS condition.

\subsection{Phenomenology of forward jets at an $ep$ collider}

Inserting the leptonic vertex of Eq.~\eqref{eq:leptimpproj} in the leptonic cross section, started in Eq.~\eqref{eq:oppartonicep2}, we get by performing the same transformations as in the Mueller-Navelet case
\begin{multline}
  \frac{d\hat\sigma}{dy\,dQ^2\,d\phi} = 
\frac{\pi^2\asbar^2}{2}\int d\nu \int d\nu' \Bigg[A_1^{(0)}\left(\nu,y,Q^2\right)\bra{0,\nu}\opf\ket{\nu',0}c_2(\nu')\\
+A_1^{(2)}\left(\nu,y,Q^2\right)\bra{2,\nu}\opf\ket{\nu',2}c_2(\nu')\cos\,2\phi\Bigg],\label{eq:oppartonicep3}
\end{multline}
where we introduced the azimuthal angle $\phi=\alpha_2-\alpha_1$ between the electron and the forward jet. Furthermore, we made use of the fact that $\bra{n,\nu}\opf\ket{\nu',n}=\bra{-n,\nu}\opf\ket{\nu',-n}$.

We  handle the $\nu$ derivative in $\opf_\omega$ as we did in Eqs.~(\ref{eq:fnloacting} - \ref{eq:fnloacting2}). The required terms are
\begin{subequations}
\label{eq:impcontributiondis}
\begin{align}
  i\frac{\partial}{\partial\nu}\ln\frac{A_1^{(0)}\left(\nu,y,Q^2\right)}{c_2(\nu)} =& -2\ln Q^2-\frac{1}{2\nu^2+\frac{1}{2}}+\frac{3\ln (2)}{5-4\cos(\nu\,\ln 4)}\non
&-i\Bigg[ \pi\frac{\cosh(2\pi\nu)-3}{\sinh(2\pi\nu)}+\frac{4\sin(\nu\,\ln 4)\ln (2)}{5-4\cos(\nu\,\ln 4)} \non
&\hphantom{-i\Bigg[}-8\nu\frac{y^2+6(1-y)}{9y^2+22(1-y)+4(y^2+6(1-y))\nu^2} \non
&\hphantom{-i\Bigg[}  +\frac{3\nu^2+1}{\nu(\nu^2+1)}-\frac{\nu}{\nu^2+\frac{1}{4}} \Bigg] , \\
  i\frac{\partial}{\partial\nu}\ln\frac{A_1^{(2)}\left(\nu,y,Q^2\right)}{c_2(\nu)} =& -2\ln Q^2-\frac{1}{2\nu^2+\frac{1}{2}}+\frac{3\ln (2)}{5-4\cos(\nu\,\ln 4)}\non
&-i\Bigg[\pi\frac{\cosh(2\pi\nu)-3}{\sinh(2\pi\nu)}+\frac{4\sin(\nu\,\ln 4)\ln (2)}{5-4\cos(\nu\,\ln 4)}\non
&\hphantom{-i\Bigg[}+ \frac{3\nu^2+1}{\nu(\nu^2+1)}-\frac{3\nu}{\nu^2+\frac{1}{4}} \Bigg] .
\end{align}
\end{subequations}

Hence we can rewrite Eq.~\eqref{eq:oppartonicep3} as
\begin{equation}
  \frac{d\hat{\sigma}(Y)}{d\phi\;dy\;dQ^2} = \frac{\pi^2\asbar^2}{2}\left[B^{(0)}\left(y,Q^2,Y\right) +B^{(2)}\left(y,Q^2,Y\right) \cos 2\phi\right],
\end{equation}
where the coefficients $B^{(n)}$ are given at LO as
\begin{equation}
  B^{(n)}_{\rm LO}\left(y,Q^2,Y\right)= \int d\nu \;A^{(n)}\left(\nu,y,Q^2\right)\,c_2(\nu) e^{Y\asbar \chi_0\left(\left|n\right|,\nu\right)},
\end{equation}
and at NLO as
\begin{multline}
 B^{(n)}_{\rm NLO}\left(y,Q^2,Y\right)= \int d\nu \;A^{(n)}\left(\nu,y,Q^2\right)\,c_2(\nu)\\
\times e^{{\bar \alpha}_s \left(Q^2\right){\rm Y} \left(\chi_0\left(\left|n\right|,\nu\right)+{\bar \alpha}_s  \left(Q^2\right) \left(\chi_1\left(\left|n\right|,\nu\right)-\frac{\beta_0}{8 N_c} \chi_0\left[2\ln Q^2+i\frac{\partial}{\partial\nu}\ln\frac{A^{(n)}\left(\nu,y,Q^2\right)}{c_2(\nu)}\right]\right)\right)}.
\end{multline}
For the coefficients with resummed kernel we can write accordingly
\begin{equation}
  B^{(n)}_{\rm resum}\left(y,Q^2,Y\right)= \int d\nu \;A^{(n)}\left(\nu,y,Q^2\right)\,c_2(\nu) e^{Y\,\omega^{\rm resum}\left(\left|n\right|,\nu\right)}.
\end{equation}

When we introduced the new modified jet vertex in Eq.~\eqref{eq:defc2mod}, we were considerate of the kinematical cuts applied in experiment for the forward jet. In addition, we use the following cuts motivated by experimental receivables \cite{Didar:private} concerning the leptonic part:
\begin{equation}
\begin{array}{cc}  
20{\rm GeV}^2  < Q^2   < 100{\rm GeV}^2 \, ,   & \hspace{2cm} .05 < y < .7\, , \\
  5\cdot 10^{-3} < x_{\rm Bj} < 4\cdot 10^{-4}\, . & 
\end{array}
\label{eq:heracuts}
\end{equation}
Finally, the cross section at the hadronic level reads
\begin{equation}
  \label{eq:disfinal}
  \frac{d\sigma}{dY\;d\phi} =: C_0(Y)+C_2(Y)\cos 2\phi ,
\end{equation}
with
\begin{equation}
  C_n(Y) = \frac{\pi^2\asbar^2}{2}\int_{\rm cuts} \hspace{-.3cm}dx_{\rm FJ}\,dQ^2\,dy\,f_{\rm eff}(x_{\rm FJ},Q^2) B^{(n)}(y,Q^2,Y) \delta\left(x_{\rm FJ}-\frac{Q^2}{ys}e^Y\right),
\end{equation}
where we performed the convolution with the effective parton distribution from Eq.~\eqref{eq:feff}. By the index at the integral sign, we indicate that the cuts of Eqs.~\eqref{eq:heracuts} are applied. The integration over the longitudinal momentum fraction $x_{\rm FJ}$ of the forward jet involves a delta function fixing the rapidity $Y=\ln x_{\rm FJ}/x_{\rm Bj}$. Let us note that an additional experimental upper cut on $x_{\rm FJ}$ would change the single coefficients $C_n$, while the change of their  ratios  is negligible.

\begin{figure}[htbp]
  \centering
  \includegraphics[width=9cm]{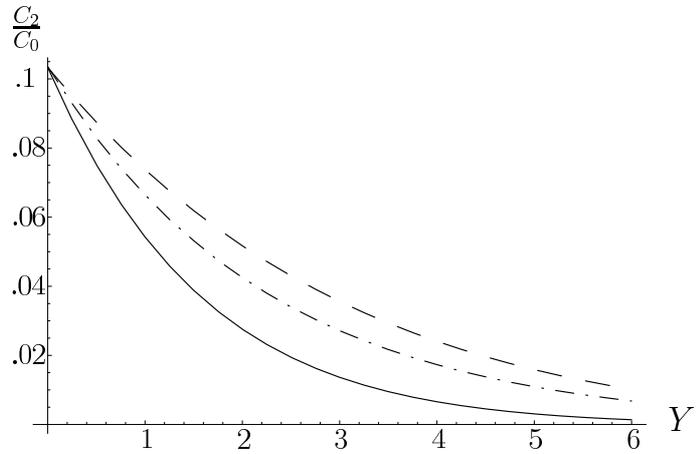}
  \caption{$<\cos 2\phi>$ at the $ep$ collider HERA at leading (solid), next to leading order (dashed), and for resummed kernel (dash-dotted). }
  \label{fig:hera}
\end{figure}

Since the structure of the electron vertex singles out the components with conformal spin 0 and 2, the number of azimuthal observables is limited when compared to the Mueller-Navelet scenario. Nevertheless, we can calculate the rapidity dependence of $<\cos 2\phi> = C_2/C_0$. The result is shown in Fig.~\ref{fig:hera}, and again the NLO calculation predicts a slower decorrelation compared to the LO calculation. Since the convolution with the effective parton density does not give just a global pre-factor, we have to numerically evaluate the kernel several times. For the resummed kernel this requires to solve a transcendental equation which  turns out to be rather time consuming. 
In  the Mueller-Navelet case we recalculated all our results with an approximation to the resummed kernel developed in Ref.~\cite{Vera:2005jt}. The agreement is very good, while the time to perform the numerical evaluation is drastically reduced. In addition, it approximates the kernel with impact factor contribution in the DIS case very well. On that account we present the results for this approximation as being representative for the resummed kernels in Fig.~\ref{fig:hera}. 

\begin{figure}[htbp]
  \centering
  \includegraphics[width=9cm]{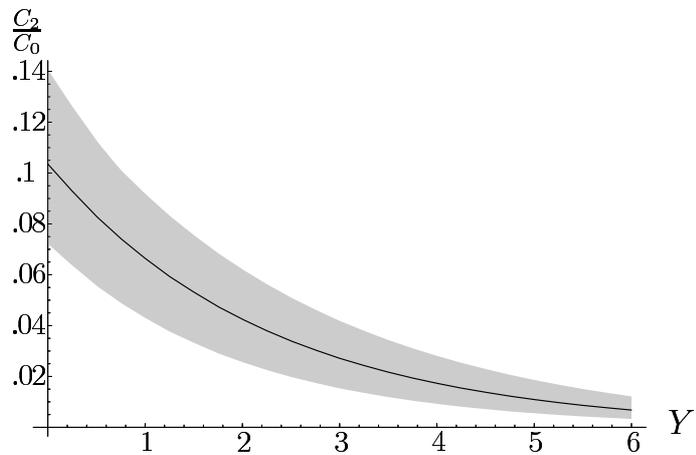}
  \caption{$<\cos 2\phi>$ at the $ep$ collider HERA for resummed kernel (dash-dotted). The gray band reflects the uncertainty in $s_0$ and in the renormalization scale $\mu$.}
  \label{fig:heras0}
\end{figure}

Because the evolution in rapidity is driven by the kernel, the $Y$ dependence of the different results is very similar to the Mueller-Navelet case. This observation gives, a posteriori,  an additional legitimation to consider the jet vertices at LO when studying the rapidity dependence. The additional inclusive quark-antiquark pair -- produced to couple the electron to the gluon evolution -- yields already in the case of no gluon emission some angular decorrelation between the forward jet and the electron.
Following the treatment of the Mueller-Navelet case, we present as our final prediction the calculation based on the resummed kernel with an error band reflecting the uncertainty in $s_0$ and in $\mu$ in Fig.~\ref{fig:heras0}. For this purpose we varied $s_0$ and $\mu$ independently by a factor of $1/2$ and $2$ respectively.

\clearpage{\pagestyle{empty}\cleardoublepage}
\chapter{Summary and Outlook}
\label{sec:summary}

In this thesis we have investigated phenomenological implications of perturbative QCD calculations on jet observables. We did so in the high energy regime, described by the BFKL equation, where our special interest lay in effects of NLO BFKL accuracy on exclusive measurements.

The first part of this work dealt with the production of a jet at central rapidity. Based on a careful study of the different contributions to the NLO BFKL equation and the different energy scales involved, we have derived a NLO jet vertex in $k_T$-factorization.
Therefore, we `opened' the NLO BFKL kernel and introduced a jet definition in a consistent way. We have shown the infrared finiteness of this jet vertex and its 
dependence on the scale $s_\Lambda$, which separates MRK from QMRK. Furthermore, we have examined the connections between the energy scale $s_0$, the Reggeon scales $s_{R;i,j}$, and the final form of the kernel. 

As a central result of this part, we have calculated the jet production vertex \eqref{eq:freeofsing}, and we have explicitly given the necessary subtraction, both at the matrix element level \eqref{eq:differentterms} as well as integrated over the corresponding phase space \eqref{eq:resultbsquare}. Additionally, we have given an alternative form of this subtraction term (\ref{eq:differenttermsalternative}, \ref{eq:resultbsquarenew}). It is not as compact as our original one, but it combines the divergent parts in a way, that it is very convenient if one wants to switch from $\overline{\rm MS}$ renormalization to gluon bremsstrahlung renormalization. The difference between the two subtraction terms, of course, concerns only the finite part of them. Our calculations also suggest that the natural scale for the running of the coupling at the jet vertex is the square of the transverse momentum of the jet \eqref{eq:runningcoupling}.

Our vertex can be used in the context of $\gamma^*\gamma^*$-scattering \eqref{eq:masterformula0} or of hadron-hadron scattering \eqref{ppfinal}. The original derivation of the BFKL equation, at LO and NLO as well, relies on the colliding objects to have a hard and similar scale. The scattering of virtual photons is a natural implementation of this requirement. Hence, for this process the embedding of our jet vertex is straight forward. We have shown that for hadron-hadron scattering special care has to be taken. The asymmetry between the soft scale of the hadrons and the hard scale of the jet induces additional evolution in $k_T$. This fact is reflected by a modified kernel \eqref{newkernel} which governs the evolution of the unintegrated gluon density, and a modification of the jet vertex itself \eqref{newemissionvertex}.  The modification to the evolution kernel can be considered as the first term of a resummation of higher order terms, induced by the scale change. Hence, for the first time, we have given an example of a complete process described in NLO $k_T$-factorization with NLO integrated gluon densities. Nevertheless, we have to state that our expression is only valid for small $x$. 

In our analysis we have been careful to retain the dependence upon the energy scale $s_\Lambda$ which appears at NLO accuracy and separates multi-Regge kinematics from quasi-multi-Regge kinematics. In the NLO calculation of the total cross section, one may be tempted to take the limit $s_\Lambda \to \infty$, thus disregarding the $1/s_\Lambda$ corrections to the NLO BFKL kernel. However, when discussing inclusive (multi-) jet production one has to remember that $s_\Lambda$ has a concrete physical meaning: it denotes the lower cutoff of rapidity gaps and thus directly enters the rapidity distribution of multi-jet final states. In a self-consistent description then also the evolution of the unintegrated gluon density has to depend upon this scale. 

In fact, to study multi-jet production is  a promising application of our jet vertex. The vertex, as we provide it,  is already well suited for the implementation into a Monte-Carlo program. There, momentum conservation can be taken into account, and together with the running of the coupling and a physical choice of  $s_\Lambda$ deeper insight into QCD multi-jet production is expected. Especially, the large available center of mass energy at the forthcoming Large Hadron Collider (LHC) calls for an appropriate description of multi-jet events. The LHC will only tap its full discovery potential if a well understanding of the background is guaranteed. Furthermore, the dependence on $s_\Lambda$ is an inherently interesting topic. Therefore, a numerical study of this feature will be advisable. For this purpose, the calculation of the inclusive jet cross section, as presented in our work,  might act as a theoretical laboratory environment.

Although the NLO unintegrated gluon density might appear as some kind of byproduct of our work, it is a significant result. Further applications of our NLO $k_T$-formalism include $W$ and $Z$ as well as heavy flavor production in the small-$x$ region. Compared to the results presented in this paper, these applications require the calculation of further production vertices; however, for the treatment of the different scales and of the unintegrated gluon density all basic ingredients have been collected in this thesis.

In the second part of this thesis, we studied the angular decorrelation between Mueller-Navelet jets, and between electron and forward jet in deep inelastic scattering (DIS). We have derived an analytical LO master formula (\ref{eq:dsigmadphi}, \ref{eq:cll}), with only one integration left to be performed numerically. Furthermore, we have obtained the corresponding expression including the NLO kernel in its angular dependent form \eqref{eq:cnll2}. In face of the scale dependent part, we have shown, that the kernel still can be exponentiated from the very beginning. In our approach we circumvent to be content with an expansion of the Green's function. As a consequence, a contribution from the impact factors entered the energy dependent exponent. This indicates that the full separation of energy dependence and impact factors does only hold at LO and is violated due to the running of the coupling, spoiling the conformal invariance.

The NLO corrections to the kernel are known to be negative and very large.  It has been shown that it even leads to instabilities which can be cured by a resummation of terms beyond NLO. Therefore, in section~\ref{sec:bfklmellin}, we extended different resummation procedures to arbitrary conformal spin. We then derived the equations describing angular decorrelation also for the resummed kernel \eqref{eq:cresummed}. We explicitly demonstrated the need for this resummation by studying the dependence on renormalization schemes and scales.

We have compared our results for Mueller-Navelet jets with data obtained at the Tevatron. The comparison is given in Figs.~\ref{fig:tevatron}, \ref{fig:tevatrondsigma}. We have reproduced the LO result that overestimates the decorrelation. In contrast, our calculation using the resummed kernels improves the theoretical prediction although it still overestimates the decorrelation. Our study of different renormalization schemes has given additional support to the necessity of resuming the kernel. We strongly suggest to study the different observables, connected to angular decorrelation of Mueller-Navelet jets, at the LHC as well. For that purpose, we give predictions in Fig.~\ref{fig:lhc} where we also have elaborated the theoretical uncertainty by the renormalization scale $\mu$ and the energy scale $s_0$.

Finally we have extended our theoretical work to the scenario of DIS where one can study the azimuthal correlation of the electron and a forward jet. Unfortunately, there are no corresponding measurements. None of the forward jet studies at HERA has studied the angular dependence. We hope that the predictions we give -- incorporating the experimental cuts usually applied in studies of forward jets -- will motivate the experimentalists to investigate this observable.
\clearpage{\pagestyle{empty}\cleardoublepage}

\renewcommand{\sectionmark}[1]{\markright{\thesection.\ #1}{}}
\renewcommand{\chaptermark}[1]{\markboth{Appendix \thechapter .\ #1}{}}

\appendix

\chapter{Alternative subtraction term}
\label{sec:alternative}

In this appendix we present an alternative subtraction term which does not 
make use of the simplifications $A_{(3)}+A_{(4)}\to 2A_{(4)}$ and 
$A_{(5)}+A_{(6)}\to 2A_{(5)}$ which we used in Eqs.~(\ref{eq:agluonssoft}, \ref{eq:agluonssoftcoll2}). These limits are valid in the kinematic regions leading to IR--divergences and hence they do provide the correct $\epsilon$ poles. However, they also alter the finite terms. Here we want to study also this finite 
part as accurately as possible and hence we do not take these limits  
but use instead the complete sum
\begin{equation}
  \label{eq:alternativebase}
  A_{(1)}+A_{(2)}+A_{(3)}+A_{(4)}+A_{(5)}+A_{(6)}\;+\;A_{\rm MRK} 
\end{equation}
as the gluonic subtraction term.

The full gluonic matrix element written in Eq.~\eqref{eq:Agluons} contains 
spurious UV--divergences which are canceled when combined with the MRK contribution. One fourth of the MRK contribution cancels the UV--divergence of $A_{(4)}$ while another fourth cancels that of $A_{(6)}$. The remaining half cancels the UV--divergence of two terms present in Eq.~\eqref{eq:Agluons}:
\begin{align}
 A_{(7)}\equiv& -\frac{\qa^2\qb^2}{4}\left(\frac{1-x}{x}\frac{1}{\ktwo^2\that}+\frac{x}{1-x}\frac{1}{\kone^2\uhat}\right)\\
 A_{(8)} \equiv& \frac{\qa^2\qb^2}{4\Sigma}\left(\frac{1-x}{x}\frac{1}{\ktwo^2}+\frac{x}{1-x}\frac{1}{\kone^2}\right),
\end{align}
which are IR--finite and hence so far not included in the subtraction term.

By doubling $A_{(4)}$ and $A_{(5)}$ in the subtraction term constructed in Eq.~\eqref{eq:differentterms} also their spurious UV--divergences are doubled and thus completely canceled by the MRK contribution. But Eq.~\eqref{eq:alternativebase} so far only contains half of the spurious UV--divergences of the full matrix element in such a way that half of the MRK contribution is not compensated. Therefore a subtraction term based on Eq.~\eqref{eq:alternativebase} which is also free from spurious UV--divergences should also include $A_{(7)}$ and $A_{(8)}$ and reads
\begin{multline}
  \widetilde{A}_{\rm gluons}^{\rm singular} = A_{(1)}+A_{(2)}+A_{(3)}+A_{(4)}+A_{(5)}+A_{(6)}+A_{\rm MRK}+A_{(7)}+A_{(8)} \\
= A_{(1)}+A_{(2)}+A_{(3)}+\big(A_{(5)}-A_{(4)}\big)+A_{(6)}+\frac{A_{\rm MRK}}{2}+A_{(7)}+A_{(8)}\label{eq:differenttermsalternative}.
\end{multline}
If we now define $\mathcal{S}_{(3,6,7,8)}$ and $\mathcal{S}_{\rm MRK}$ as we 
did in Eq.~\eqref{eq:allJs} we get a new integrated subtraction term from 
the previous Eq.~\eqref{eq:resultbsquare} by replacing
\begin{equation}
\mathcal{S}_{III}+\mathcal{S}_{IV} = \frac{1}{\delt^2}\frac{\pi\Gamma(1-\epsilon)}{(4\pi)^\epsilon}\left(\frac{\delt^2}{\mu^2}\right)^\epsilon\left[\frac{1}{\epsilon^2}-\frac{5\pi^2}{6}+\order{\epsilon}\right]
\end{equation}
with
\begin{equation}
 \frac{1}{2}\left(\mathcal{S}_{III}+\mathcal{S}_{IV}\right)+\mathcal{S}_{(3)}+\mathcal{S}_{(6)}+\frac{\mathcal{S}_{\rm MRK}}{2}+\mathcal{S}_{(7)}+\mathcal{S}_{(8)} .
\end{equation}
The results for $\mathcal{S}_{(3)}$ and $\mathcal{S}_{(6)}$ can be easily obtained from Eqs.~(C.43) and (C.40) of Ref.~\cite{Kotsky:1998ug}:
\begin{align}
  \mathcal{S}_{(3)} =\;& \frac{1}{\delt^2}\frac{\pi\Gamma(1-\epsilon)}{(4\pi)^{\epsilon}}\left(\frac{\delt^2}{\mu^2}\right)^\epsilon\Bigg[\frac{1}{2\epsilon^2}+\frac{1}{2\epsilon}\ln\frac{\qa^2\qb^2}{\delt^4}-\frac{\pi^2}{12}+\frac{1}{4}\ln^2\frac{\qa^2}{\qb^2}\non
&+\frac{\qa^2\qb^2\big(\delt(\qa-\qb)\big)}{\delt^2(\qa-\qb)^2}\Bigg\{\frac{1}{2}\ln\left(\frac{\qa^2}{\qb^2}\right)\ln\left(\frac{\qa^2\qb^2\delt^4}{(\qa^2+\qb^2)^4}\right)\non
&-\text{Li}_2\left(-\frac{\qa^2}{\qb^2}\right)+\text{Li}_2\left(-\frac{\qb^2}{\qa^2}\right)\Bigg\}
-\frac{\qa^2\qb^2}{2}\left(1-\frac{\big(\delt(\qa-\qb)\big)^2}{\delt^2(\qa-\qb)^2}\right)\non
&\times\left(\int_0^1-\int_1^\infty\right)dz\frac{\ln\left(\frac{(z\qa)^2}{\qb^2}\right)}{(\qb+z \qa)^2}+\order{\epsilon}\Bigg],\\
\mathcal{S}_{(6)} =\;& \frac{1}{\delt^2}\frac{\pi\Gamma(1-\epsilon)}{(4\pi)^{\epsilon}}\left(\frac{\delt^2}{\mu^2}\right)^\epsilon\left[\frac{1}{\epsilon^2}-\frac{\pi^2}{6}+\order{\epsilon}\right].
\end{align}
Due to the UV--singularity of $A_{\rm MRK}$ we regularize the $x$ integration by a cutoff $\delta$ to obtain
\begin{align}
  \mathcal{S}_{\rm MRK} =&   -\int_\delta^{1-\delta} \frac{dx}{x(1-x)}\int\frac{\dktwo}{\mu^{2\epsilon}(2\pi)^{D-4}}  \frac{1}{\ktwo^2(\delt-\ktwo)^2}\non
=&\frac{1}{\delt^2}\frac{\pi\Gamma(1-\epsilon)}{(4\pi)^{\epsilon}}\left(\frac{\delt^2}{\mu^2}\right)^\epsilon\frac{\Gamma^2(\epsilon)}{\Gamma(2\epsilon)}2\ln\frac{\delta}{1-\delta}.
\end{align}
Making use of $2\qa\kone-\qa^2=\that+\kone^2/x$ we can decompose Eq.~(C.41) of Ref.~\cite{Kotsky:1998ug} into one integration very similar to that of $\mathcal{S}_{\rm MRK}$ and another one which can be transformed to give $\mathcal{S}_{(7)}$.
\begin{align}
\mathcal{S}_{(7)} =&  \frac{1}{\delt^2}\frac{\pi\Gamma(1-\epsilon)}{(4\pi)^{\epsilon}}\left(\frac{\delt^2}{\mu^2}\right)^\epsilon\Bigg[-\frac{1}{2}\frac{\Gamma^2(\epsilon)}{\Gamma(2\epsilon)}\ln\frac{\delta}{1-\delta}-\frac{1}{2\epsilon^2}-\frac{1}{2\epsilon}\ln\frac{\qa^2\qb^2}{\delt^4}\non
&\hphantom{\frac{1}{\delt^2}\frac{\pi\Gamma(1-\epsilon)}{(4\pi)^{\epsilon}}\left(\frac{\delt^2}{\mu^2}\right)^\epsilon\Bigg[}-\frac{1}{4}\ln^2\frac{\qa^2}{\qb^2}+\frac{\pi^2}{12}+\order{\epsilon}\Bigg].\\
\intertext{The two parts forming $A_{(8)}$ can be obtained from each other by 
the exchange $k_1\leftrightarrow k_2$ and we only need to double the 
calculation of one:}
   \mathcal{S}_{(8)} =&  2\int_\delta^{1-\delta} \frac{dx}{x(1-x)} \int\frac{\dkone}{\mu^{2\epsilon}(2\pi)^{D-4}}\frac{x}{4(1-x)}\frac{1}{\Sigma\kone^2}\non
=&2\int_\delta^{1-\delta} \frac{dx}{x(1-x)} \int\frac{\dkone}{\mu^{2\epsilon}(2\pi)^{D-4}}\non
&\times\frac{1}{4}\int_0^1 d\xi \frac{x^2}{\Big\{[\kone-\xi x\delt]^2+\xi(1-\xi)x^2\delt^2+\xi x(1-x)\delt^2\Big\}^2}\non
=&\frac{1}{2} \frac{\pi\Gamma(1-\epsilon)}{(4\pi)^\epsilon}\frac{1}{\delt^2}\left(\frac{\delt^2}{\mu^2}\right)^\epsilon\int_\delta^{1-\delta} dx\frac{1}{1-x}B_{x}(\epsilon,\epsilon)\non
 =& \frac{1}{\delt^2}\frac{\pi\Gamma(1-\epsilon)}{(4\pi)^{\epsilon}}\left(\frac{\delt^2}{\mu^2}\right)^\epsilon\left[-\frac{1}{2}\frac{\Gamma^2(\epsilon)}{\Gamma(2\epsilon)}\ln\delta-\frac{1}{2\epsilon^2}-\frac{\pi^2}{12}+\order{\epsilon}\right] .
\end{align}
When we add up these new contributions, the spurious UV--divergences indeed cancel and we can safely take the $\delta\to 0$ limit. Furthermore, the new subtraction term has the same pole structure and differs only in the finite parts when compared to that in Eq.~\eqref{eq:differentterms} and its integrated form in Eq.~\eqref{eq:resultbsquare}. To complete the calculation we combine it with the 
corresponding unmodified quark part and obtain
\begin{multline}
\int\dktwo\int dy_2 \btssquare{\qa}{\qb}{\kjet-\ktwo}{\ktwo} =\\
\frac{\bar{g}_\mu^4 \mu^{-2\epsilon}}{\pi^{1+\epsilon}\Gamma(1-\epsilon)}\frac{4}{\kjet^2}\left(\frac{\kjet^2}{\mu^2}\right)^\epsilon 
\Bigg\{\frac{2}{\epsilon^2}-\frac{\beta_0}{N_c}\frac{1}{\epsilon}+\frac{67}{9}-\frac{10n_f}{9N_c}-\frac{4\pi^2}{3}\\
+\frac{2\qa^2\qb^2\big(\delt(\qa-\qb)\big)}{\delt^2(\qa-\qb)^2}\Bigg[\frac{1}{2}\ln\left(\frac{\qa^2}{\qb^2}\right)\ln\left(\frac{\qa^2\qb^2\delt^4}{(\qa^2+\qb^2)^4}\right)\\
-\text{Li}_2\left(-\frac{\qa^2}{\qb^2}\right)+\text{Li}_2\left(-\frac{\qb^2}{\qa^2}\right)\Bigg]\\
-\qa^2\qb^2\left(1-\frac{\big(\delt(\qa-\qb)\big)^2}{\delt^2(\qa-\qb)^2}\right)
\left(\int_0^1-\int_1^\infty\right)dz\frac{\ln\left(\frac{(z\qa)^2}{\qb^2}\right)}{(\qb+z \qa)^2}+\order{\epsilon}\Bigg\}
\label{eq:resultbsquarenew}.
\end{multline}

\clearpage{\pagestyle{empty}\cleardoublepage}

\chapter[Resummation]{Resummation of sub-leading corrections to the BFKL kernel}
\label{sec:resummation}

In this appendix we want to elaborate the different schemes of resummation of formally sub-leading corrections to the BFKL kernel, as it is introduced in section ~\ref{sec:bfklresummation} and used in chapter \ref{sec:decorrelation}.

The first section describes different concrete schemes to implement the resummation procedure. We then briefly show how a different renormalization scheme effects the resummation. We have shown in chapter \ref{sec:decorrelation} how the running of the coupling effectively leads to an contribution of the impact factors to the kernel (see Eqs.~(\ref{eq:actionofnloktoto}-\ref{eq:fnloacting2})). In section \ref{sec:impfactor} we explain how the resummation changes if this contribution is included. In the last section of this appendix we discuss some consequences of the resummation in detail.

\section{Different schemes of implementations}

\subsection{Scheme 1}

For scheme 1 we choose $D_k(\gamma)$ to be
\begin{equation}
  \label{eq:mschemeoneD}
  D_k(\gamma) = \frac{(-1)^{k-1}}{(k-1)!}\frac{d^{k-1}}{d\gamma^{k-1}}[\psi(1)-\psi(\gamma)].
\end{equation}
From Eq.~\eqref{eq:mchinode} we get
\begin{equation}
  \label{eq:mchinodeschemeone}
  \chi^{(0)}(m,\gamma) = 2\psi(1)-\psi\left(\gamma+\frac{m}{2}+\frac{\omega}{2}\right)-\psi\left(1-\gamma+\frac{m}{2}+\frac{\omega}{2}\right).
\end{equation}
By an $\omega$-expansion of Eq.~\eqref{eq:mchinodeschemeone} we obtain
\begin{equation}
  \chi^{(0)}_1(m,\gamma) = -\frac{1}{2}\chi_0(m,\gamma)\left[\psi'\left(\gamma+\frac{m}{2}\right)+\psi'\left(1-\gamma+\frac{m}{2}\right)\right] .
\end{equation}
From Eq.~\eqref{eq:mchiNexpanded} we calculate  $d^{(0)}_{n,k}(m) =\bar{d}^{(0)}_{n,k}(m) $ and
\begin{subequations}
\begin{align}
  d^{(0)}_{1,1}(m) =& -\Psi'(1+m), \\
  d^{(0)}_{1,2}(m) =& \phantom{-}\frac{1}{2}\Big(\Psi(1+m)-\Psi(1)\Big) , \\
  d^{(0)}_{1,3}(m) =& -\frac{1}{2} .
\end{align}
\end{subequations}
We can now use master formula \eqref{eq:mchiN} to write
\begin{multline}
\label{eq:mmasterschemeone}
    \chi^{(1)}(m,\gamma) = \chi^{(0)}(m,\gamma)+\asbar\left(\chi_1(m,\gamma)-\chi^{(0)}_1(m,\gamma)\right)\\
+\asbar\sum_{k=1}^{2}\left(d_{1,k}(m)-d^{(0)}_{1,k}(m)\right)\Bigg[D_k\left(\gamma+\frac{m}{2}+\frac{\omega}{2}\right)-D_k\left(\gamma+\frac{m}{2}\right)\\
+D_k\left(1-\gamma+\frac{m}{2}+\frac{\omega}{2}\right)-D_k\left(1-\gamma+\frac{m}{2}\right)\Bigg].
\end{multline}
The sum in Eq.~\eqref{eq:mmasterschemeone} can be truncated after $k=2$ since $d_{1,3}(m)=d^{(0)}_{1,3}(m)$.

\subsection{Scheme 2}

For scheme 2 we choose a simpler $D_k$ namely
\begin{equation}
  D_k(\gamma)=\frac{1}{\gamma^k}.
\end{equation}
From Eq.~\eqref{eq:mchinode} we get
\begin{multline}
  \label{eq:mchinodeschemetwo}
 \chi^{(0)}(m,\gamma) = \chi_0(m,\gamma)-\frac{1}{\gamma+\frac{m}{2}}-\frac{1}{1-\gamma+\frac{m}{2}}\\
+\frac{1}{\gamma+\frac{m}{2}+\frac{\omega}{2}}+\frac{1}{1-\gamma+\frac{m}{2}+\frac{\omega}{2}},
\end{multline}
and 
\begin{equation}
 \chi^{(0)}_1(m,\gamma) = -\frac{\chi_0(m,\gamma)}{2}\left(\frac{1}{\left(\gamma+\frac{m}{2}\right)^2}+\frac{1}{\left(1-\gamma+\frac{m}{2}\right)^2}\right) .
\end{equation}
\begin{subequations}
Also for this choice we have $d^{(0)}_{n,k}(m) =\bar{d}^{(0)}_{n,k}(m) $ which explicitly read
\begin{align}
  d^{(0)}_{1,1} =& -\frac{1}{2}\left[\Psi'(1+m)-\Psi'(1)+\frac{1}{(1+m)^2}\right], \\
  d^{(0)}_{1,2} =& \phantom{-} \frac{1}{2}\Big(\Psi(1+m)-\Psi(1)\Big), \\
  d^{(0)}_{1,3} =& -\frac{1}{2} .
\end{align}
\end{subequations}

\subsection{Scheme 3} 

Scheme 3 is based on scheme 1. It is then modified to hold $d^{(0)}_{1,k}(m)=d_{1,k}(m)$ not just for $k=3$, but also for $k=1,2$. To do so, we modify Eq.~\eqref{eq:mchinodeschemeone} by two terms of higher order labeled $A(m)$ and $B(m)$:
\begin{align}
  \label{eq:mchinodeschemethree}
  \chi^{(0)}(m,\gamma) =& \big(1-\asbar A(m)\big)\Bigg[2\psi(1)-\psi\left(\gamma+\frac{m}{2}+\frac{\omega}{2}+\asbar B(m)\right)\non
&\quad-\psi\left(1-\gamma+\frac{m}{2}+\frac{\omega}{2}+\asbar B(m)\right)\Bigg].\\
\intertext{From $\omega$-expansion we get:}
 \chi^{(0)}_1(m,\gamma) =& -\left(B(m)+\frac{1}{2}\chi_0(m,\gamma)\right)\left[\psi'\left(\gamma+\frac{m}{2}\right)+\psi'\left(1-\gamma+\frac{m}{2}\right)\right]\non
&-A(m)\chi_0(m,\gamma) .
\end{align}
\begin{subequations}
The coefficients $d^{(0)}$ now depend on these newly introduced variables $A$ and $B$ in the following way
\begin{align}
  d^{(0)}_{1,1}(m) =& -A(m) -\Psi'(1+m),  \\
  d^{(0)}_{1,2}(m) =& -B(m) +\frac{1}{2}\Big(\Psi(1+m)-\Psi(1)\Big) ,  \\
  d^{(0)}_{1,3}(m) =& -\frac{1}{2} .
\end{align}
\end{subequations}
The  requirements $d^{(0)}_{1,1}(m)\stackrel{!}{=} d_{1,1}(m)$ and $d^{(0)}_{1,1}(m)\stackrel{!}{=} d_{1,1}(m)$ can be satisfied by an appropriate choice for $A(m)$ and $B(m)$:
\begin{subequations}
\begin{align}
A(m) =& -d_{1,1}(m)-\Psi'(1+m) \\
B(m) =& -d_{1,2}(m)+\frac{1}{2}\Big(\Psi(1+m)-\Psi(1)\Big).
\end{align}
\end{subequations}

Due to that extra requirement the master formula simply reads
\begin{equation}
\label{eq:mmasterschemethree}
    \chi^{(1)}(m,\gamma) = \chi^{(0)}(m,\gamma)+\asbar\left(\chi_1(m,\gamma)-\chi^{(0)}_1(m,\gamma)\right) .
\end{equation}

\subsection{Scheme 4}

Scheme 4 uses the same functions $D_k$ as scheme 2, but is modified similar to scheme 3 to hold $d^{(0)}_{1,k}(m)=d_{1,k}(m)$ not just for $k=3$, but also for $k=1,2$:
\begin{multline}
   \label{eq:mchinodeschemefour}
\chi^{(0)}(m,\gamma) =\quad\chi_0(m,\gamma)-\frac{1}{\gamma+\frac{m}{2}}-\frac{1}{1-\gamma+\frac{m}{2}}+\big(1-\asbar A(m)\big)\\
 \times\left[\frac{1}{\gamma+\frac{m}{2}+\frac{\omega}{2}+\asbar B(m)}+\frac{1}{1-\gamma+\frac{m}{2}+\frac{\omega}{2}+\asbar B(m)}\right] .
\end{multline}
From $\omega$-expansion we obtain:
\begin{multline}
\chi^{(0)}_1(m,\gamma) = -\frac{(1+m)A(m)}{\left(\gamma+\frac{m}{2}\right)\left(1-\gamma+\frac{m}{2}\right)} \\
 -\left(B(m)+\frac{1}{2}\chi_0(m,\gamma)\right)\left[\frac{1}{\left(\gamma+\frac{m}{2}\right)^2}+\frac{1}{\left(1-\gamma+\frac{m}{2}\right)^2}\right].
\end{multline}

The result of calculating the coefficients $d^{(0)}$ is
\begin{subequations}
\begin{align}
  d^{(0)}_{1,1}(m) =&  -A(m)-\frac{1}{2}\left[\Psi'(1+m)-\Psi'(1)+\frac{1}{(1+m)^2}\right]\\
  d^{(0)}_{1,2}(m) =& - B(m)  +\frac{1}{2}\Big(\Psi(1+m)-\Psi(1)\Big)\\
  d^{(0)}_{1,3}(m) =& -\frac{1}{2}
\end{align}
\end{subequations}
The  requirements $d^{(0)}_{1,1}(m)\stackrel{!}{=} d_{1,1}(m)$ and $d^{(0)}_{1,1}(m)\stackrel{!}{=} d_{1,1}(m)$ can be satisfied by an appropriate choice for $A(m)$ and $B(m)$:
\begin{subequations}
\begin{align}
 A(m) =& -d_{1,1}(m) -\frac{1}{2}\left[\Psi'(1+m)-\Psi'(1)+\frac{1}{(1+m)^2}\right],\\
 B(m) =& -d_{1,2}(m) +\frac{1}{2}\Big(\Psi(1+m)-\Psi(1)\Big) .
\end{align}
\end{subequations}
The resulting master formula equals that of scheme 3 given in Eq.~\eqref{eq:mmasterschemethree}, but with the different $\chi^{(0)}(m,\gamma)$, $\chi^{(0)}_1(m,\gamma)$.

\section{Renormalization schemes}

To perform the renormalization in the gluon-bremsstrahlung (GB) scheme instead of the $\overline{\rm MS}$ scheme is straight forward. The redefinition of the Landau pole
\begin{equation}
  \Lambda_{\overline{\rm MS}} \quad\rightarrow\quad
  \Lambda_{\rm GB} = \Lambda_{\overline{\rm MS}}\; e^{\frac{2N_c}{\beta_0}\mathcal{S}}
\end{equation}
is accompanied by canceling the $\mathcal{S}$-term in $\chi_1(m,\gamma)$ and, as a consequence, in $d_{1,1}(m)$ as well.

\section{Impact factor contribution}
\label{sec:impfactor}

Including the impact factor contribution of Eq.~\eqref{eq:impcontributionmn} to the kernel in the Mueller-Navelet case effectively  changes $\chi_1$:
\begin{equation}
  \chi_1(m,\gamma) \quad\rightarrow\quad   \chi_1(m,\gamma) - \frac{\beta_0}{8N_c}\frac{\chi_0(m,\gamma)}{\gamma(1-\gamma)} , 
\end{equation}
leading to a change in $d_{1,k}(m)$:
\begin{subequations}
\begin{align}
d_{1,1}(m)\quad\rightarrow\quad &d_{1,1}(m) +\frac{\beta_0}{2N_c}\left(\frac{1-\delta_{m,0}}{m(m+2)}-\frac{\delta_{m,0}}{4}\right) \\
d_{1,2}(m)\quad\rightarrow\quad &d_{1,2}(m) -\frac{\beta_0}{8N_c}\delta_{m,0} .
\end{align}
\end{subequations}

In the DIS case the the additional contribution of Eq.~\eqref{eq:impcontributiondis} effectively modifies $\chi_1$ and hence the coefficients $d_{1,k}(m)$:
\begin{subequations}
\begin{align}
d_{1,1}(0)\quad\rightarrow\quad &d_{1,1}(0) -\frac{\beta_0}{8N_c}\left(\frac{7}{6}+\frac{1-y}{y\left(\frac{y}{2}-1\right)+1}\right) \\
d_{1,1}(2)\quad\rightarrow\quad &d_{1,1}(2) -\frac{\beta_0}{8N_c}\left(\frac{107}{30}+\frac{5\ln 2}{3}\right) \\
d_{1,2}(m)\quad\rightarrow\quad &d_{1,2}(m) +\frac{\beta_0}{4N_c} .
\end{align}
\end{subequations}

\section{Discussion}

\begin{figure}[htbp]
  \centering
\includegraphics[width=8cm]{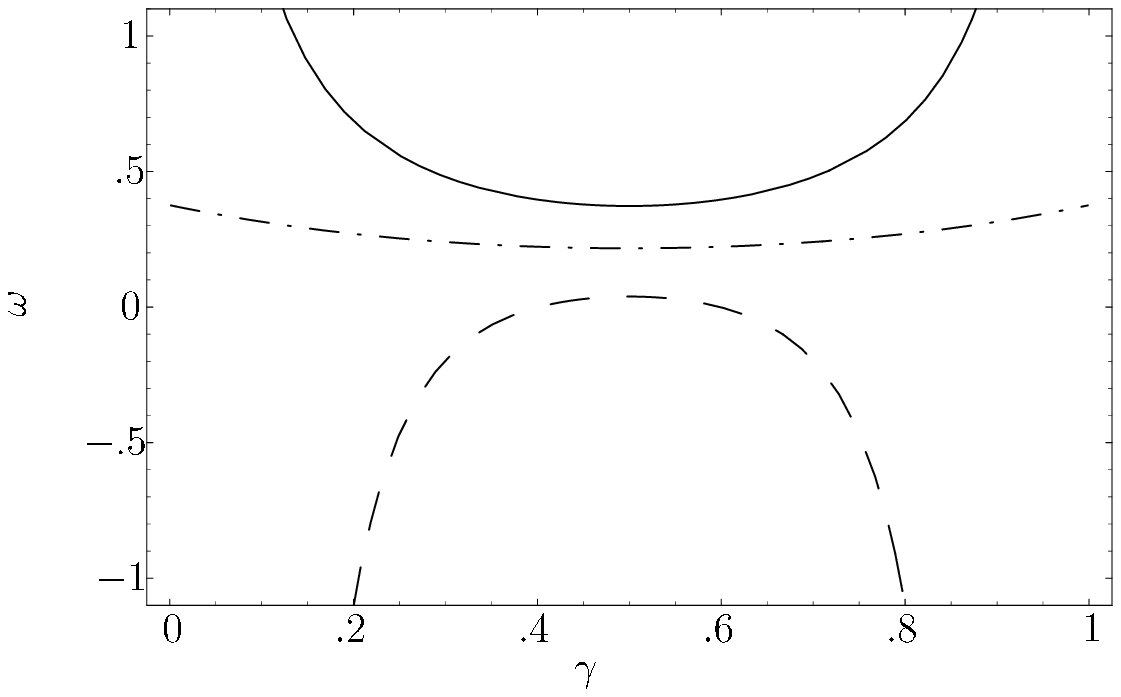}\\
\includegraphics[width=8cm]{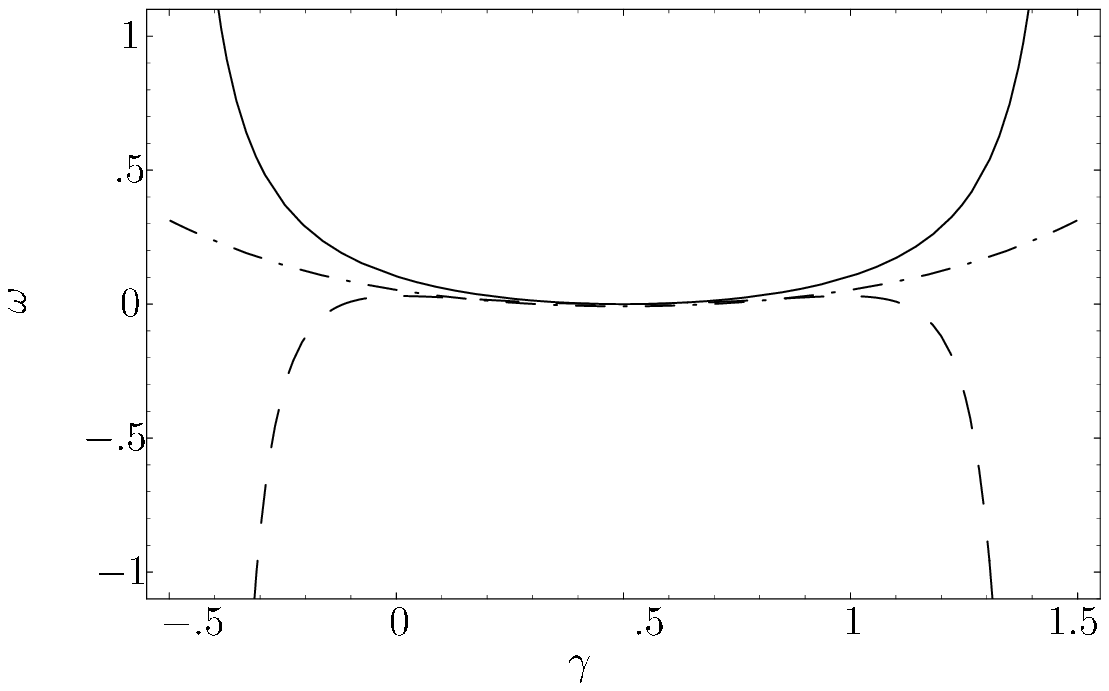}\\
\includegraphics[width=8cm]{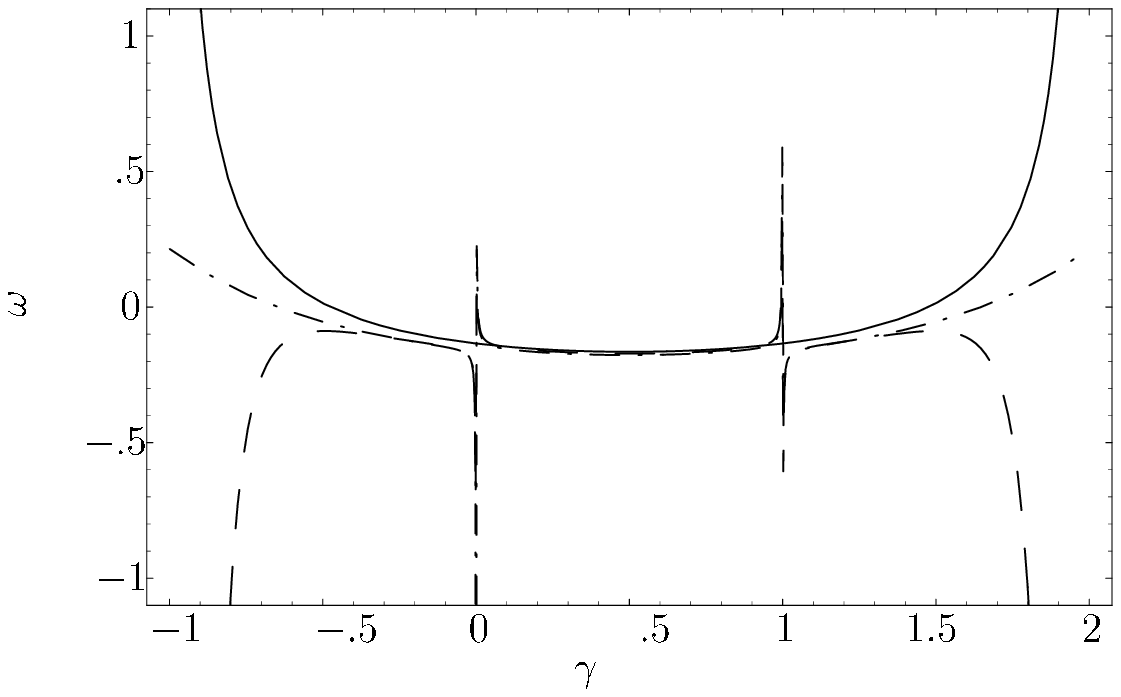}  
  \caption{Comparison of kernels $\omega(m=0,\gamma)$ at a scale $\mu=30{\rm GeV}$ as a function of $\gamma$. Shown are the LO expression $\asbar\chi_0$ (solid line), NLO expression $\asbar\chi_0+\asbar^2\chi_1$ (dashed line), and the resummed one for scheme 3 (dashed-dotted line).} 
\label{fig:gamma}
\end{figure}

In Fig.~\ref{fig:gamma} we show the kernel for conformal spin $m=0$ to $m=2$ in dependence on $\gamma$. An important common feature is, that the the LO kernel has simple poles at $\gamma=-m/2$ and $\gamma=1+m/2$, and the NLO kernel simple, double, cubic poles at these points. After the resummation procedure these poles are removed. For conformal spin $m=2$, we observe simple poles at $\gamma=0$ and $\gamma=1$ for the NLO and resummed kernel, originating from the term $\sim\frac{\gamma(1-\gamma)}{\sin^2(\pi\gamma)}\delta_{m,2}$ in the NLO kernel (see Eq.~\eqref{eq:nlokernel}). Similar poles arise for all conformal spins $m$ at $\gamma=0$ and $\gamma=1$ when the impact factor contribution, shown in the previous section, is included. These are not resummed as well. The large difference between LO and NLO kernel around $\gamma=1/2$ is specific to $m=0$. Therefore, also the resummed does not deviate visibly from LO and NLO kernel around $\gamma=1/2$ for $m>0$. This indicates that a resummation of the poles at $\gamma=0$ and $\gamma=1$ for higher conformal spin is not imperative.

\begin{figure}
  \centering
  \includegraphics[width=5.5cm]{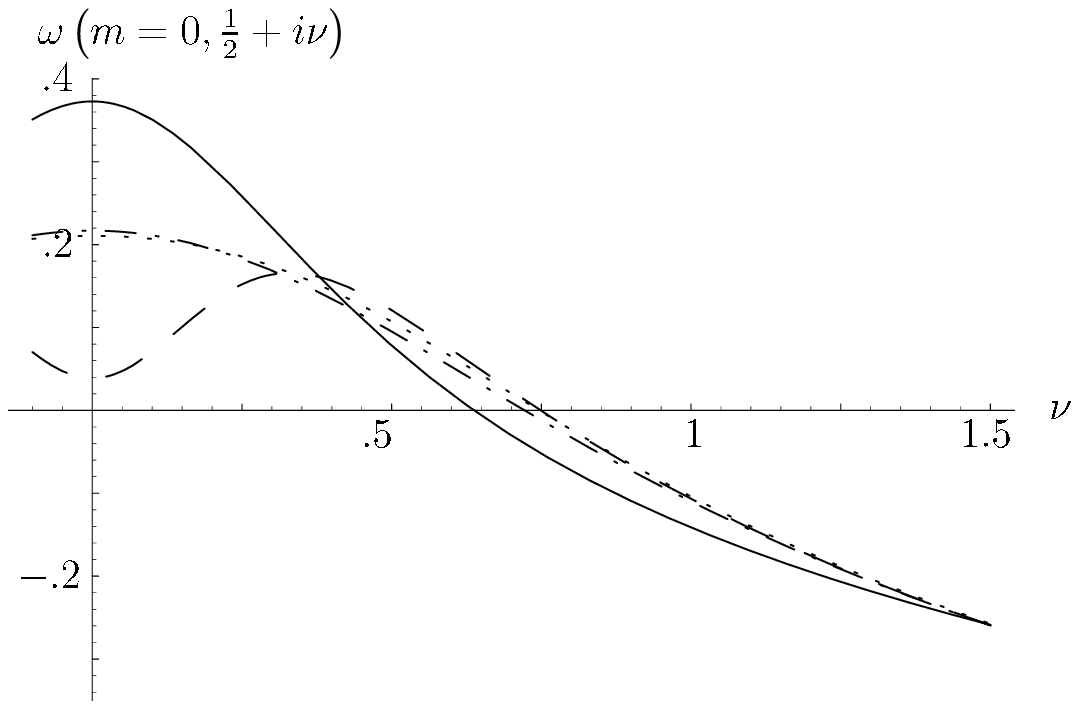}\hspace{1cm}
  \includegraphics[width=5.5cm]{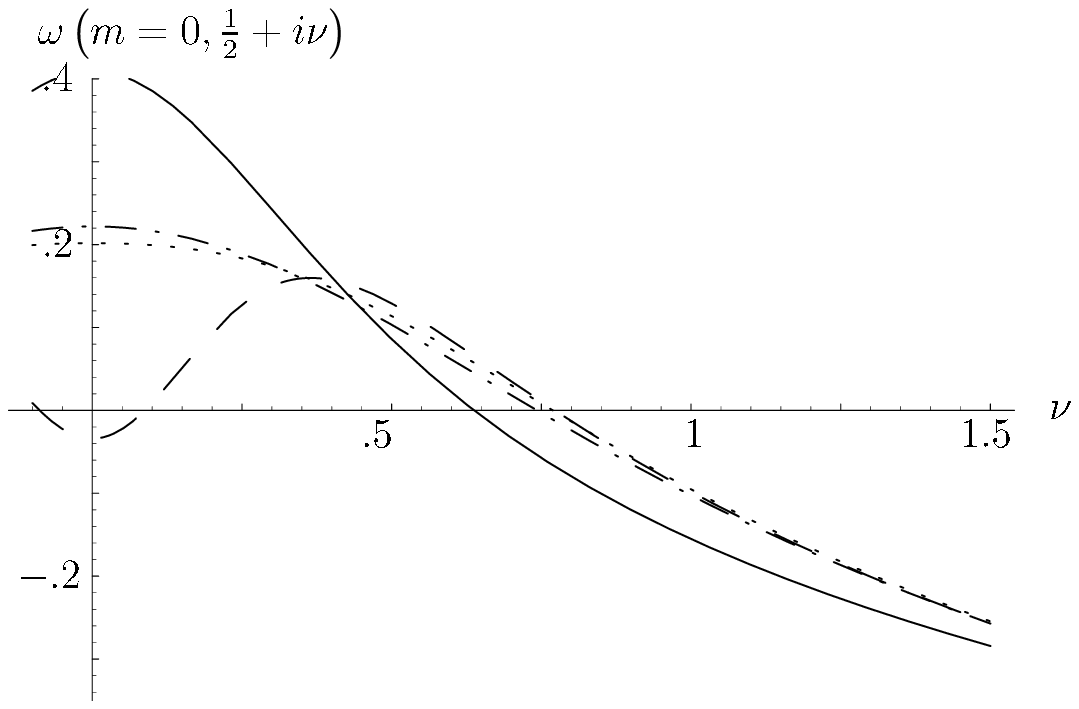}\\
  \vspace{1cm}
  \includegraphics[width=5.5cm]{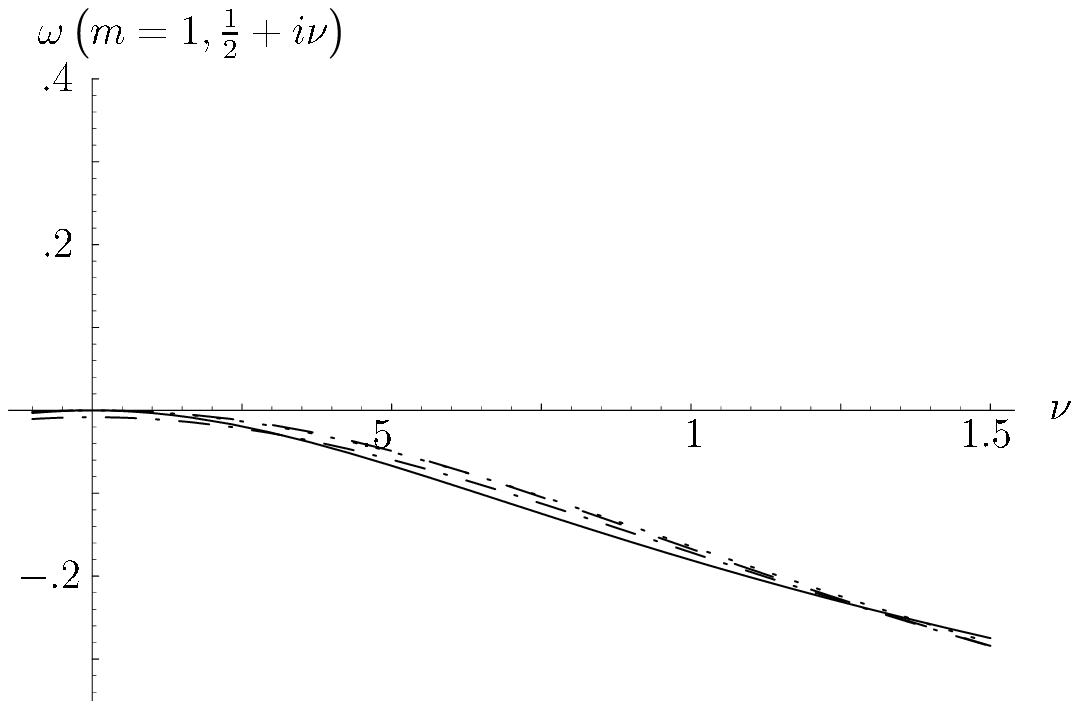}\hspace{1cm}
  \includegraphics[width=5.5cm]{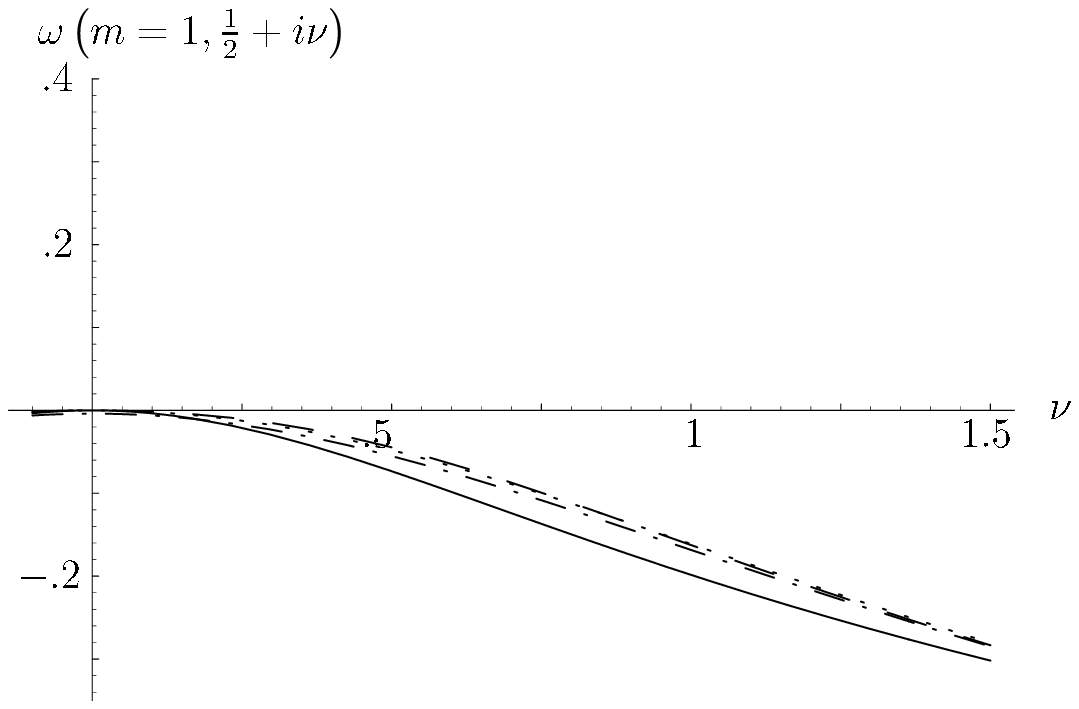}\\
  \vspace{1cm}
  \includegraphics[width=5.5cm]{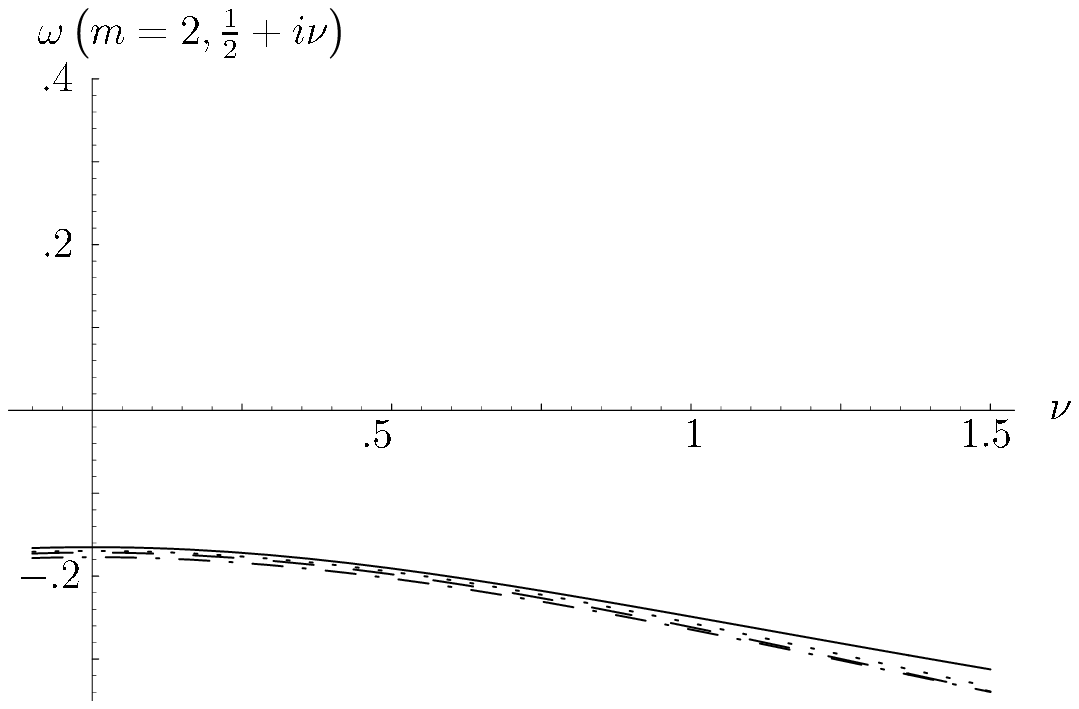}\hspace{1cm}
  \includegraphics[width=5.5cm]{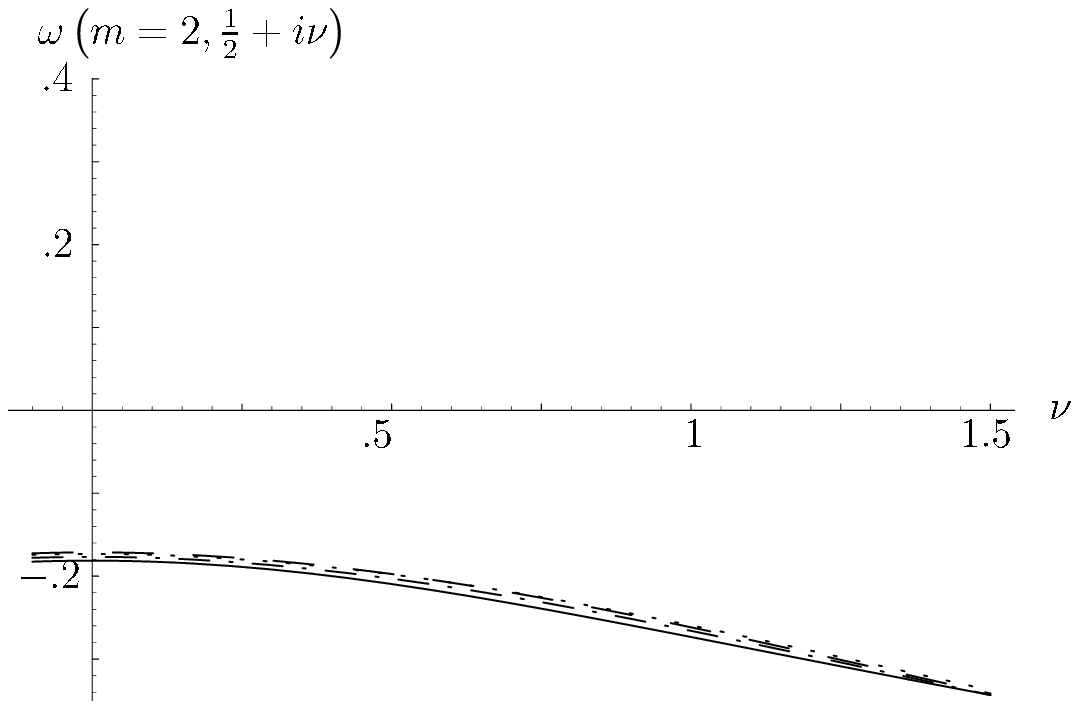}
  \caption{Comparison of kernels $\omega(m=0,\frac{1}{2}+i\nu)$ at a scale $\mu=30{\rm GeV}$ as a function of $\nu$. The left diagram shows the results in $\overline{\rm MS}$-renormalization and the right one the results in GB-renormalization.  Shown are the LO expression (solid line), NLO expression (dashed line), and the resummed ones for scheme 1 (dotted line) and scheme 3 (dashed-dotted line). }
  \label{fig:purem2mu30}
\end{figure}

In Fig.~\ref{fig:purem2mu30} we show the kernel for conformal spin $m=0$. The completely different shape around $\nu=0$ for the LO and NLO version of the kernel can be clearly seen. Since the results of the resummation scheme 2 (4) is hardly distinguishable from scheme 1 (3), we focus on schemes 1 and 3 for  the plots. In deed, the resummation procedure leads to  one maximum at $\nu=0$ and agrees with the NLO results for larger $|\nu|$. A second striking feature of the resummed kernel is its stability against a change of the renormalization scheme. We have checked that for larger conformal spin LO, NLO and resummed kernels behave in the same way. For this reason  we do not present the according plots.

\begin{figure}
  \centering
  \includegraphics[width=5.5cm]{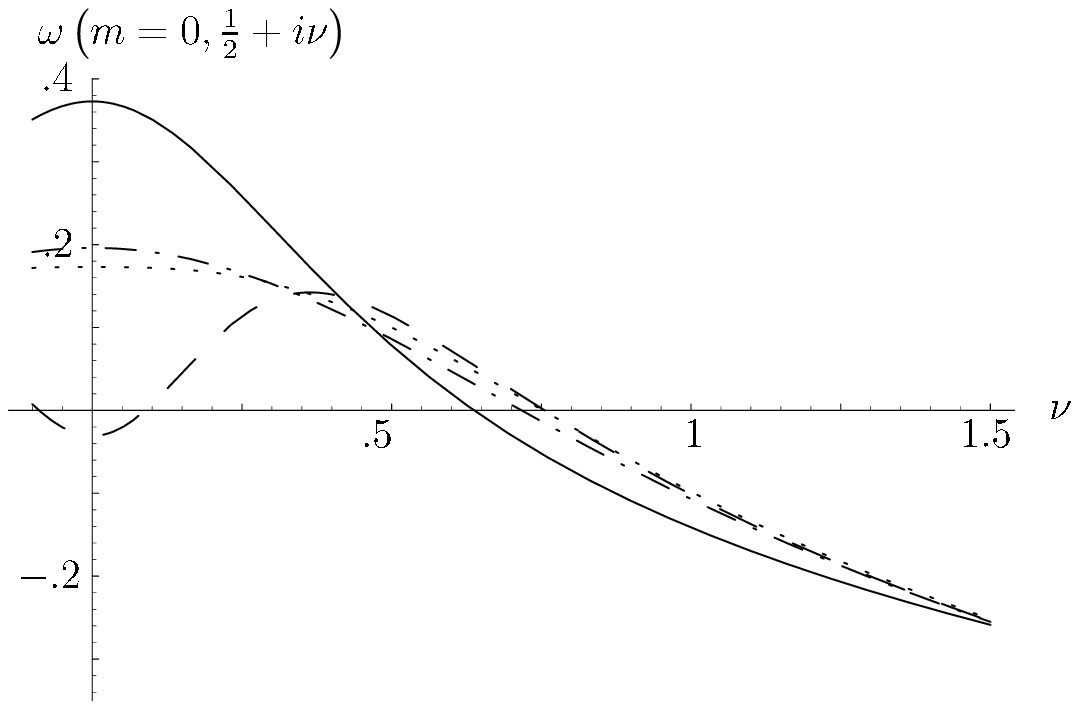}\hspace{1cm}
  \includegraphics[width=5.5cm]{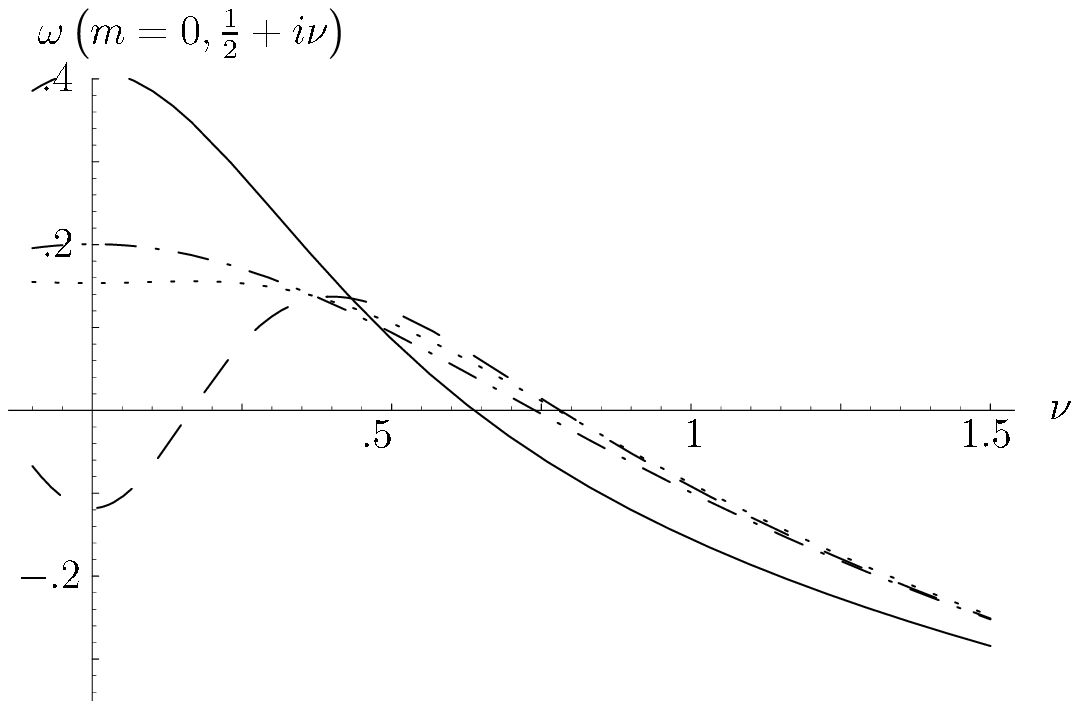}\\
  \vspace{1cm}
  \includegraphics[width=5.5cm]{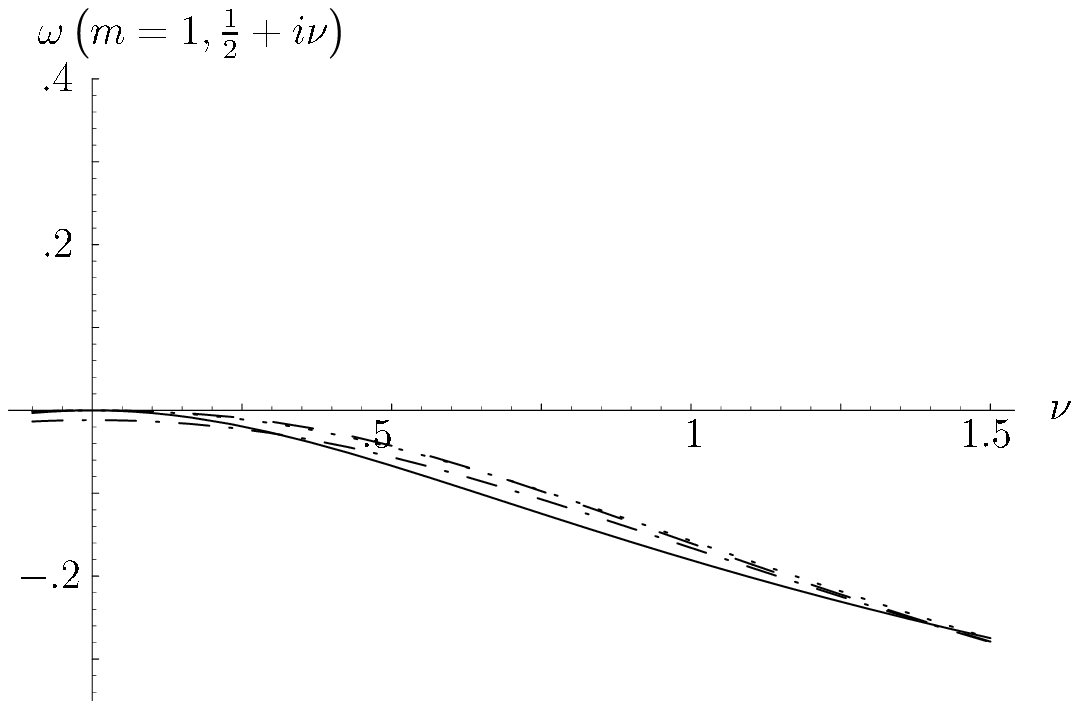}\hspace{1cm}
  \includegraphics[width=5.5cm]{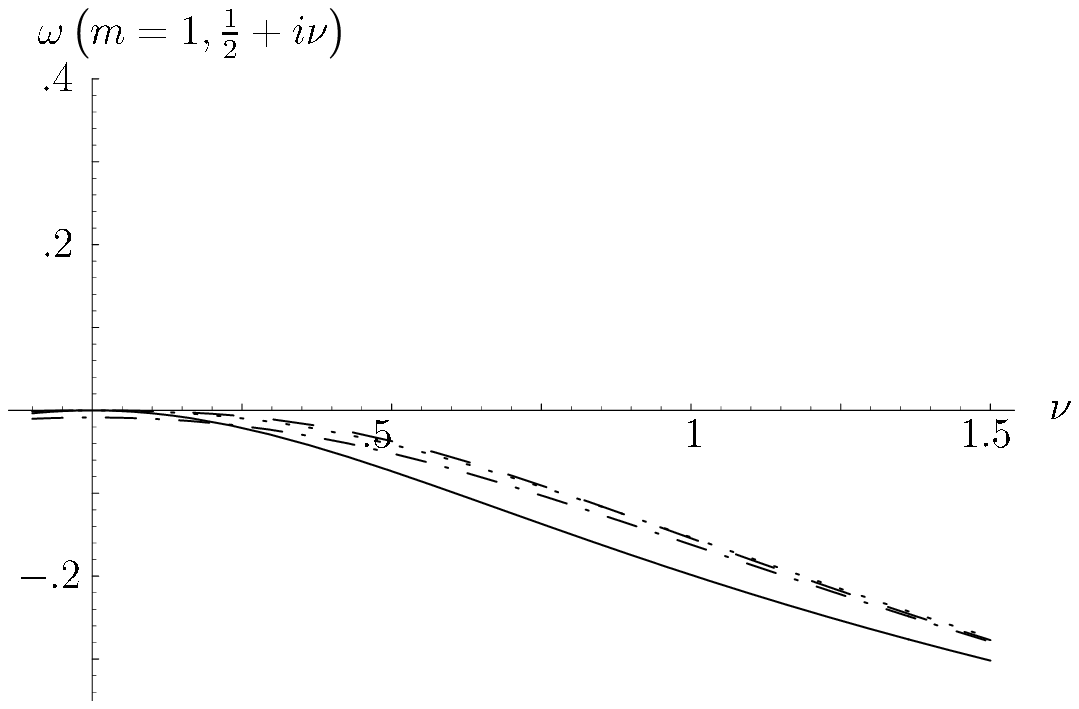}\\
  \vspace{1cm}
  \includegraphics[width=5.5cm]{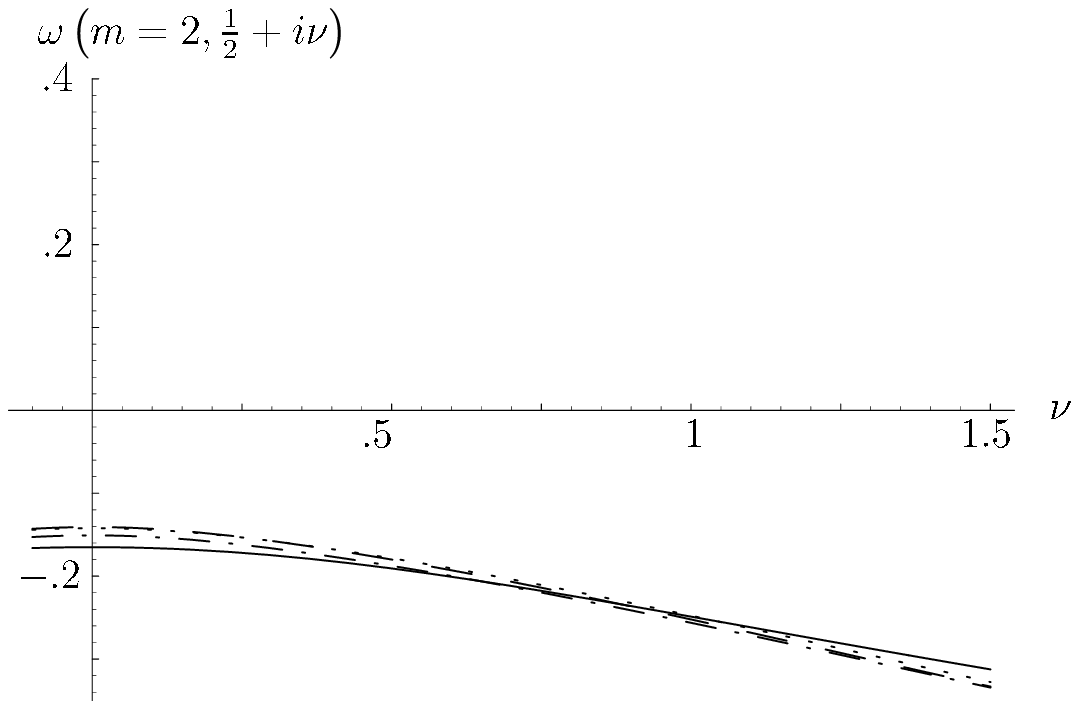}\hspace{1cm}
  \includegraphics[width=5.5cm]{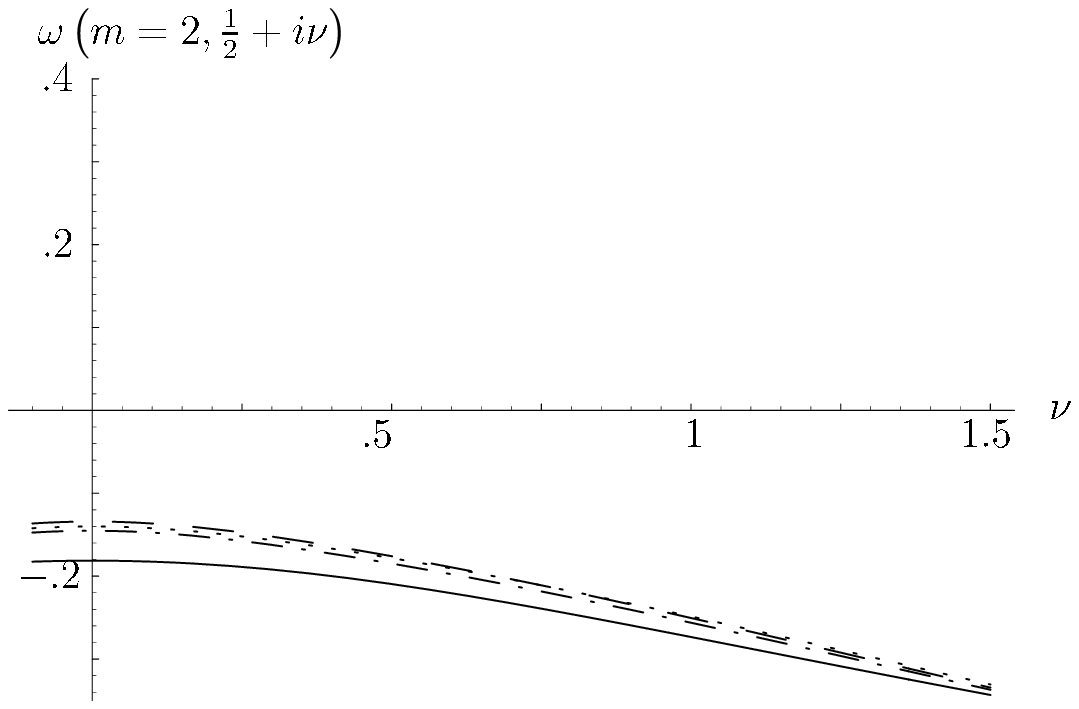}
  \caption{Comparison of kernels $\omega(m,\frac{1}{2}+i\nu)$ including the impact factor contribution at a scale $\mu=30{\rm GeV}$ as a function of $\nu$. The left column shows the results in $\overline{\rm MS}$-renormalization and the right one the results in GB-renormalization.  Shown are the  NLO expression (dashed), and the resummed ones for scheme 1 (dotted) and scheme 3 (dashed-dotted). }
  \label{fig:impm2mu30}
\end{figure}

When we extend the resummation prescription to the exponent of the coefficients $\mathcal{C}_n$ defined in Eq.~\eqref{eq:cnll2}, the kernel is augmented by an impact factor contribution $\sim \chi_0/(\gamma(1-\gamma))$ with its implications shown in section \ref{sec:impfactor}. The shapes of the LO, NLO, and the resummed version of the kernel remain the same compared to the pure form without impact factor contribution (see Fig.~\ref{fig:impm2mu30}), but scheme 1 and 2 are more sensible to the depth of the NLO-dip around $\nu=0$, because in these schemes terms such as $\asbar^n/\gamma^{n+1}$ have not been resummed \cite{Salam:1998tj}. For scheme 1 and to this can even change the maximum at $\nu=0$ back to a local minimum. Since schemes 3 and 4 are more stable, we consider them to be the best candidates for a meaningful resummed kernel.

\clearpage{\pagestyle{empty}\cleardoublepage}

\backmatter
\addcontentsline{toc}{chapter}{Literature}

\fancyhead{} \fancyfoot{} \fancyhead[LO]{Literature} \fancyhead[RE]{Literature} \fancyhead[RO,LE]{\thepage} \pagestyle{fancy}

\bibliographystyle{hunsrt}

\clearpage{\pagestyle{empty}\cleardoublepage}
\chapter*{Acknowledgments}
\addcontentsline{toc}{chapter}{Acknowledgments}

First of all, I would like to thank my supervisor Prof. Jochen Bartels for his guidance and support. His constructive criticism put me on the right track, and his explanations greatly sharpened my understanding of physics.

I am also very grateful to Agust\'in Sabio Vera for very fruitful and pleasant collaborations. I greatly benefited from instructive and motivating discussions with him.

Furthermore, I wish to thank Prof. Lev N. Lipatov for very helpful discussions.

Diverse conversations broadened and deepened my physical knowledge. Hence I am greatly indebted to a number of colleagues, in particular
Grigorios Chachamis, 
Didar Dobur, 
Frank Fugel, 
Martin Hentschinski, 
Hannes Jung, 
Torben Kneesch, 
Krzysztof Kutak, 
Leszek Motyka, 
Falk Neugebohrn, 
Michael Olschewsky, 
and
Jan Piclum.

Moreover, I am grateful to the members of the II.~In\-sti\-tut f\"ur Theo\-re\-ti\-sche Phy\-sik and the DESY theory group for creating a very pleasant and stimulating working atmosphere.

This work was supported by the Graduiertenkolleg ``Zuk\"unftige Entwicklungen in der Teilchenphysik''.

Finally, I would like to thank my parents and brothers for their vital backing and support.

\end{document}